\documentclass[aps,12pt,final,notitlepage,oneside,onecolumn,nobibnotes,%
nofootinbib,superscriptaddress,noshowpacs,centertags]%
{revtex4}
\usepackage{psfig}
\usepackage{afterpage}

\def\apj{Astrophys. J}

\def\apjs{Astrophys. J. Supp.}

\def\aa{Astro. \& Astrophys.}

\def\l{\ell}
\def\lmax{\mbox{$\ell_{\rm max}$}}

\begin{document}

\title{Statistics of the Planck CMB signal in direction of gamma-ray
 bursts from the BATSE and BeppoSAX catalogs}

\author{\firstname{M.~L.}~\surname{Khabibullina}}
\email{rita@sao.ru}
\affiliation{Special Astrophysical Observatory of RAS, Nizhnij Arkhyz, 369167 Russia}
\author{\firstname{O.~V.}~\surname{Verkhodanov}}
\email{vo@sao.ru}
\affiliation{Special Astrophysical Observatory of RAS, Nizhnij Arkhyz, 369167 Russia}
\author{\firstname{V.~V.}~\surname{Sokolov}}
\email{sokolov@sao.ru}
\affiliation{Special Astrophysical Observatory of RAS, Nizhnij Arkhyz, 369167 Russia}

\begin{abstract}
Distribution of gamma-ray bursts (GRBs) from
catalogs of the BATSE and BeppoSAX space observatories relative to the
cosmic microwave background (CMB) data by Planck space mission is
studied. Three methods were applied for data analysis: 1) a histogram
of CMB signal values in GRB directions, 2) mosaic correlation maps
calculated for GRB locations and CMB distribution, 3) calculation of an
average response in the area of ``an average
population GRB'' on the CMB map. A correlation
between GRB locations and CMB fluctuations was detected which can be
interpreted as systematic effects in the process of observations.
Besides, in averaged areas of CMB maps, a difference between
distributions of average fluctuations for short and long GRBs was
detected, which can be caused by different natures of these events.
\end{abstract}

\maketitle

\section{Inriduction}

The quality of sky surveys carried out
in the recent decade in different wavelength ranges permits us studying
the matter distribution in the observable part of the Universe on basis
of many observational effects. Beside the direct measurement of
parameters of galaxy distribution and reconstruction of the large-scale
structure, as was done in the SDSS survey
\cite{sdssIII},
there are many effects
enabling the restoration of matter distribution. Among them there are
effects of the secondary CMB anisotropy: the integrated Sachs--Wolfe
effect
\cite{swe}
caused by alteration of frequency of CMB photons in variable
gravitational potential of forming galaxy clusters and prevailing on
the scales $>$10\degr, the Zeldovich--Syunaev effect
\cite{zs}
on the
scales $<$10\arcmin\, arising in interactions between hot
electrons in galaxy clusters with CMB photons (the inverse Compton
effect), effects of scattering in the reionization epoch, and simply
the obstructive factors in the form of microwave emission of radio
sources and galaxy clusters. Gamma-ray bursts are also an independent
sign of the Large Scale Structure (LSS) allowing us tracing the matter
distribution at cosmologic distances.

On the other hand, the observable uniform
distribution of gamma-ray bursts on the projection celestial sphere, as
well as the distribution of the main bulk of radio sources (excepting
the faintest ones related to the nearest galaxies) demonstrates the
cosmological principle requiring the Universe to be uniform and
isotropic irrespective of an observer's location
\cite{princ_cosm}.
Observations show that the size of the largest structures is of order
of 400\,Mpc \cite{cosm_princ}.
On lower scales, especially at low redshifts ($z < 0.1$),
the matter is distributed anisotropically and
inhomogeneously. However, the search for such structures continues at
$z < 1$ also (e.g., see \cite{rudnick,springel}). Note that integral and
statistical characteristics of CMB distribution determined from
correlation maps with SDSS galaxy locations show the presence of the
distinguished scales of 2--3 degrees within the redshift range
$z=0.8-2$,
which corresponds to the linear scale 60\,Mpc and can be
interpreted as the maximum size of a heterogeneity cell
\cite{yadav,sarkar,labini_baryshev,jubilee,cmb_sdss_cell}.
This agrees with the model of activity of radio sources in the range
$z\sim1-2$ \cite{hierar,bh_rita},
where variations of gravitational potential in forming
clusters are expected. In this respect the comparative distribution of
CMB maps extrema and GRBs on the celestial sphere is interesting as a
new indicator of LSS signature on CMB maps at different redshifts.
The uniform observable distribution of GRBs also enables the testing of the
cosmological principle. Besides, it is assumed that they can be used as
standard candles for estimation of distance to
objects under consideration
\cite{amati1,amati2}.
The available rather large
catalogs -- BeppoSAX\footnote{\tt http://www.asdc.asi.it/bepposax/}
(Satellite per Astronomia X, ``Beppo'' in honor of Guiseppe Occhialini)
 \cite{bepposax}
and BATSE\footnote{\tt http://www.batse.msfc.nasa.gov/batse/}
(Burst and Transient Source Experiment) \cite{batse}
comprising such objects
allow us studying the spatial distribution of these objects.

\begin{figure*} [!th]
\setcaptionmargin{5mm}
\onelinecaptionstrue
\centerline{
\psfig{figure=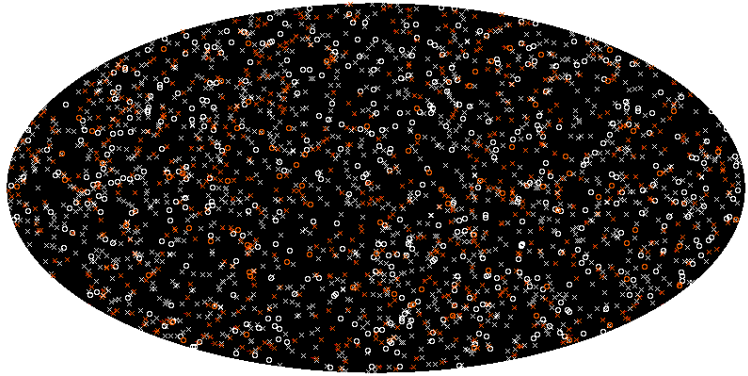,angle=-90,width=10cm}
}
\caption{
Distribution of gamma-ray bursts over the celestial sphere. BeppoSAX
data are shown by black, those of BATSE -- by white. Short bursts are
denoted by circles, long bursts -- by crosses.
}
\label{fig_grb_sphere}
\end{figure*}

\begin{figure*} [!th]
\setcaptionmargin{5mm}
\onelinecaptionstrue
\centerline{\vbox{
\hbox{
\psfig{figure=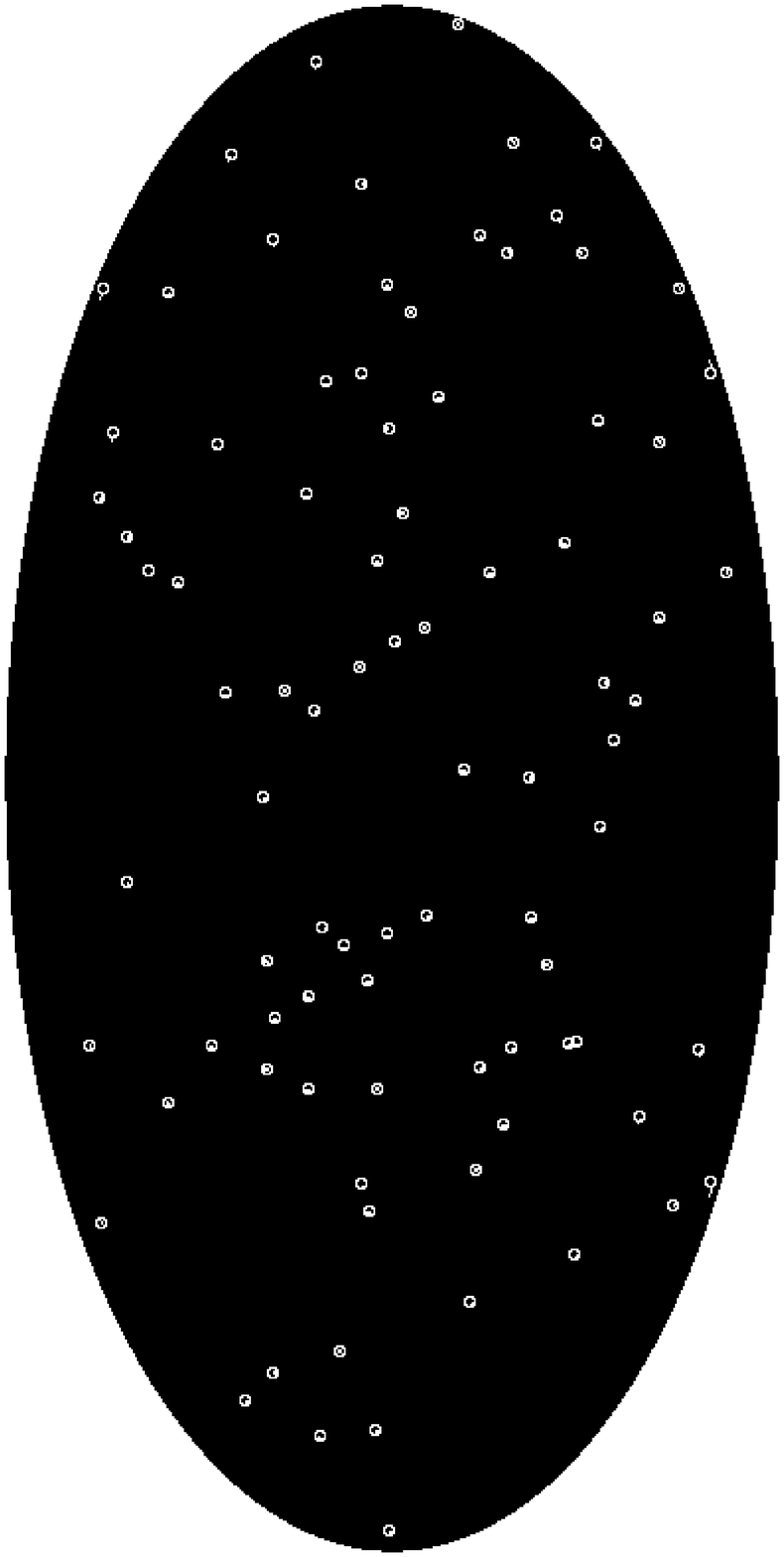,angle=-90,width=7cm}
\psfig{figure=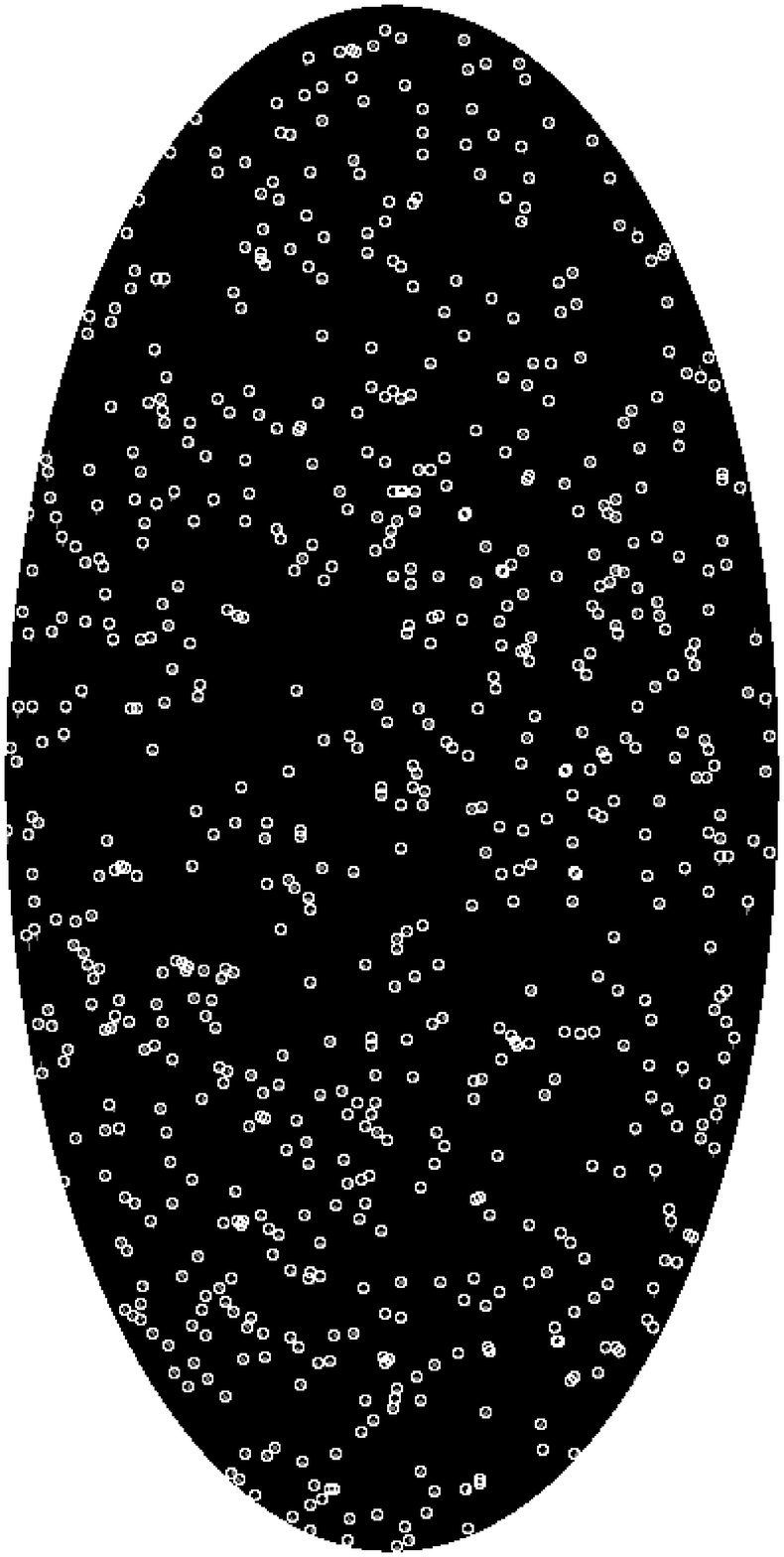,angle=-90,width=7cm}
}
\hbox{
\psfig{figure=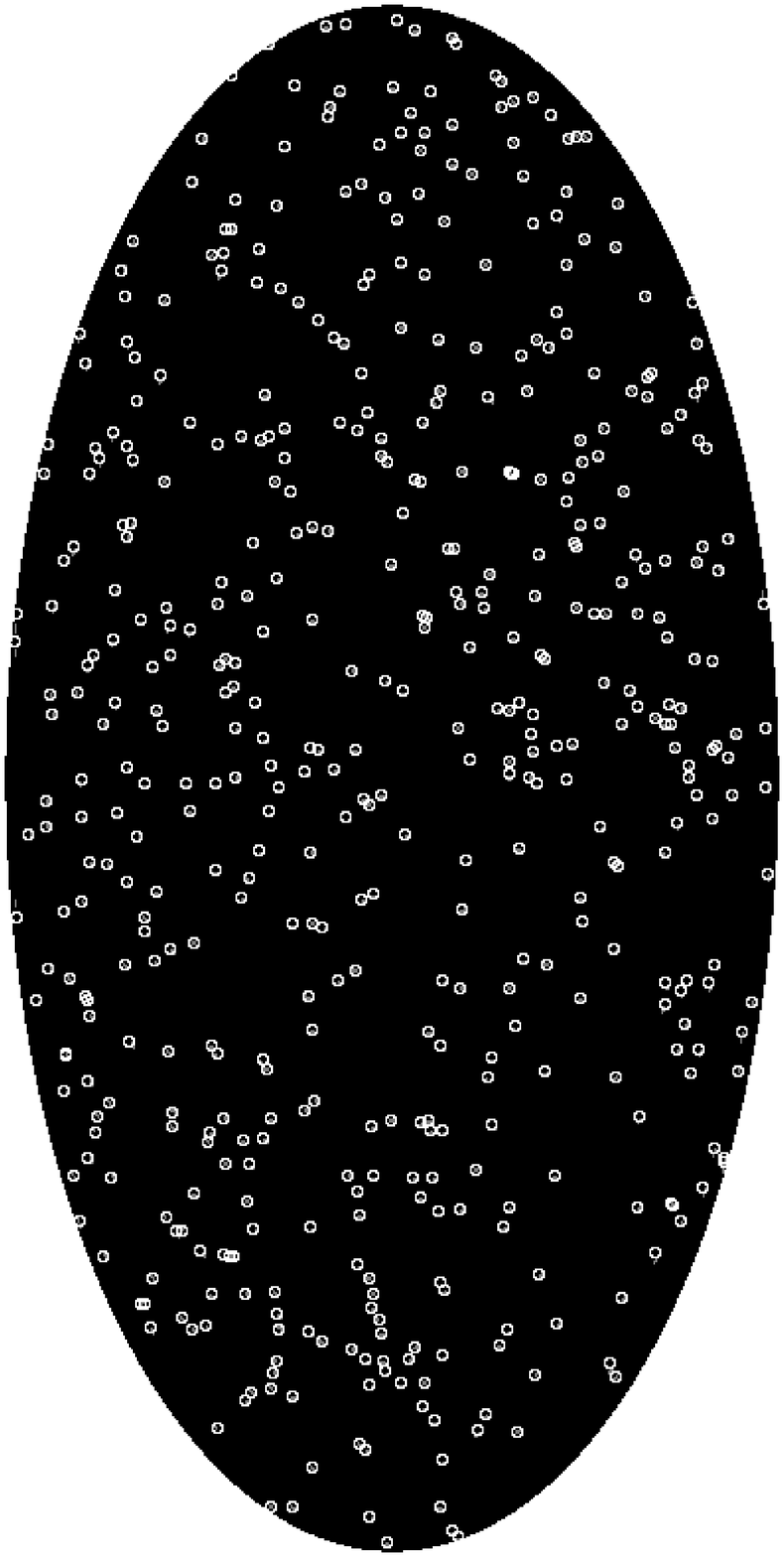,angle=-90,width=7cm}
\psfig{figure=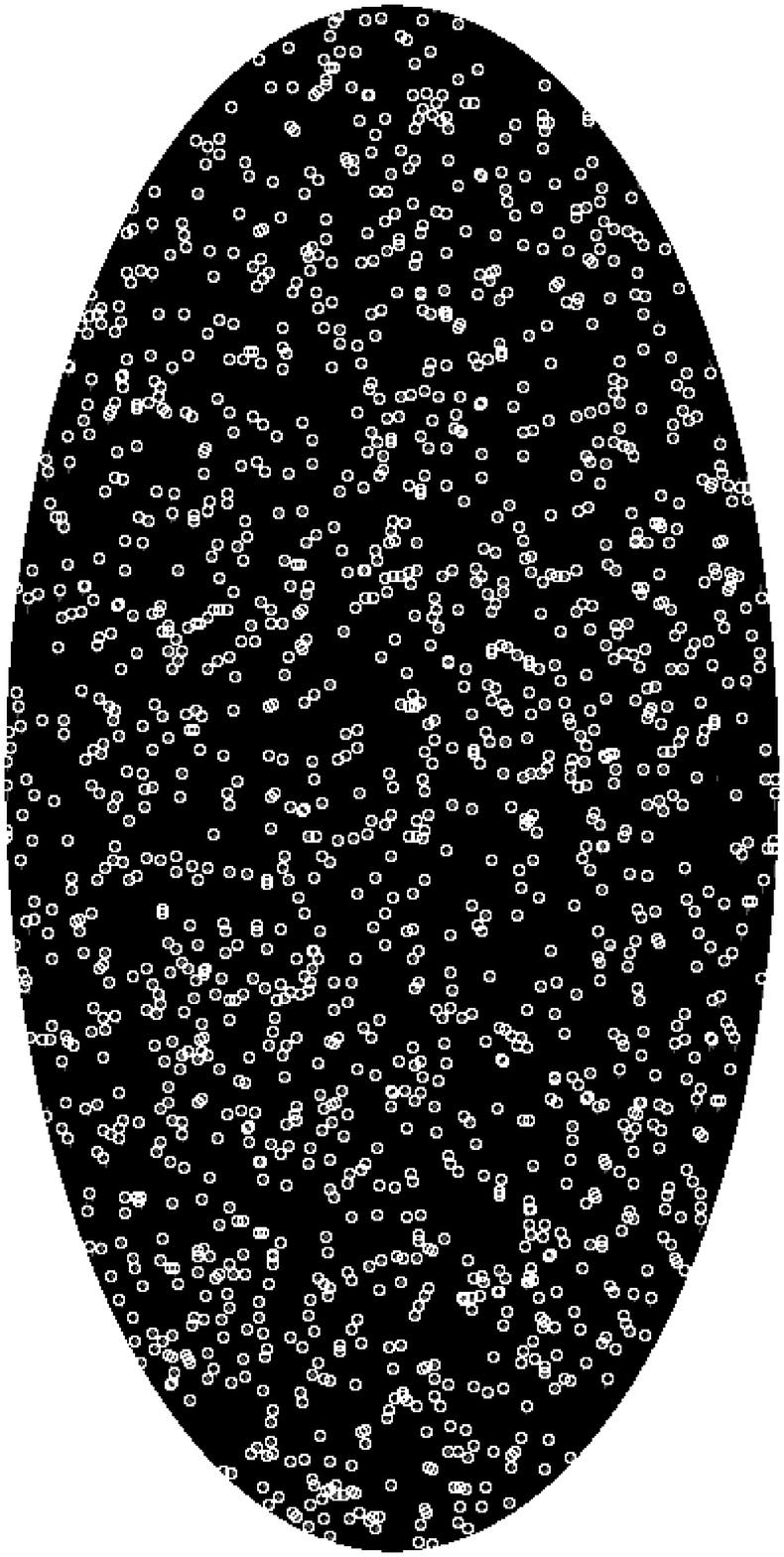,angle=-90,width=7cm}
}}}
\caption{
Distribution of GRB catalog subsamples over the celestial sphere. The
top left image shows the Beppo SAX data, $t<2$\,sec.
The top right image shows the BeppoSAX data, $t>2$\,sec.
The bottom left image presents the BATSE data, $t<2$\,sec.
The bottom right image is for the BATSE data, , $t>2$\,sec.
}
\label{fig_grbsub_sphere}
\end{figure*}
In recent years, many authors
investigated the gamma-ray distribution by many methods
\cite{vavrek,grb_apm,grg_anis,Bernui,grb_vor,v_grb,raikov_orlov}.
Paper
\cite{grb_vor}
can be marked out among them. Its authors studied GRBs of short
($t<2$\,sec),
medium (($2<t<10$\,sec) and
long ($t>10$\,sec) duration from the BATSE catalog by different
methods (by means of Voronoi tesselations, minimum spanning tree,
multifractal spectrum). For the first two groups, they discovered
deviations from homogeneity as compared with model data. On this basis,
they discuss the satisfiability of the cosmological principle. In paper
 \cite{raikov_orlov}
the locations of supernovae bursts with $z<1.4$,
as well as gamma-ray bursts, were used as probing
objects. For supernovae data, a deviation from uniform distribution on
the diagram ``CMB temperature in a source direction --- z'' was discovered
in contrast to the similar diagram
for gamma-ray bursts. The authors explain this difference by
contribution of the integral Sachs-Wolfe effect.

\begin{figure*} [!th]
\setcaptionmargin{5mm}
\onelinecaptionstrue
\centerline{\vbox{
\hbox{
\psfig{figure=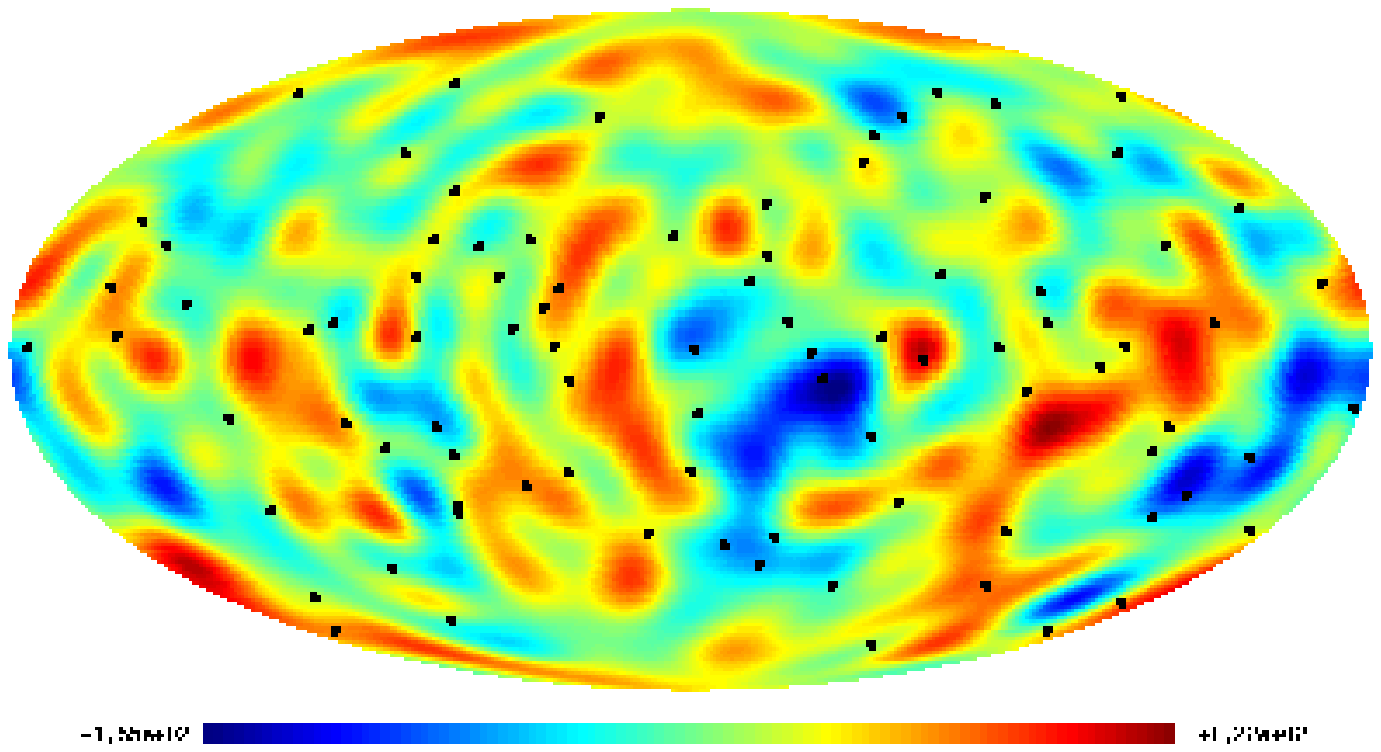,width=7cm}
\psfig{figure=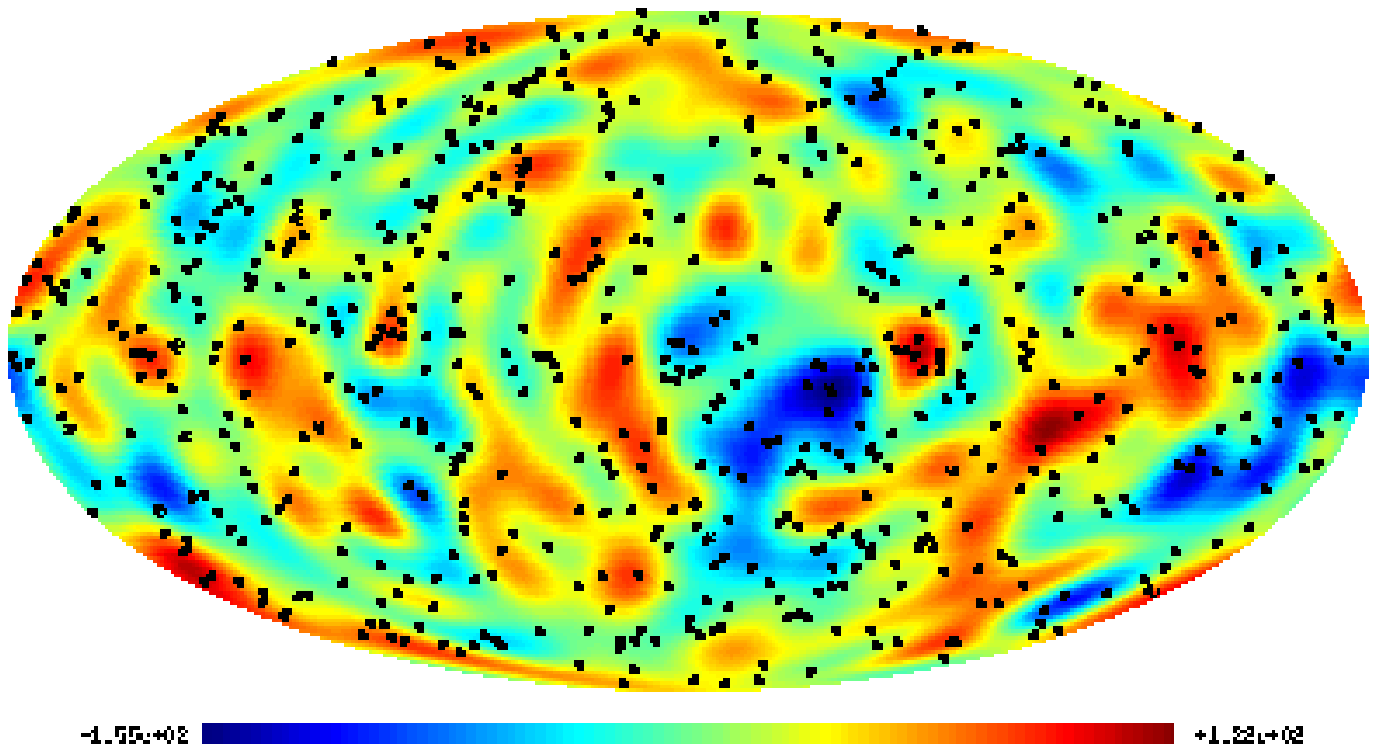,width=7cm}
}
\hbox{
\psfig{figure=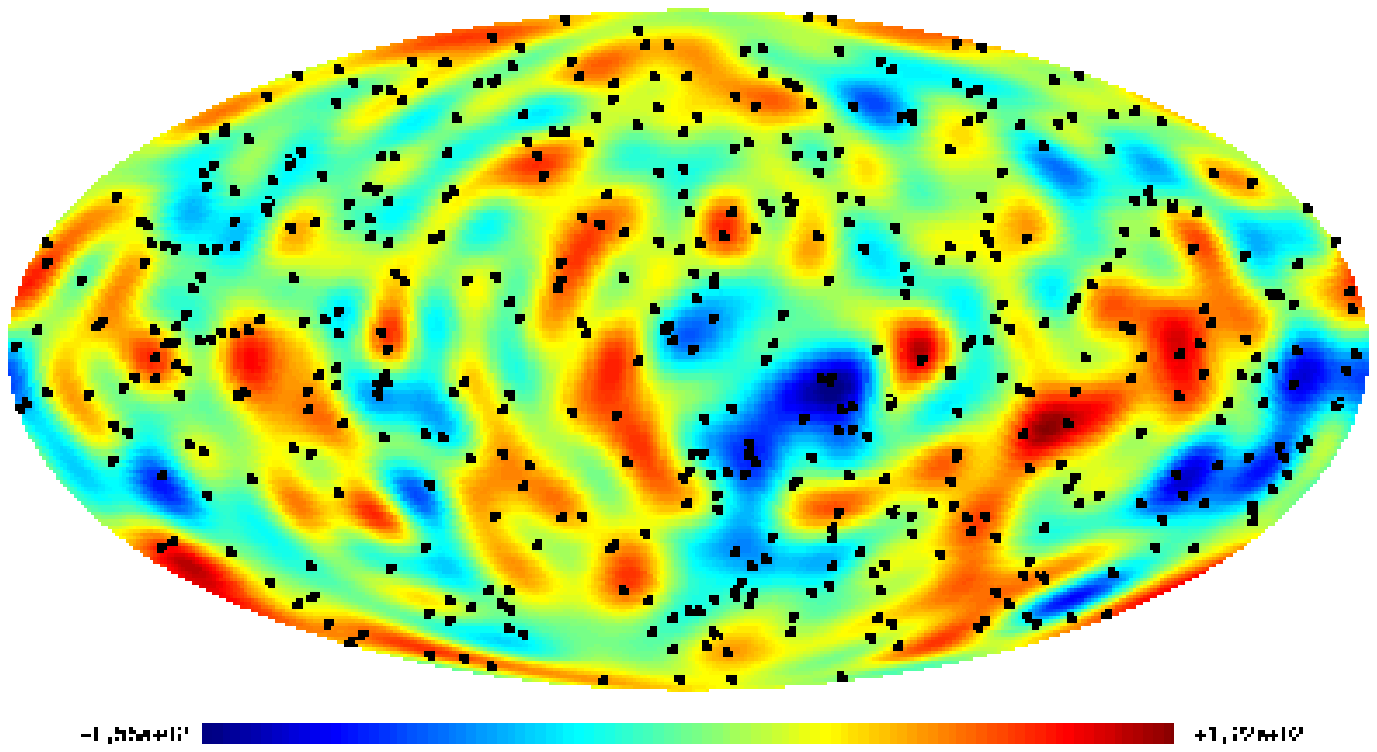,width=7cm}
\psfig{figure=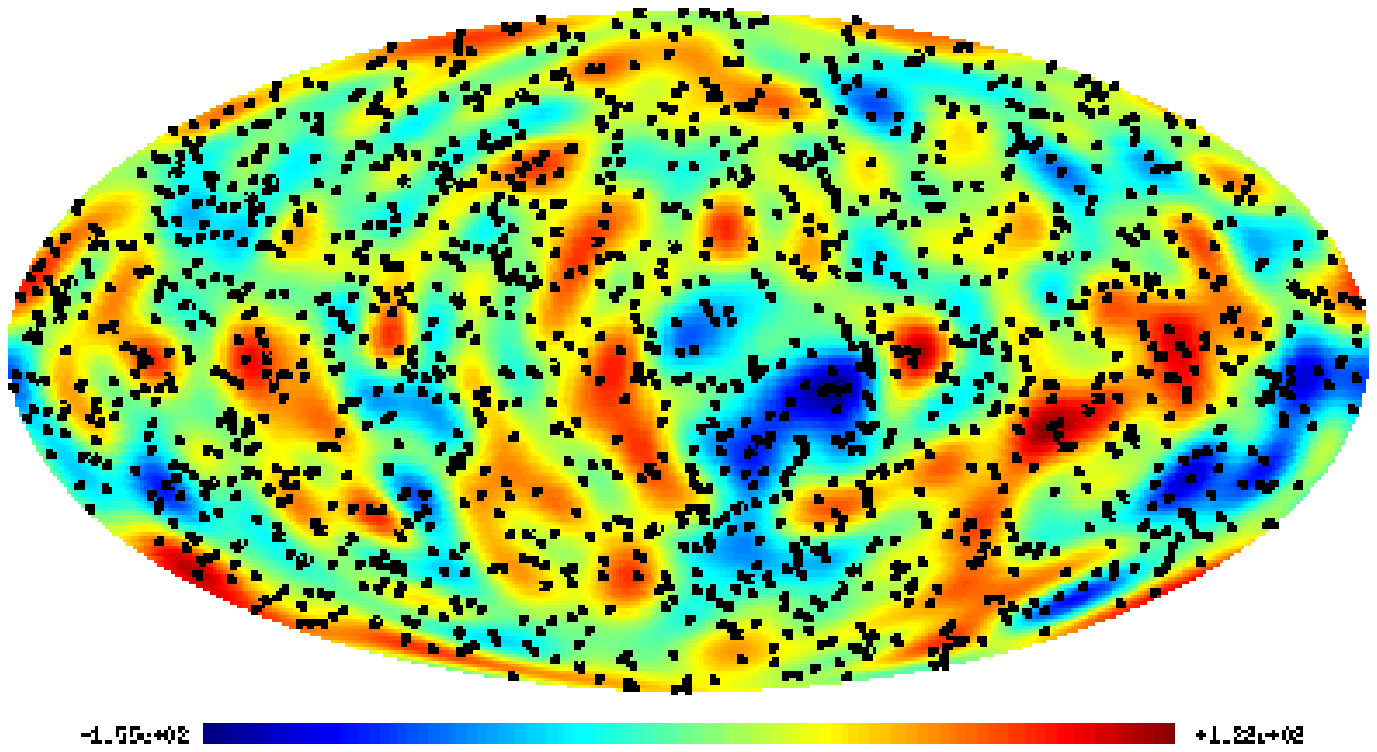,width=7cm}
}}}
\caption{
Location of GRBs from different samples on CMB maps with resolution
 $\lmax=20$.
The top left image presents the BeppoSAX data, $t<2$\,sec.
The top right image shows the BeppoSAX data, $t>2$\,sec.
The bottom left image presents the BATSE data, $t<2$\,sec.
2 sec. The bottom right image shows the BATSE data, $t>2$\,sec.
}
\label{grb_cmbL20}
\end{figure*}

\begin{figure*} [!th]
\setcaptionmargin{5mm}
\onelinecaptionstrue
\centerline{\vbox{
\hbox{
\psfig{figure=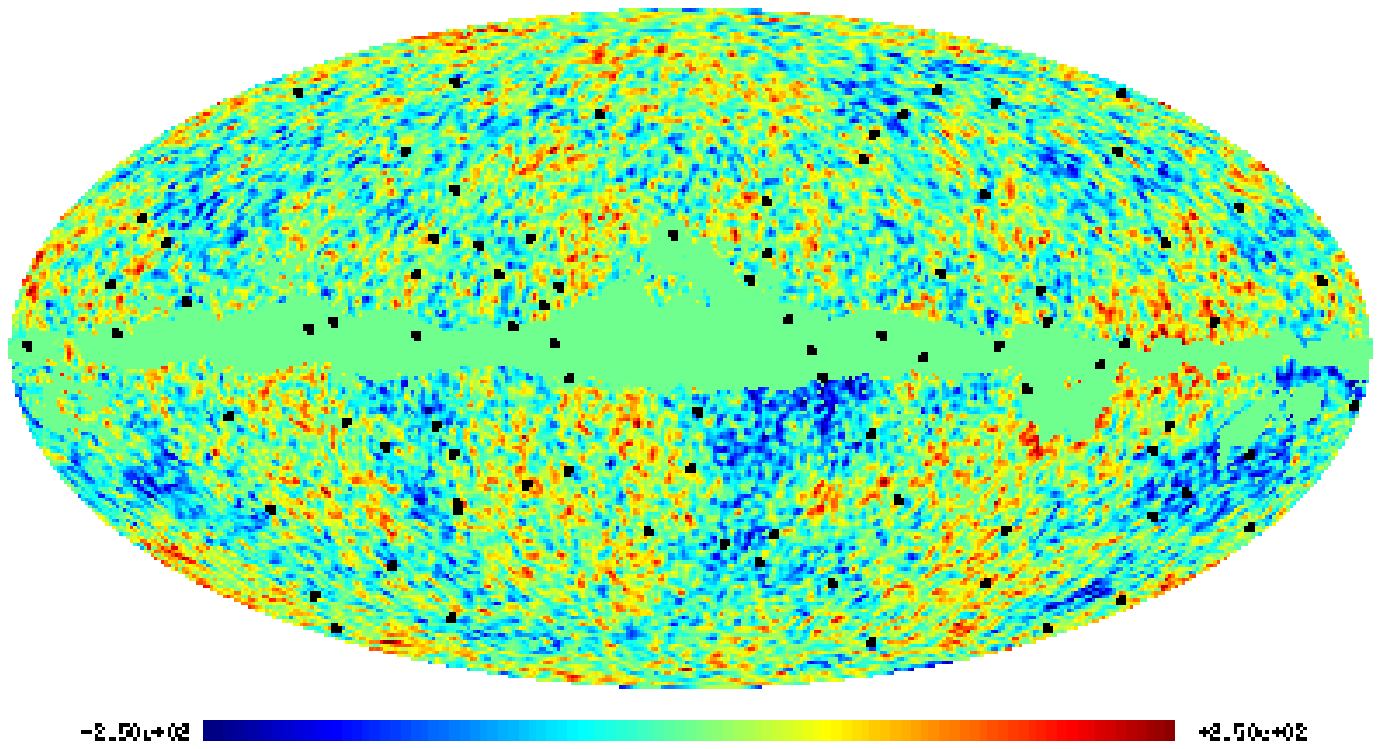,width=7cm}
\psfig{figure=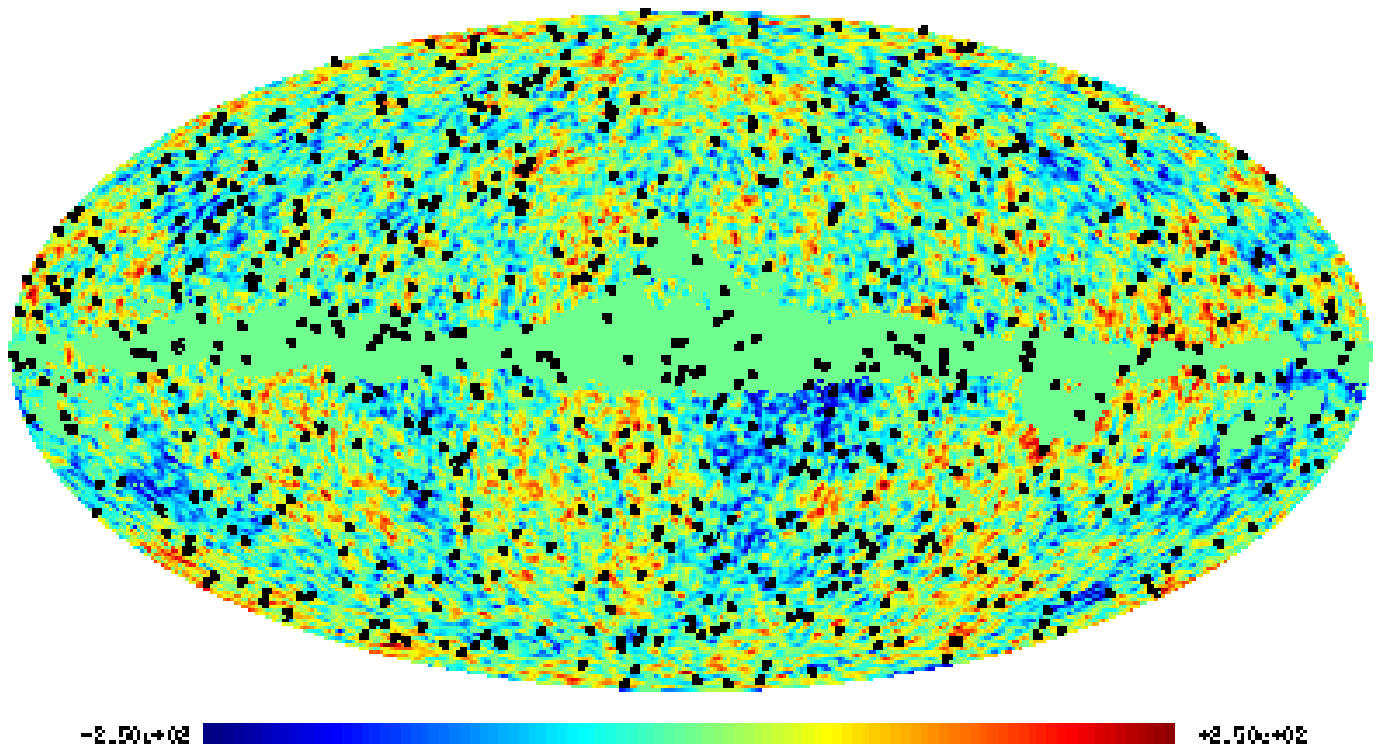,width=7cm}
}
\hbox{
\psfig{figure=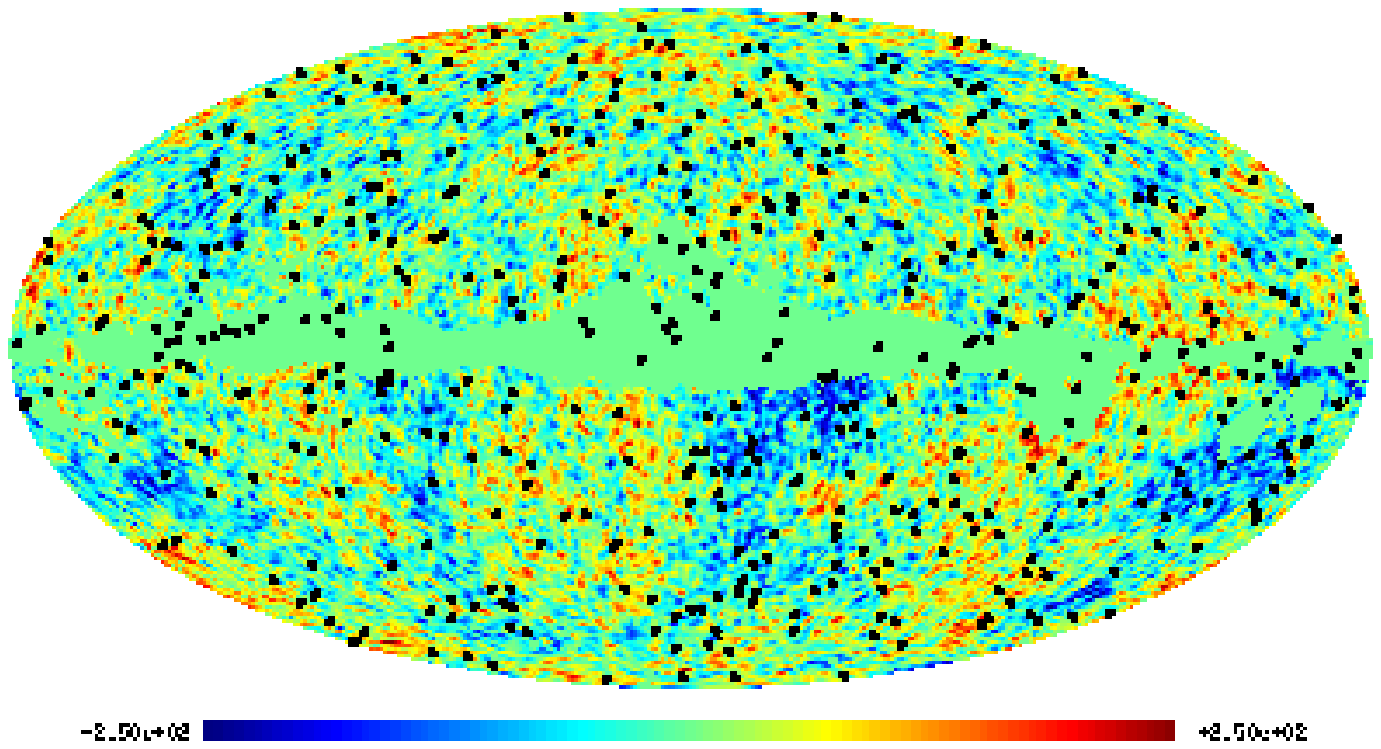,width=7cm}
\psfig{figure=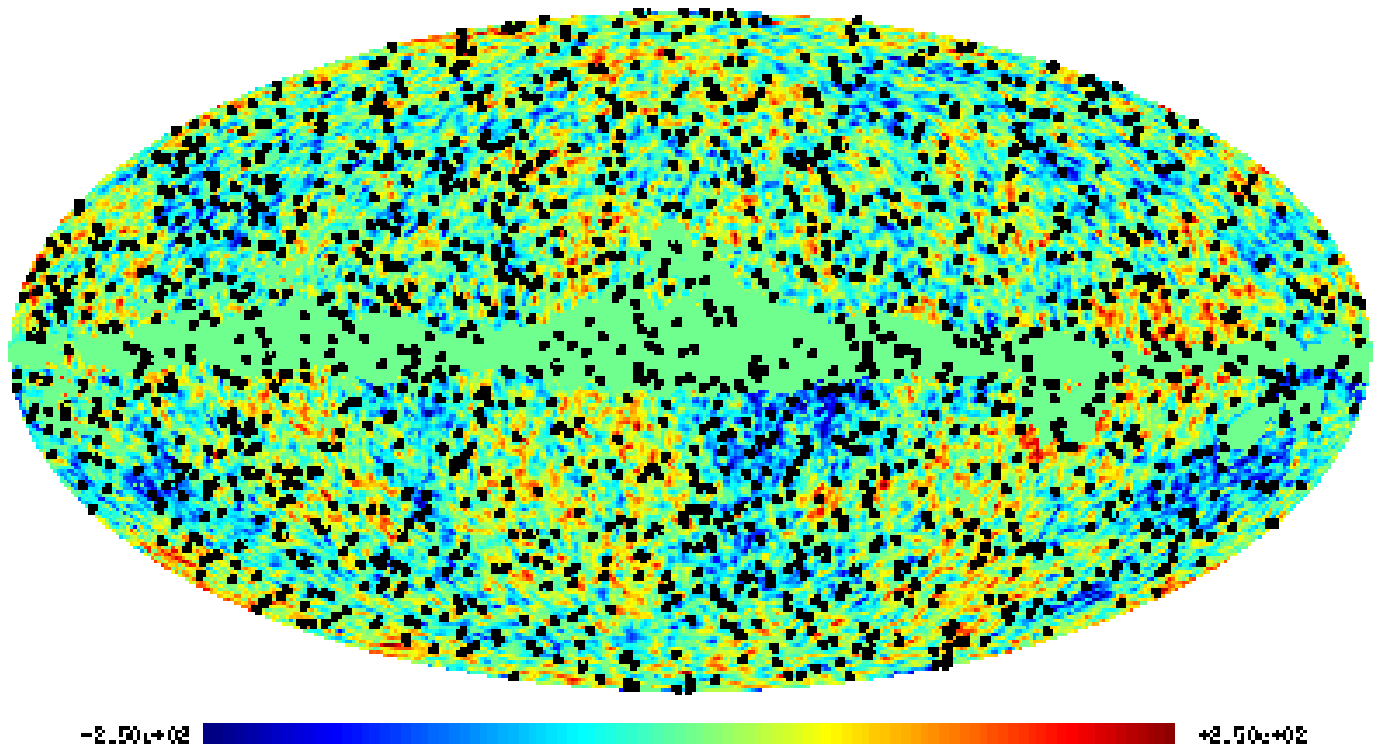,width=7cm}
}}}
\caption{
Location of gamma-ray bursts from different subsamples on CMB maps with
the resolution $\lmax=150$.
The CMB map is masked. The top left image shows the BeppoSAX data, $t<2$\,sec.
The top right image presents the BeppoSAX data, $t>2$\,sec.
The bottom left image -- the BATSE data, $t<2$\,sec.
The bottom right image -- the BATSE data, $t>2$\,sec.
}
\label{grb_cmbL150}
\end{figure*}

\begin{figure*} [!th]
\setcaptionmargin{5mm}
\onelinecaptionstrue
\centerline{\vbox{
\hbox{
\psfig{figure=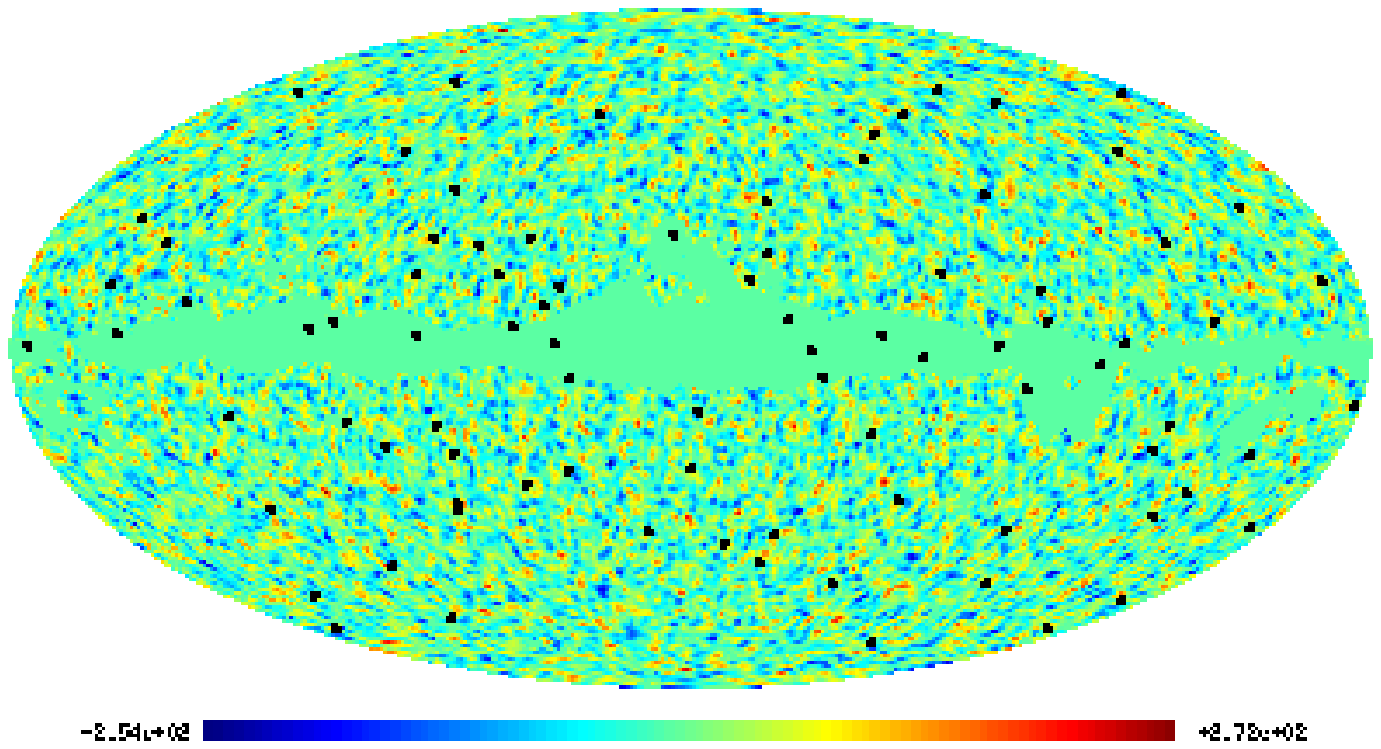,width=7cm}
\psfig{figure=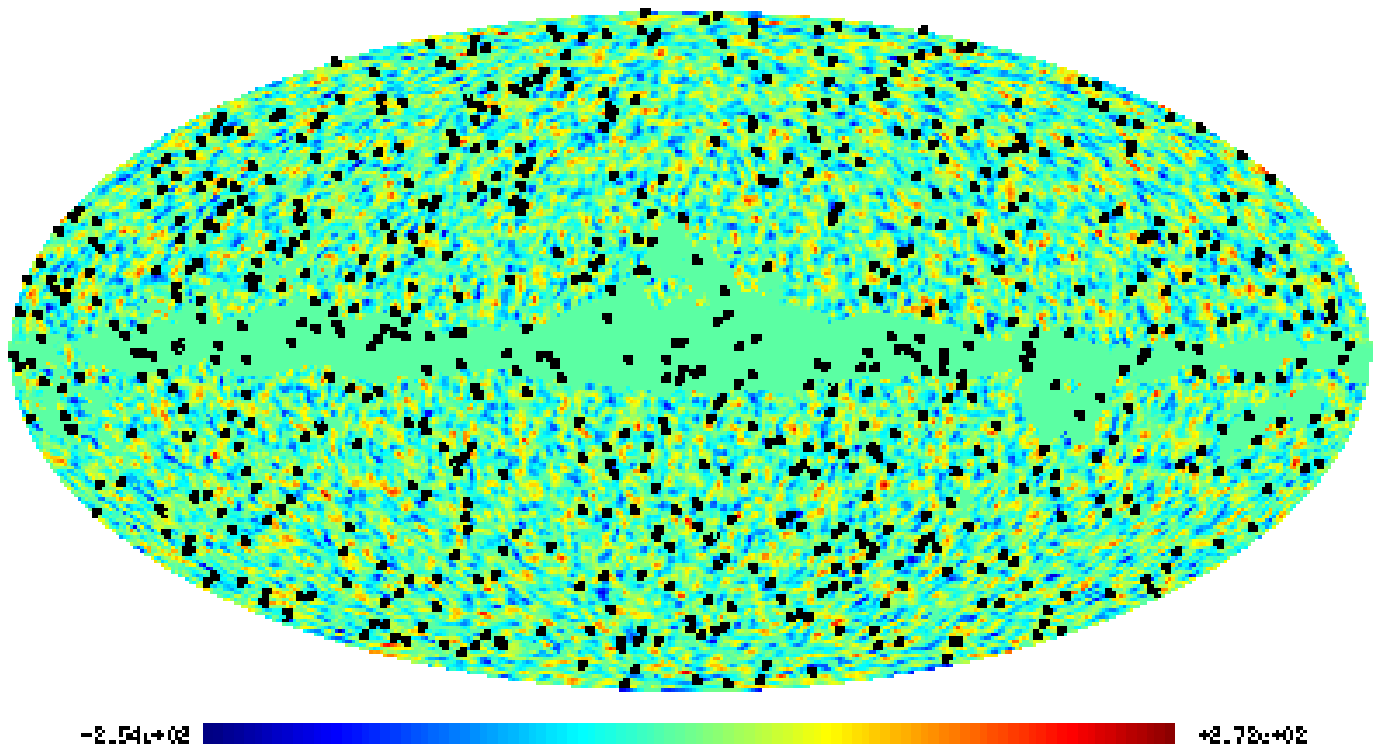,width=7cm}
}
\hbox{
\psfig{figure=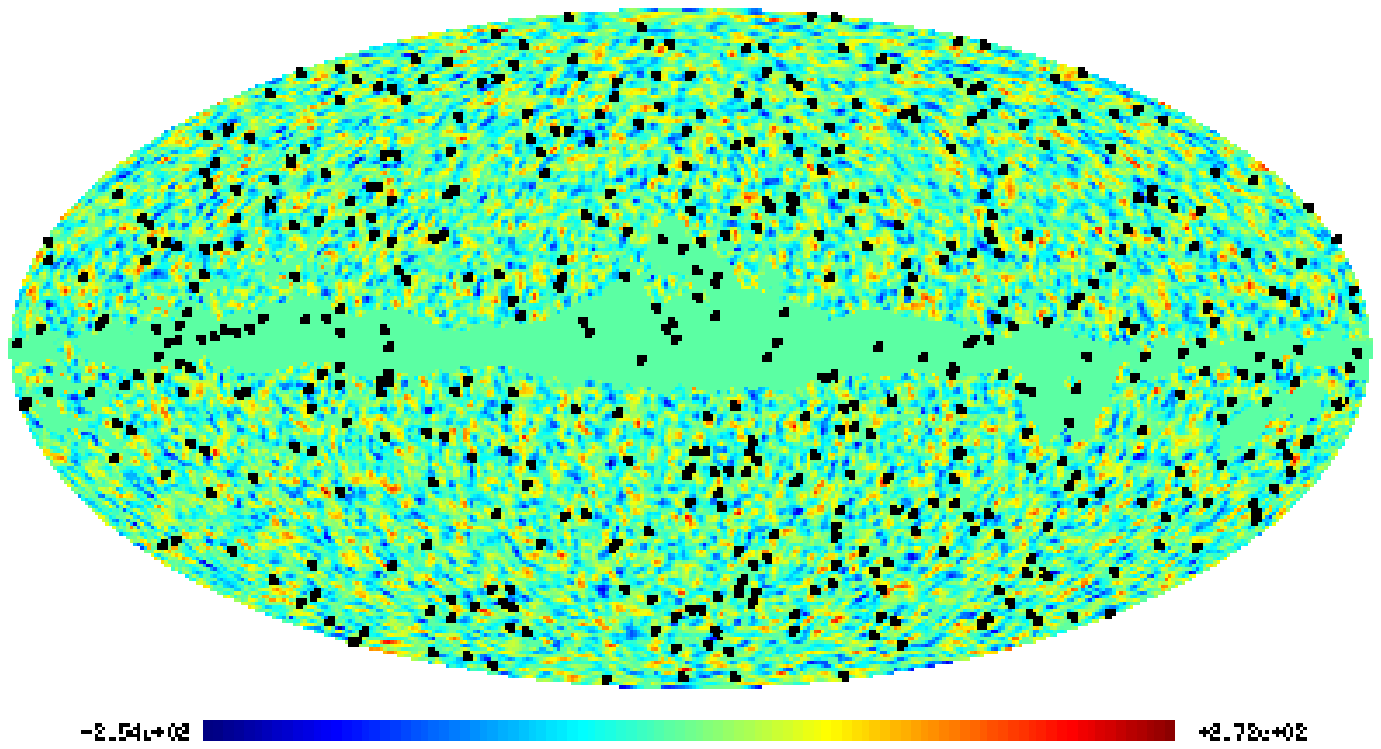,width=7cm}
\psfig{figure=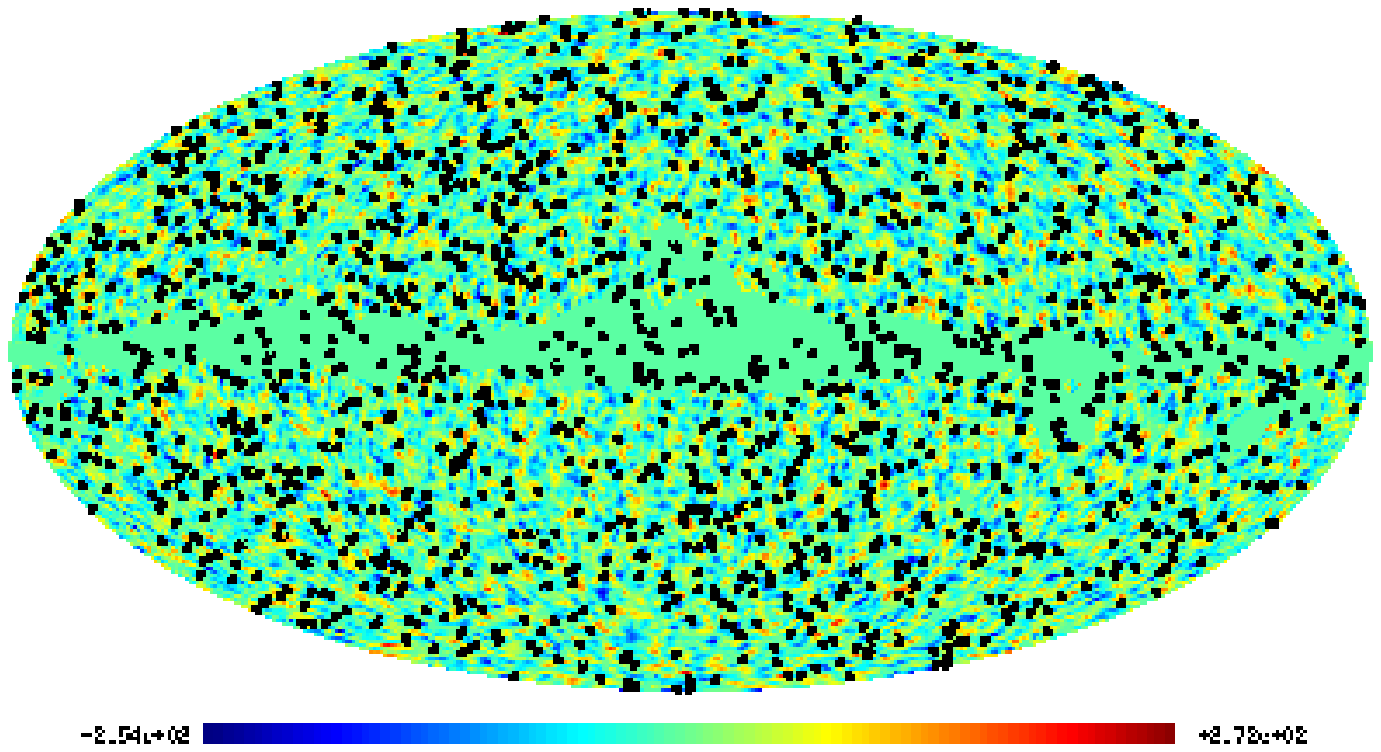,width=7cm}
}}}
\caption{
Location of GRBs from different subsamples on CMB maps in the multipole
range $20<\ell\le150$.
The CMB map is masked. The top left image shows the BeppoSAX data, $t<2$\,sec.
The top right image presents the BeppoSAX data, $t>2$\,sec.
The bottom left image -- the BATSE data, $t<2$\,sec.
The bottom right image -- the BATSE data, $t>2$\,sec.
}
\label{grb_cmbL150_20}
\end{figure*}

In our previous paper \cite{v_grb}, we studied
statistical correlation properties of sky distribution of GRBs relative
to CMB by the mosaic correlation mapping method
\citep{cormap,corr_ecl}.
The study involved WMAP\footnote{\tt http://lambda.gsfc.nasa.gov}
(Wilkinson Microwave Anisotropy Probe) data \cite{wmap7ytem},
data by the Italian--Dutch
satellite BeppoSAX (the energy range 0.1-200 keV, 781 sources) and
results of the BATSE experiment (20 keV -- 2 MeV, 2037 sources). Each
catalog was divided into two subsamples containing short (of duration
$t<2$\,sec)
and long ($t>2$\,sec) events.
Figure\,\ref{fig_grb_sphere} shows location of all gamma-ray
bursts of the BeppoSAX and BATSE catalogs. Figure\,\ref{fig_grbsub_sphere}
presents locations
of short and long GRBs from the BeppoSAX and BATSE catalogs.

If we assume that GRBs are related to massive
spiral (for long GRBs) or elliptic (for short bursts) galaxies, and,
respectively, their location is related to the large-scale structure,
then one can study statistics of CMB inhomogeneities arising due to
effects of the secondary anisotropy. Thus, the GRB locations can be
related to distribution of CMB fluctuations (e.g., revealing themselves
by deviations from statistical anisotropy) in projection to the sphere
of regions where GRBs were registered. Since in most cases the main
problem when studying GRBs is the large size of error boxes in
determination of source coordinates (of order of 1\degr$\times$1\degr), we
worked with maps smoothed to 1\degr. Our previous work with the WMAP data
resulted in discovery of a correlation between CMB peaks and GRB
locations, which, in particular, can be caused by systematic effects.
The detected correlation of GRB locations and CMB distribution is
sensitive to the equatorial coordinate system and can be caused, for
example, by the fact that the microwave radiation of the Earth gets to
far side lobes of the antenna beam.

This paper continues our previous one \cite{v_grb}, in
which the data from the WMAP archives \cite{wmap7ytem} were used.
In this work, we
applied and developed our approach for data of the Planck space
mission
PlanÓk\footnote{http://www.sciops.esa.int/wikiSI/planckpla/index.php?titl
e\=Main\_Page\&instance=Planck\_Public\_PLA}
\cite{planck_rev},
specifically, for the SMICA map  \cite{planck_sep}.

Below we apply several statistical approaches
to study distribution of gamma-ray bursts over the sphere.
Section \ref{cmb_grb}
deals with CMB signal statistics in the region of GRBs.
In Section \ref{cmb_corr_grb}, we
investigate mosaic correlations of CMB maps (Planck SMICA) and GRB
locations. Further (Section \ref{cmb_stack_grb}) we use the averaging
procedure
(stacking) of CMB map fields in direction of a gamma-ray burst to
estimate an average ``population''
microwave signal. The obtained results are discussed in Conclusion.

\section{Statistics of CMB signal in region of gamma-ray bursts}
\label{cmb_grb}

The SMICA map \cite{planck_sep} of the Planck experiment
CMB was restored from multifrequency observations obtained with the
High Frequency Instrument (HFI) in bands at 100, 143, 217, 353, 545,
857 GHz and with the Low Frequency Instrument (LFI) in bands at 30, 44,
70 GHz. Resolution of the CMB map was $\sim$5\arcmin. In spite of the fact
that the Planck mission is secondary with respect to another NASA space
mission --- WMAP (Wilkinson Microwave Anisotropy Probe), its
observational characteristics are better. Among them one can mark out a
higher resolution (by 3 times), which gave an opportunity to measure
the angular power spectrum to higher harmonics (i.e. to higher values
of $\l$), a higher sensitivity (by 10 times) and 9 frequency bands
improving the
procedure of separation of background components. These Planck
parameters allowed us obtaining new, practically independent (of WMAP)
observational data. In this paper we used the SMICA map smoothed to
$\lmax= 150$, and in a number of cases we applied the mask
Mask-RulerMinimal\_2048\_R1
\cite{planck_rev}.

\begin{figure*} [!th]
\setcaptionmargin{5mm}
\onelinecaptionstrue
\centerline{\vbox{
\hbox{
\psfig{figure=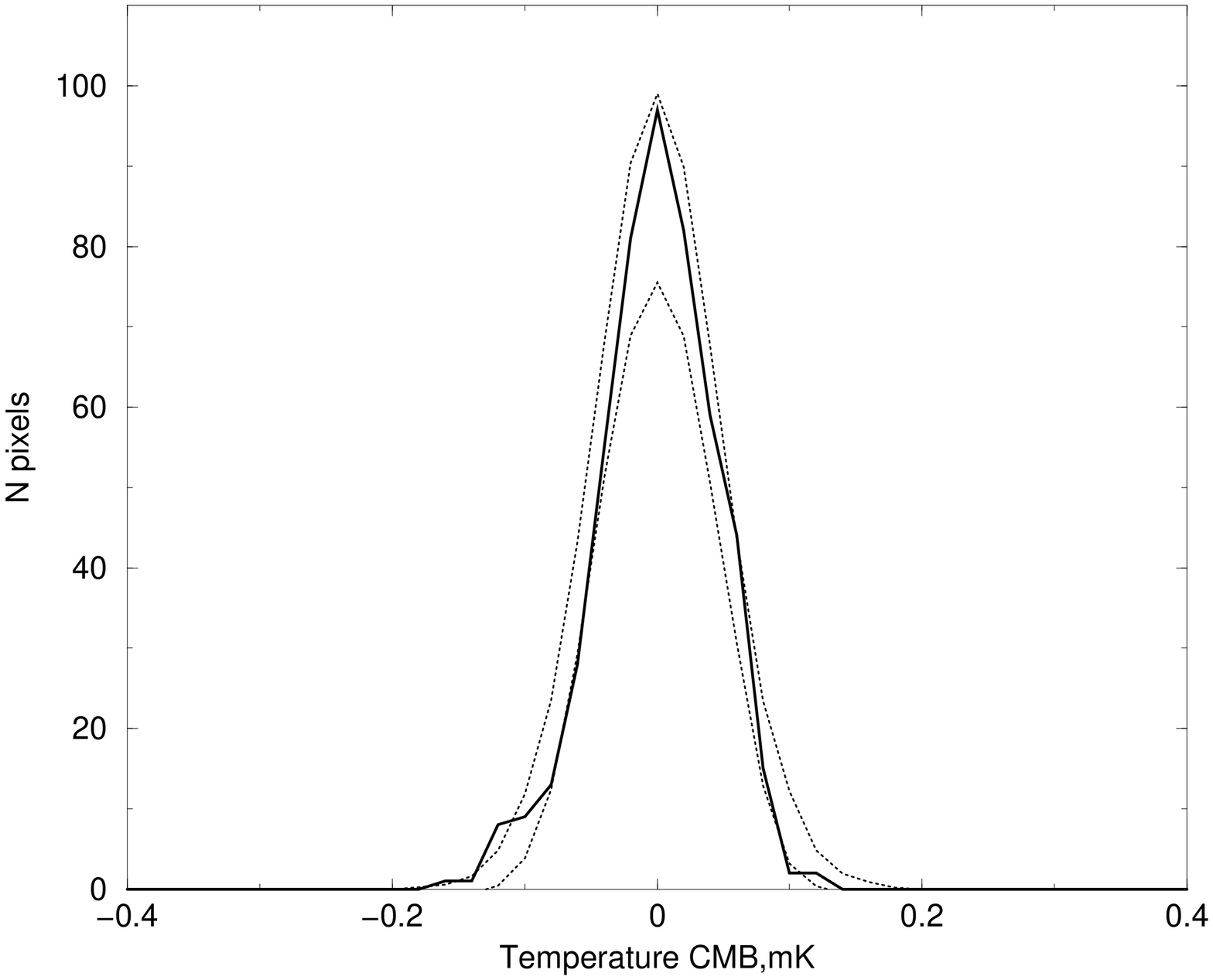,angle=-90,width=5.5cm}
\psfig{figure=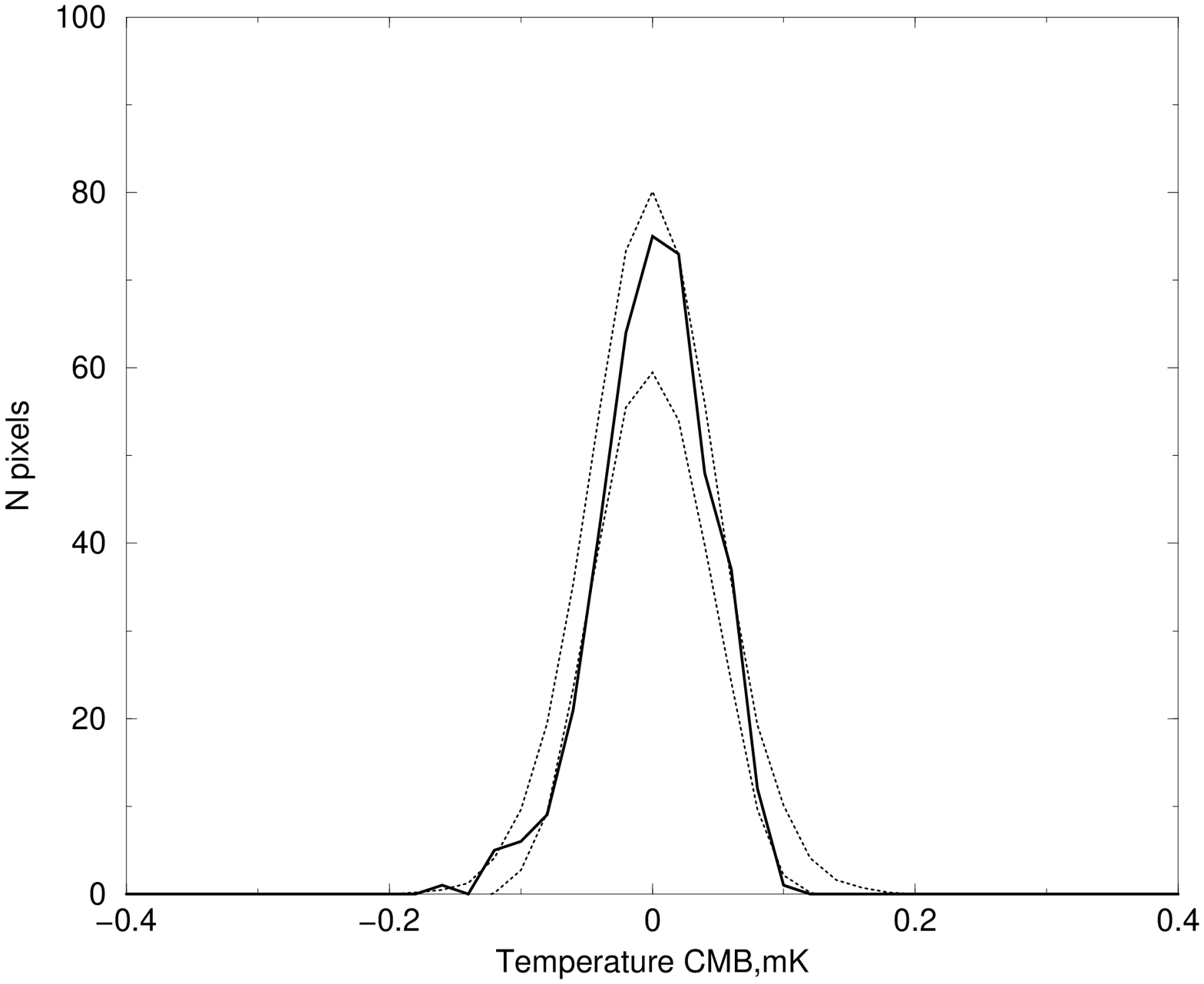,angle=-90,width=5.5cm}
}
\hbox{
\psfig{figure=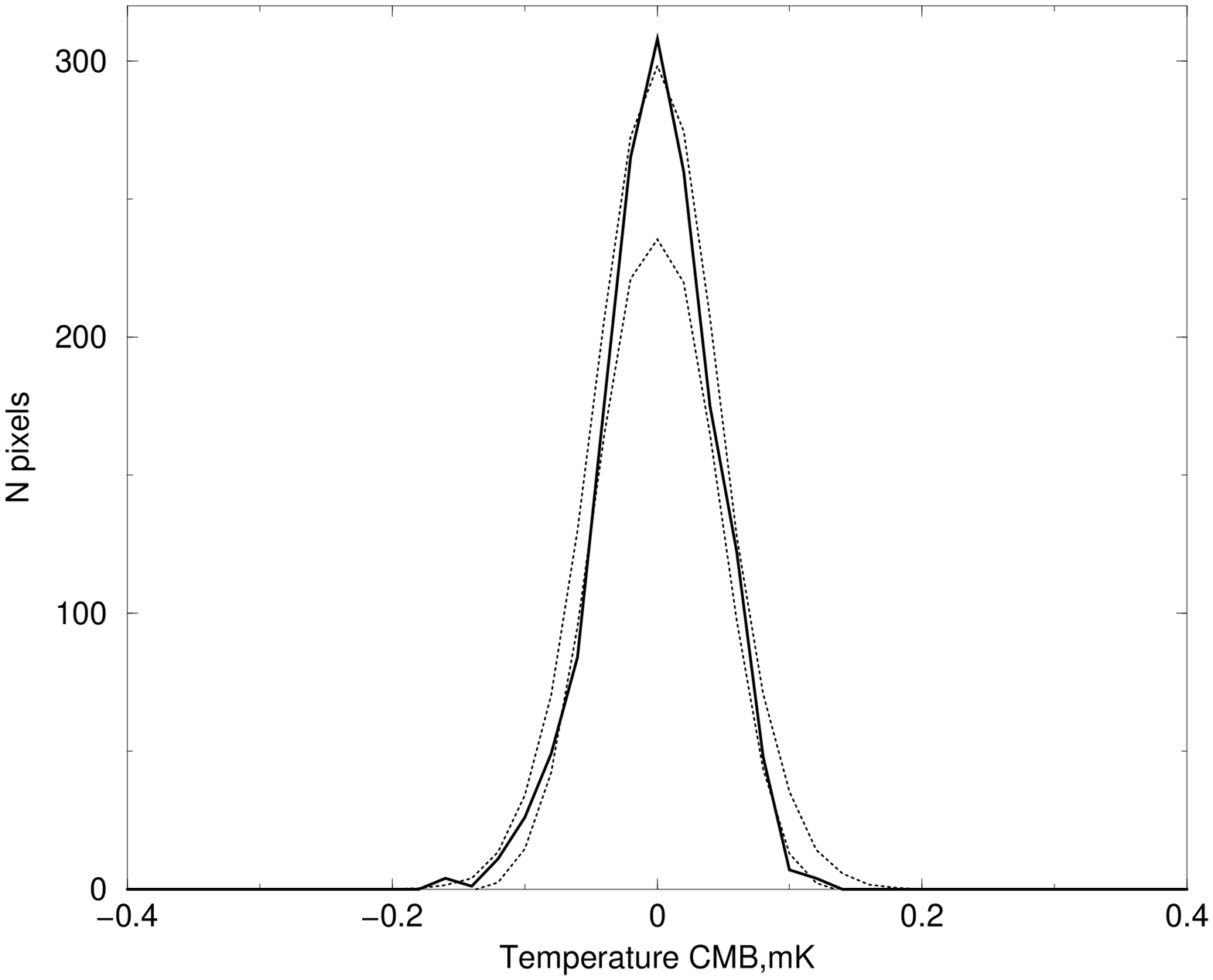,angle=-90,width=5.5cm}
\psfig{figure=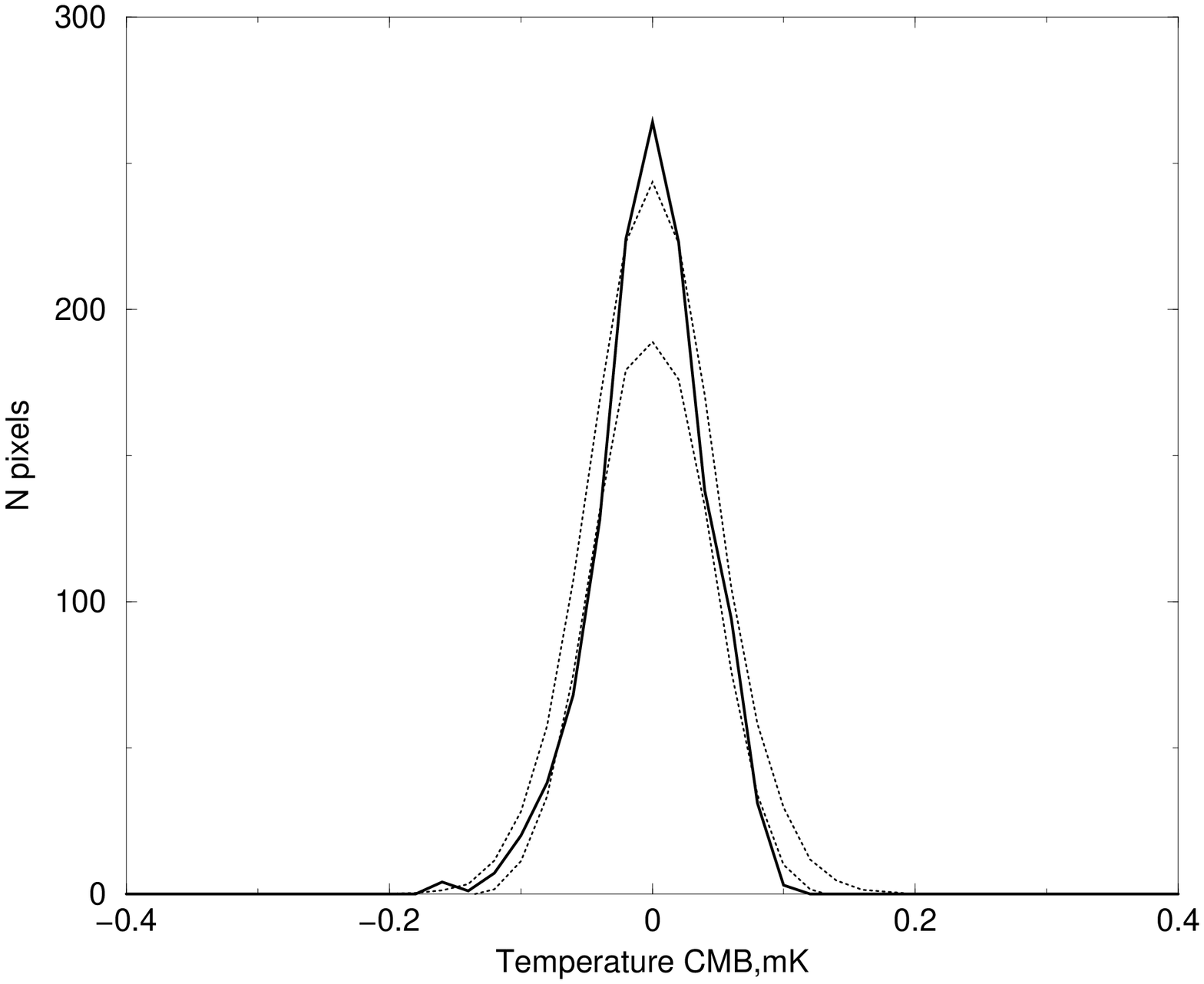,angle=-90,width=5.5cm}
}
\hbox{
\psfig{figure=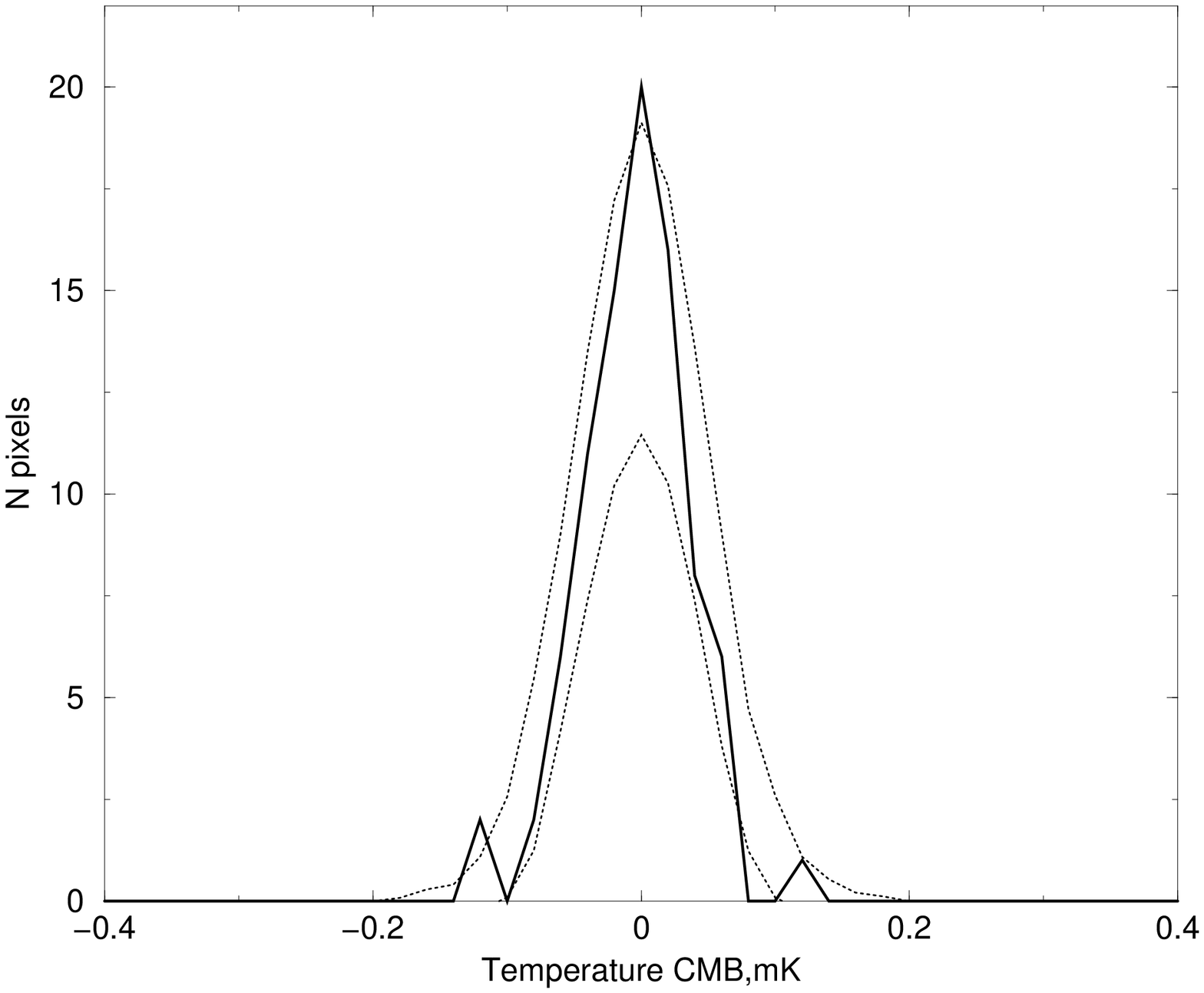,angle=-90,width=5.5cm}
\psfig{figure=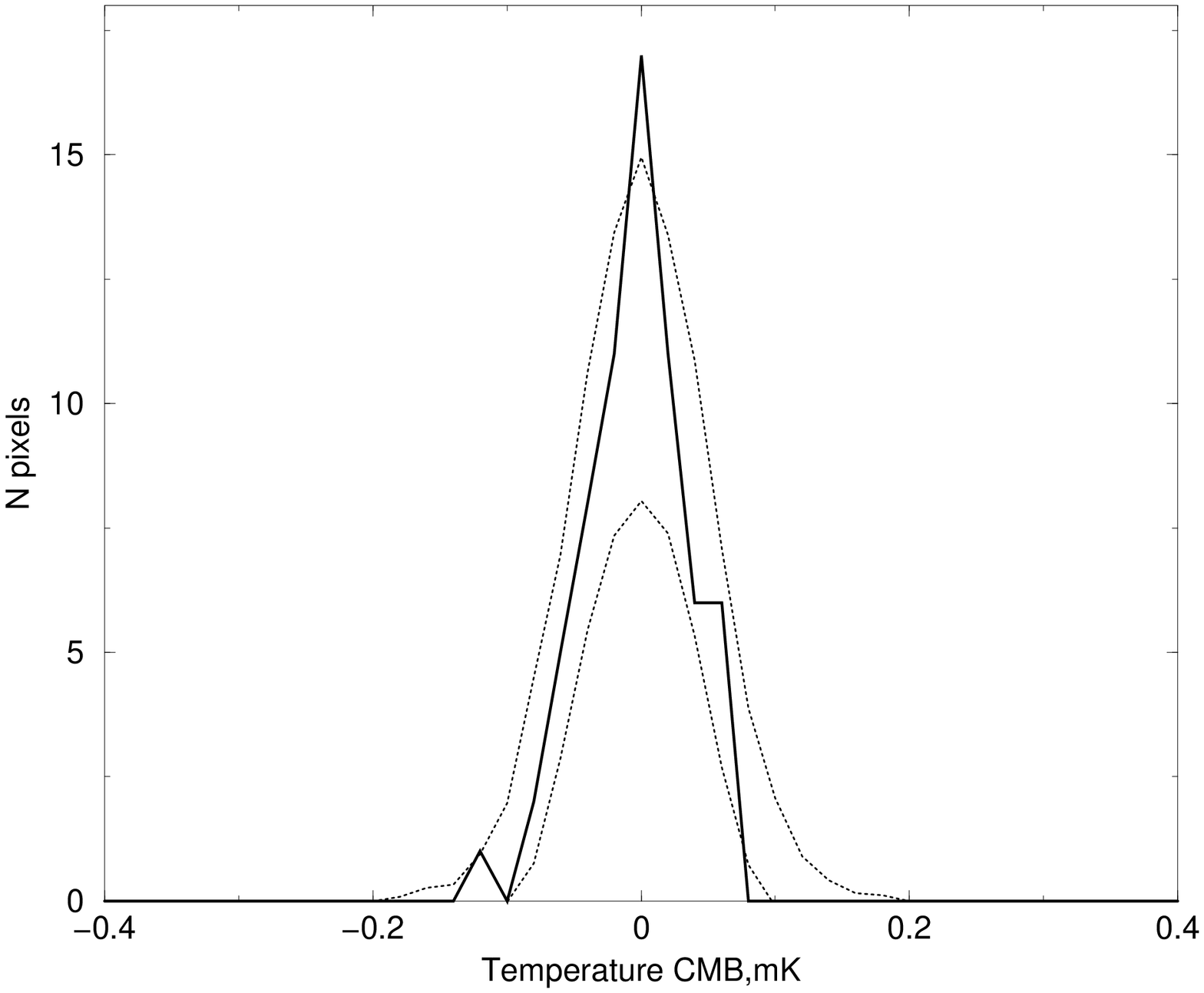,angle=-90,width=5.5cm}
}
\hbox{
\psfig{figure=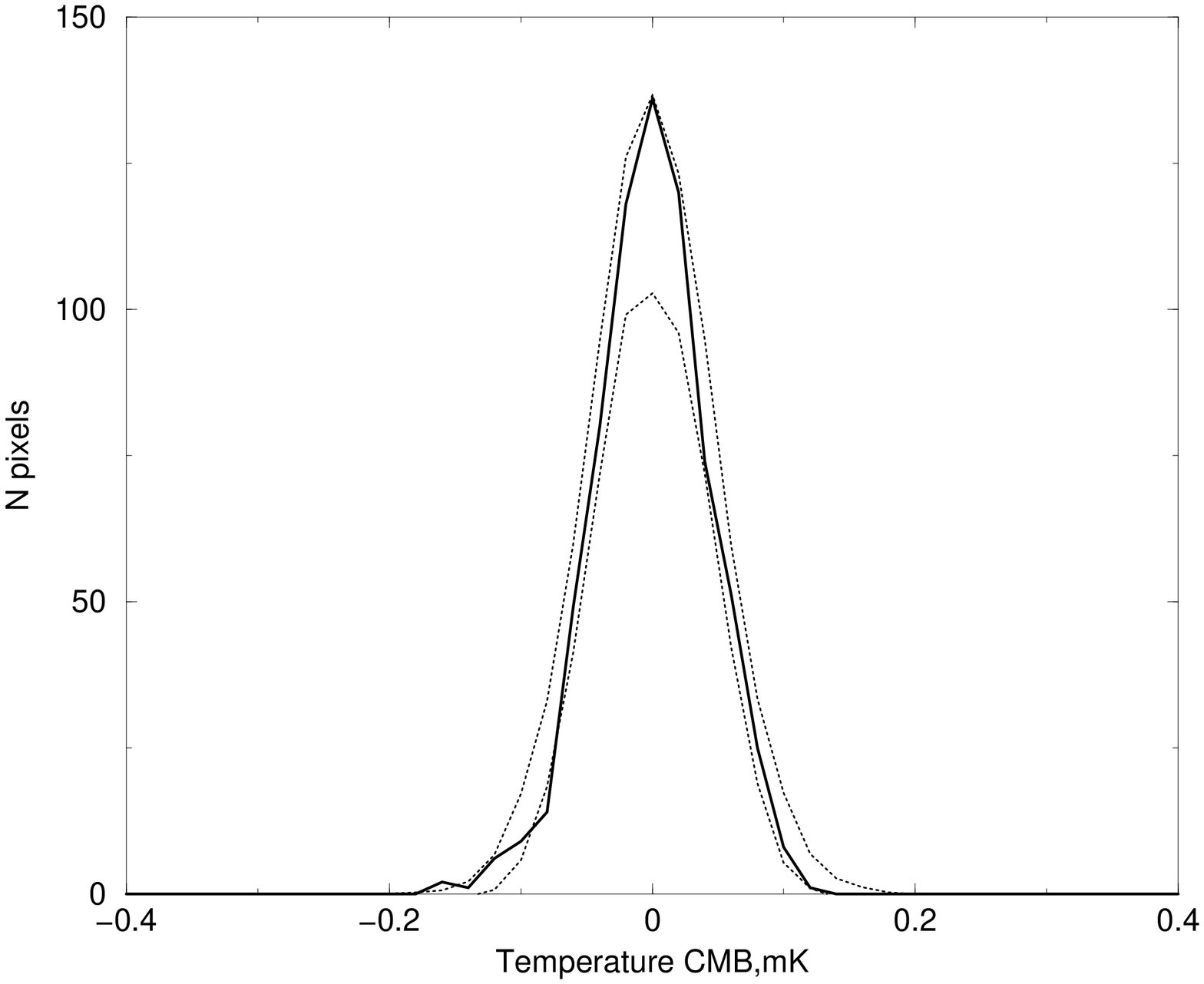,angle=-90,width=5.5cm}
\psfig{figure=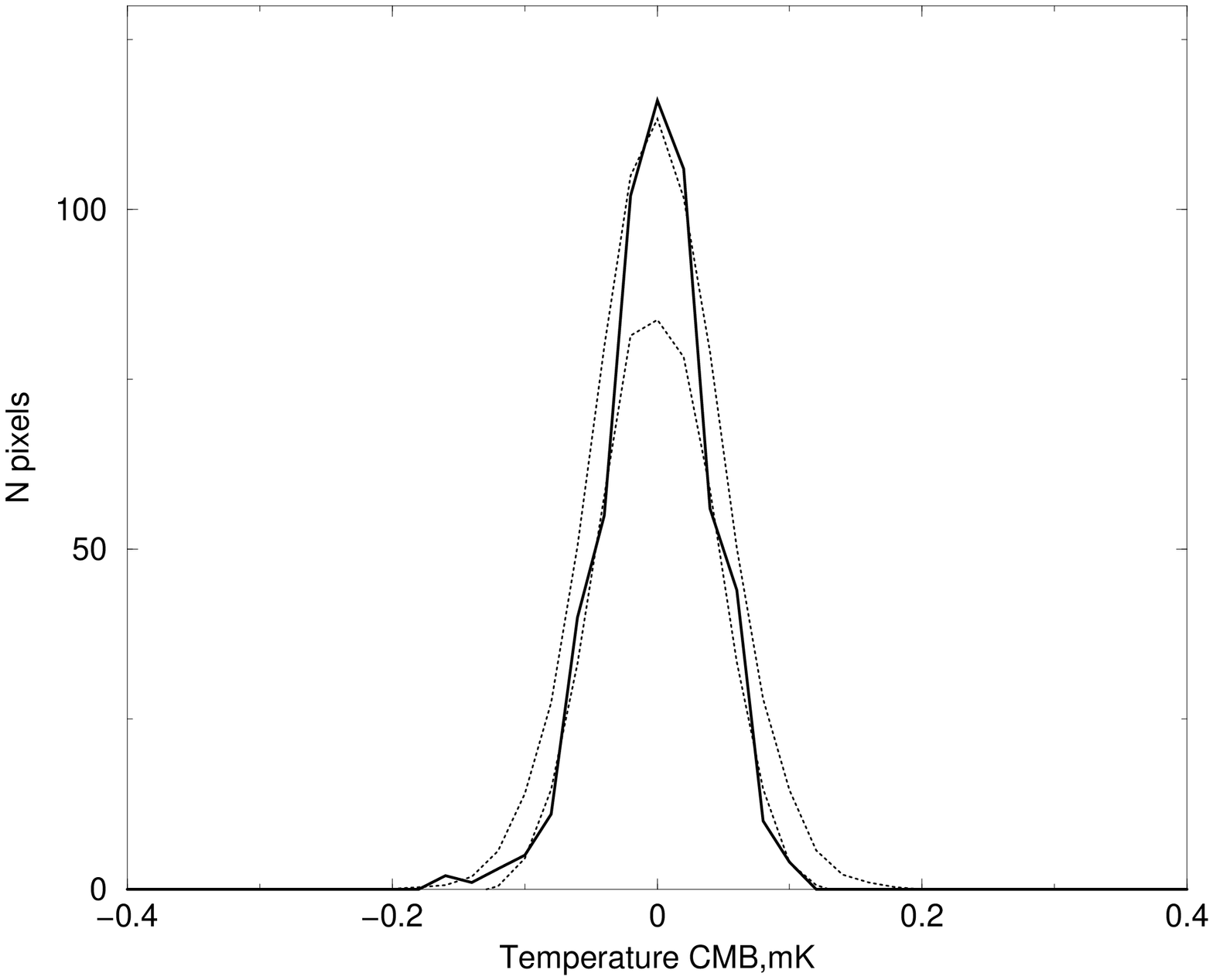,angle=-90,width=5.5cm}
}
}}
\caption{
Distribution of CMB fluctuations values in SMICA map pixels
corresponding to GRB locations with the map resolution
$\lmax=20$ for different GRB subsamples. The results without account for
a mask in the SMICA map are given on the left; those with the mask are on the
right. The upper pair of diagrams shows the distribution of short ($t<2$\,sec)
BATSE GRBs. The second pair presents the distribution of long ($t>2$\,sec)
BATSE GRBs. The third pair gives the distribution of signals for short
BeppoSAX GRBs. The lowest pair --- those for long BeppoSAX GRBs. The
dashed lines show the 1$\sigma$-dispersion of CMB values in the
$\Lambda$CDM cosmological model.
}
\label{hist_grb_cmb_L20}
\end{figure*}

\afterpage{\clearpage}

\begin{figure} [!th]
\setcaptionmargin{5mm}
\onelinecaptionstrue
\centerline{\vbox{
\hbox{
\psfig{figure=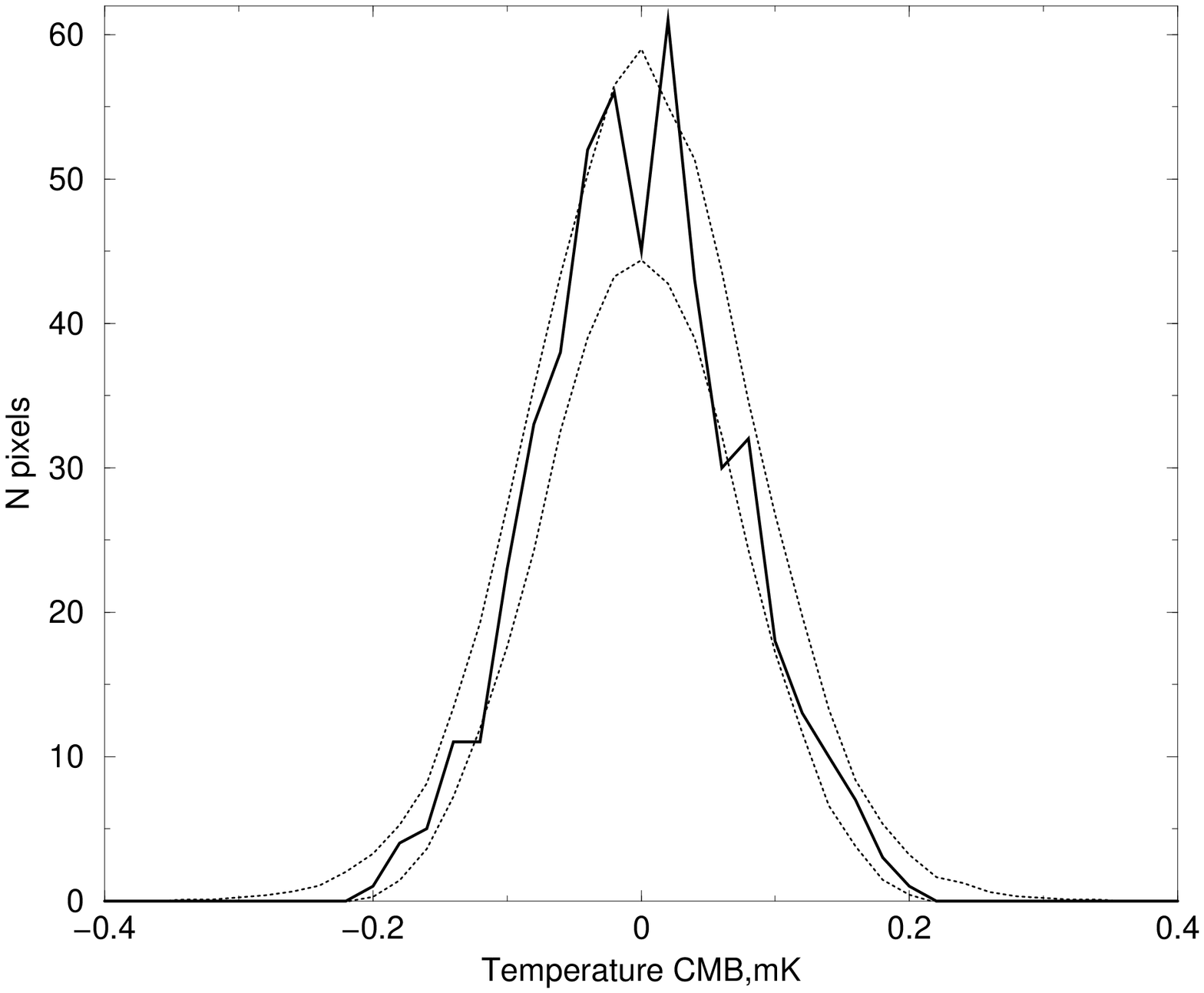,angle=-90,width=5.5cm}
\psfig{figure=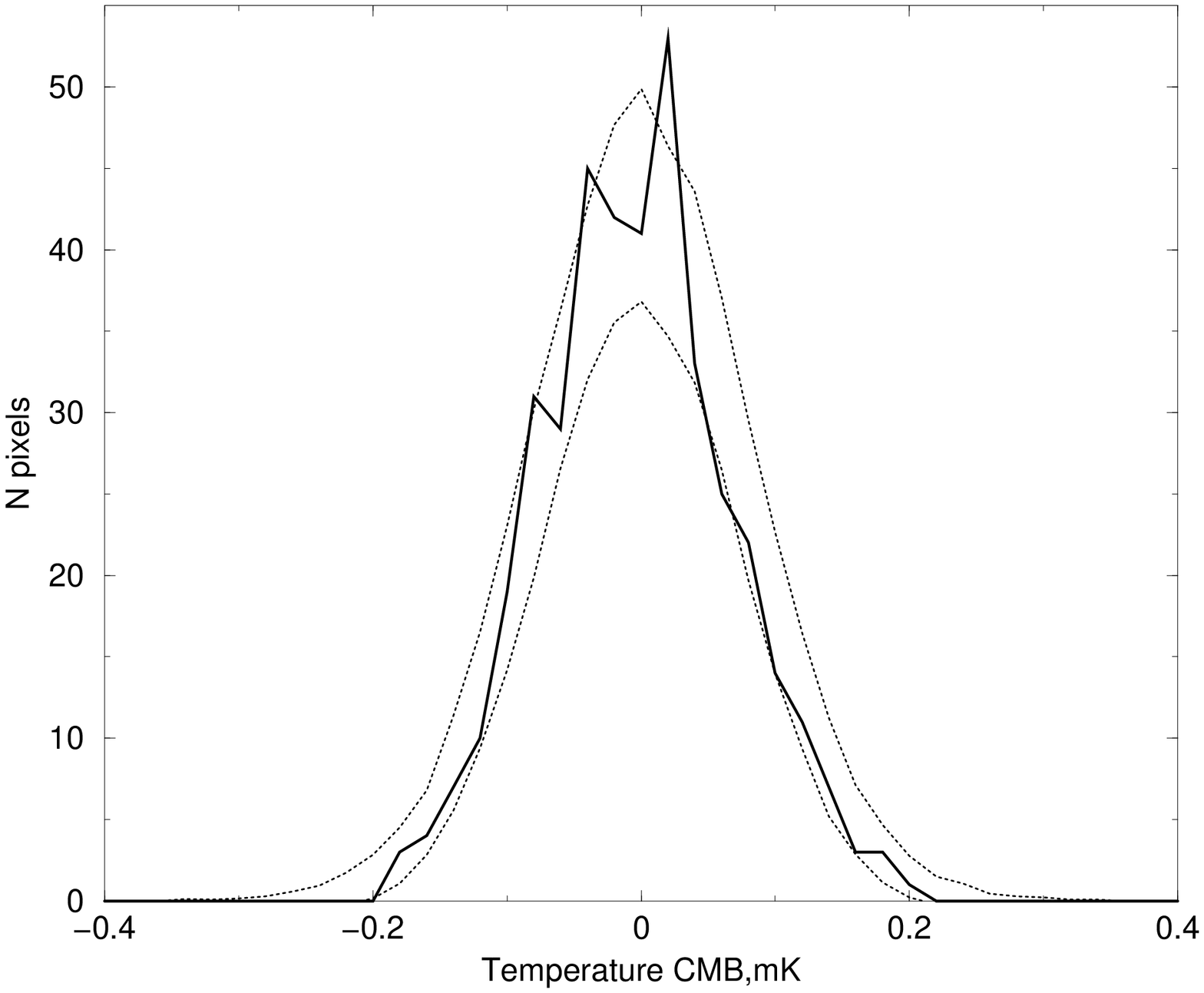,angle=-90,width=5.5cm}
}
\hbox{
\psfig{figure=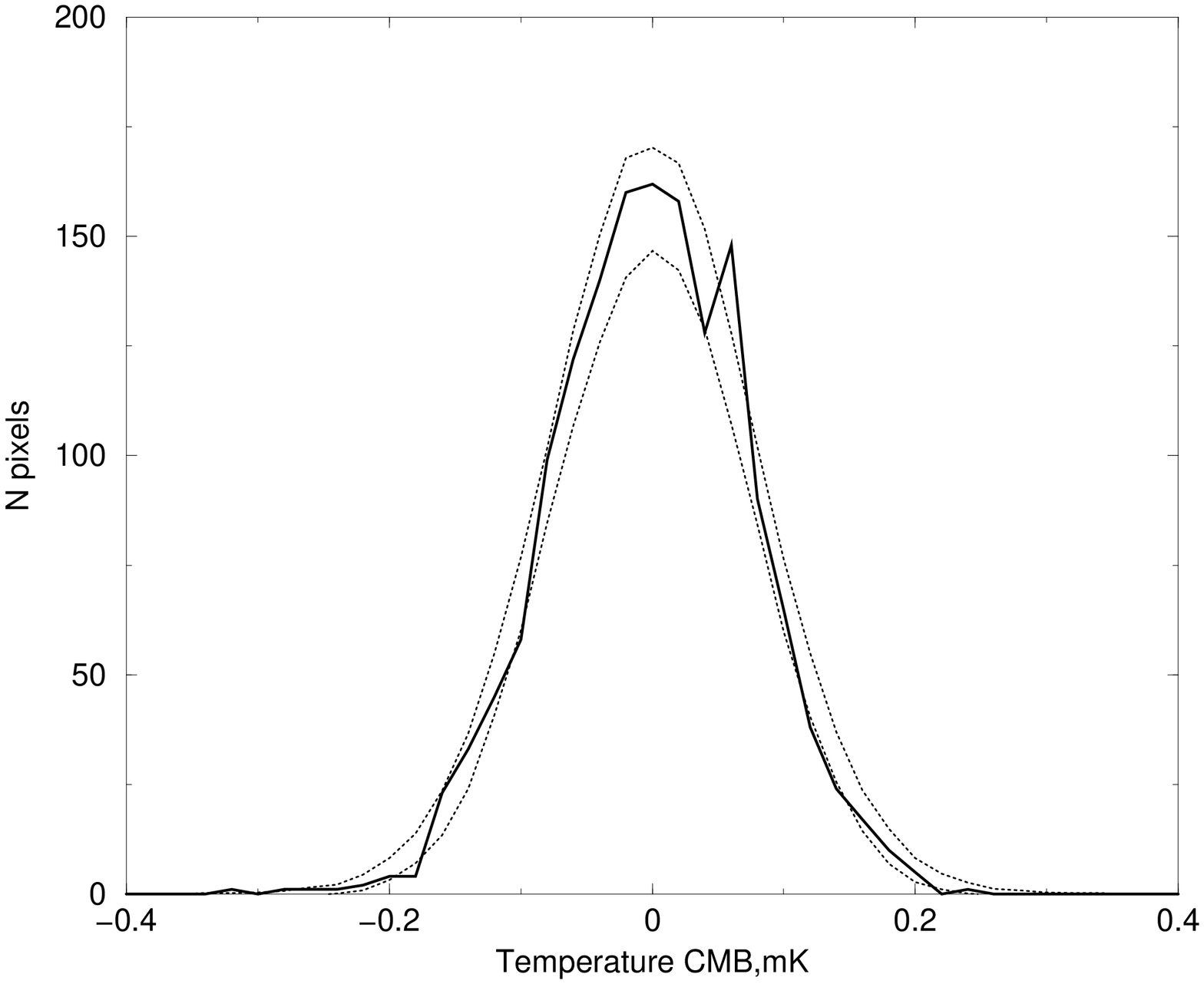,angle=-90,width=5.5cm}
\psfig{figure=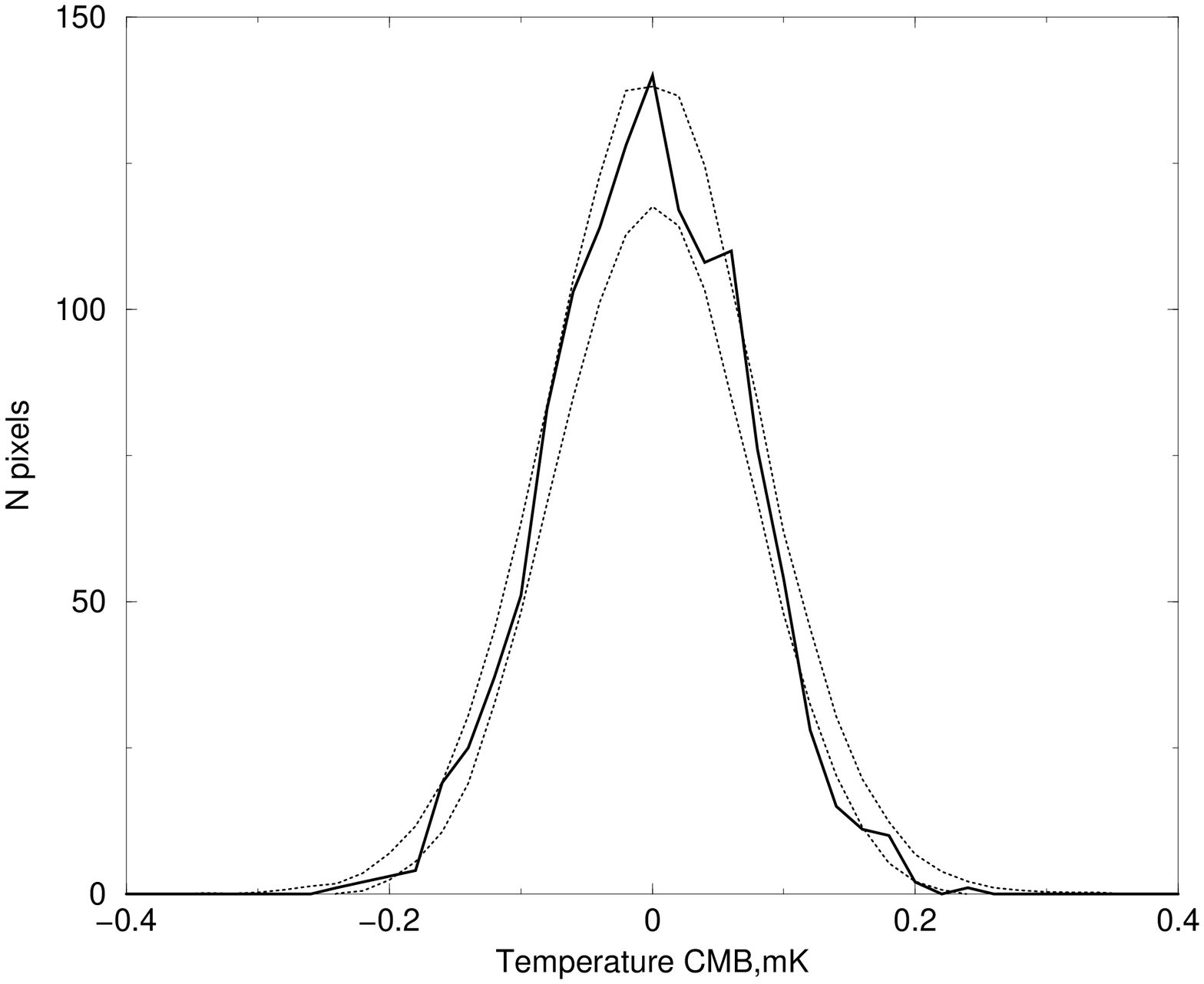,angle=-90,width=5.5cm}
}
\hbox{
\psfig{figure=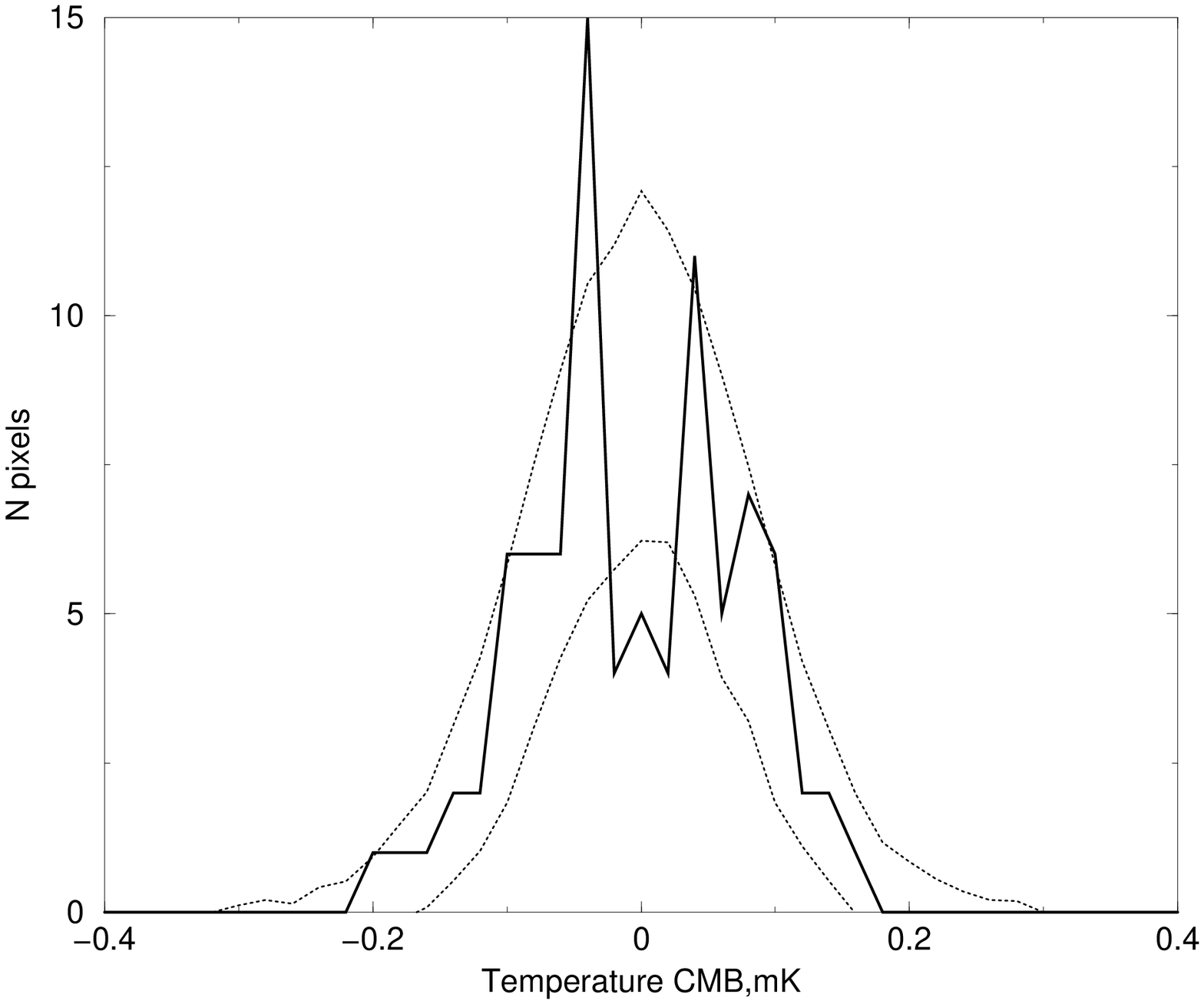,angle=-90,width=5.5cm}
\psfig{figure=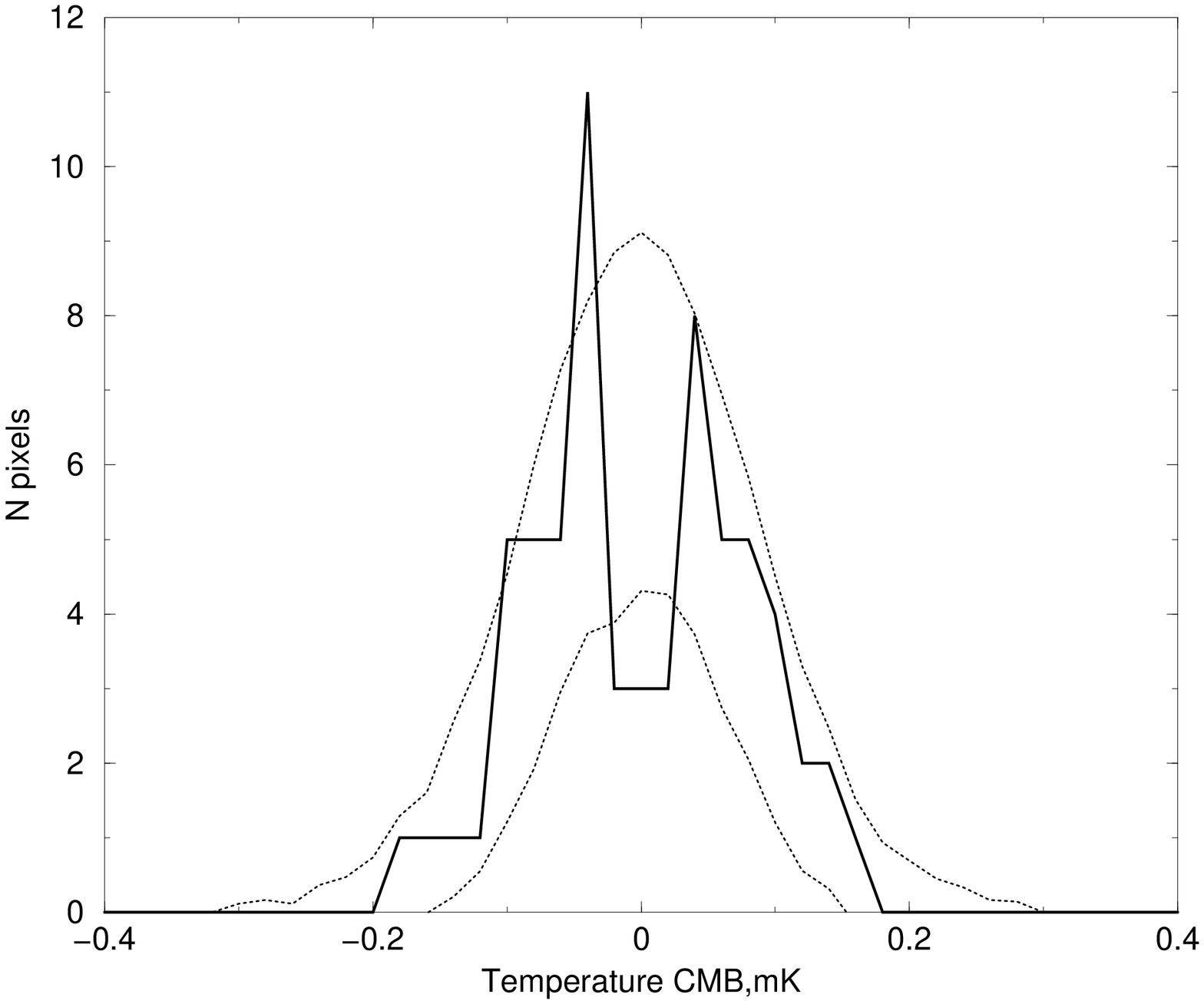,angle=-90,width=5.5cm}
}
\hbox{
\psfig{figure=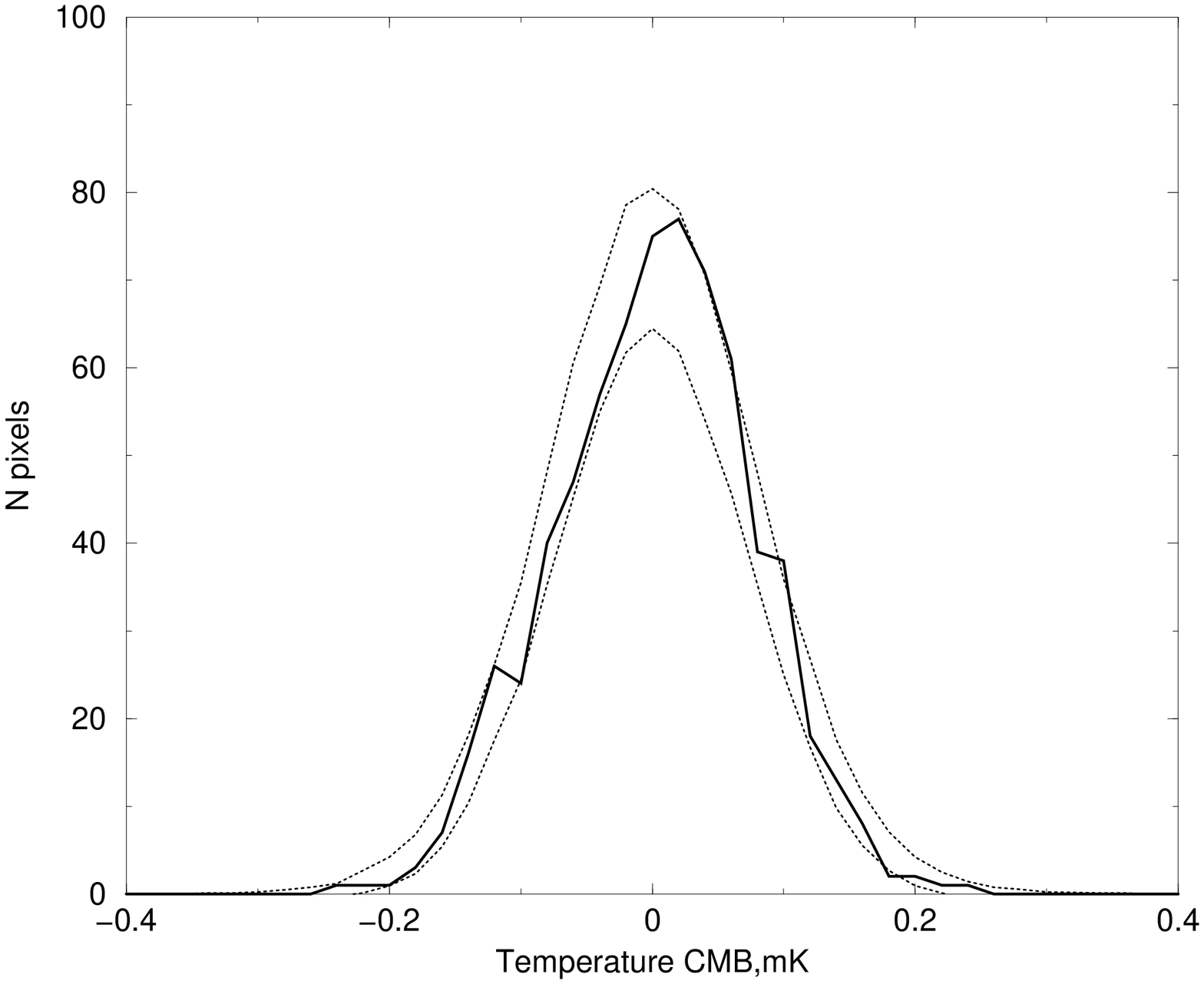,angle=-90,width=5.5cm}
\psfig{figure=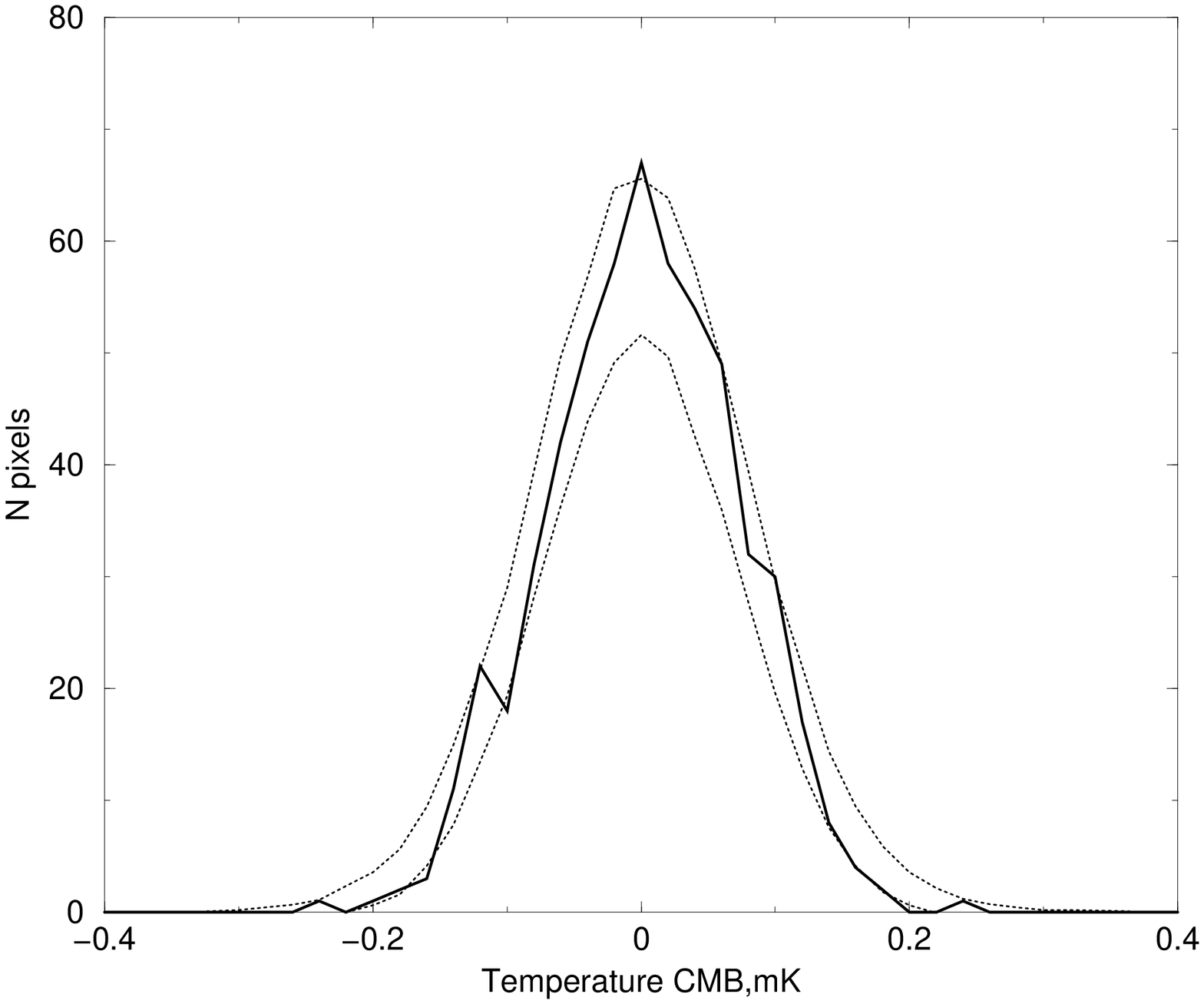,angle=-90,width=5.5cm}
}
}}
\caption{
Distribution of CMB fluctuations in SMICA map pixels corresponding to
GRB locations with the map resolution $\lmax=150$
for different GRB subsamples. The results without account for a mask in
the SMICA map are given on the left; those with the mask are on the
right. The upper pair of diagrams shows the distribution of short ($t<2$\,sec)
BATSE GRBs. The second pair presents the distribution of long ($t>2$\,sec)
BATSE GRBs. The third pair gives the distribution of signals for short
BeppoSAX GRBs. The lowest pair shows those for long BeppoSAX GRBs. The
dashed lines demonstrate the 1$\sigma$-dispersion of CMB values in the
$\Lambda $CDM cosmological model.
}
\label{hist_grb_cmb_L150}
\end{figure}

\begin{figure} [!th]
\setcaptionmargin{5mm}
\onelinecaptionstrue
\centerline{\vbox{
\hbox{
\psfig{figure=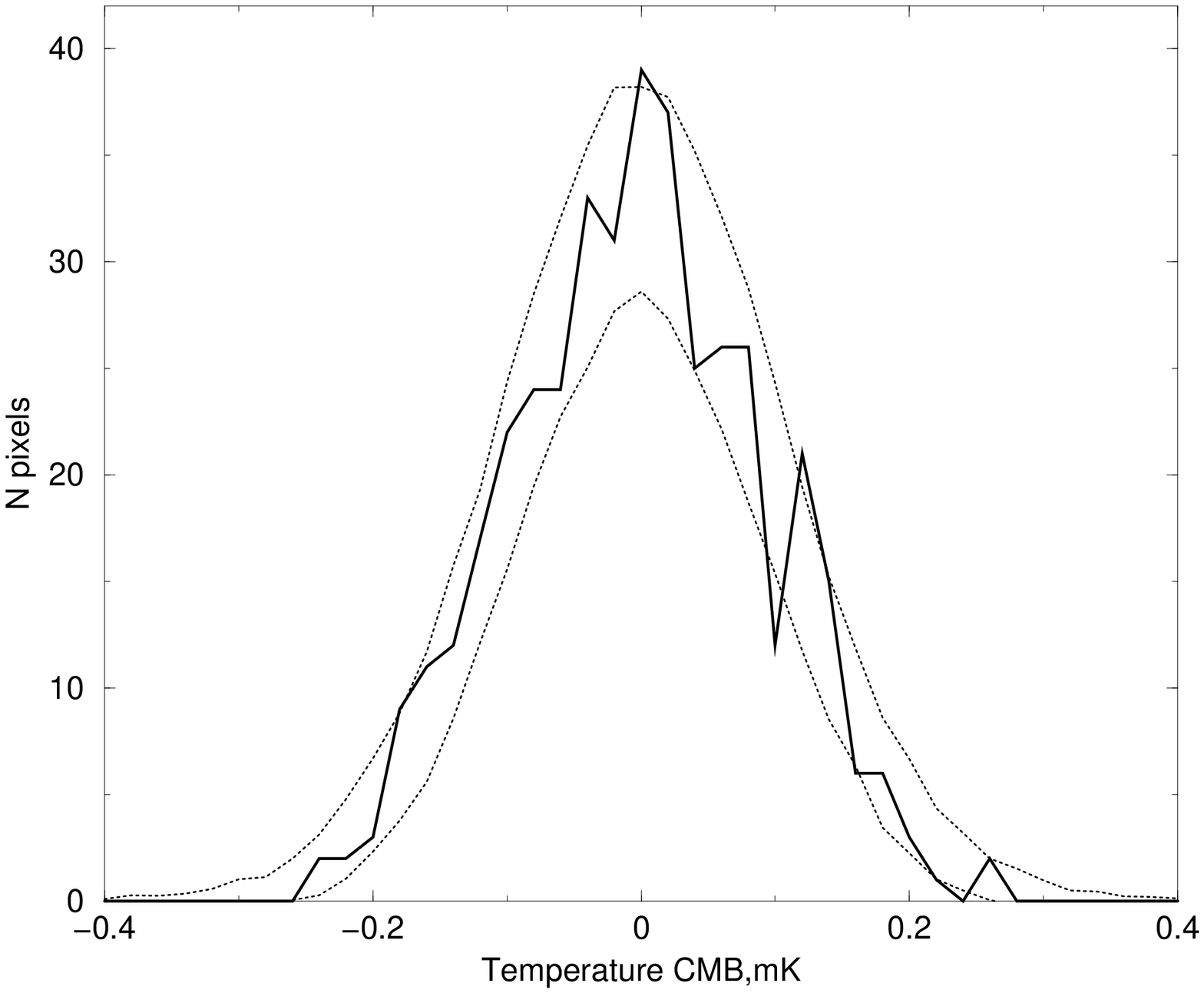,angle=-90,width=7cm}
\psfig{figure=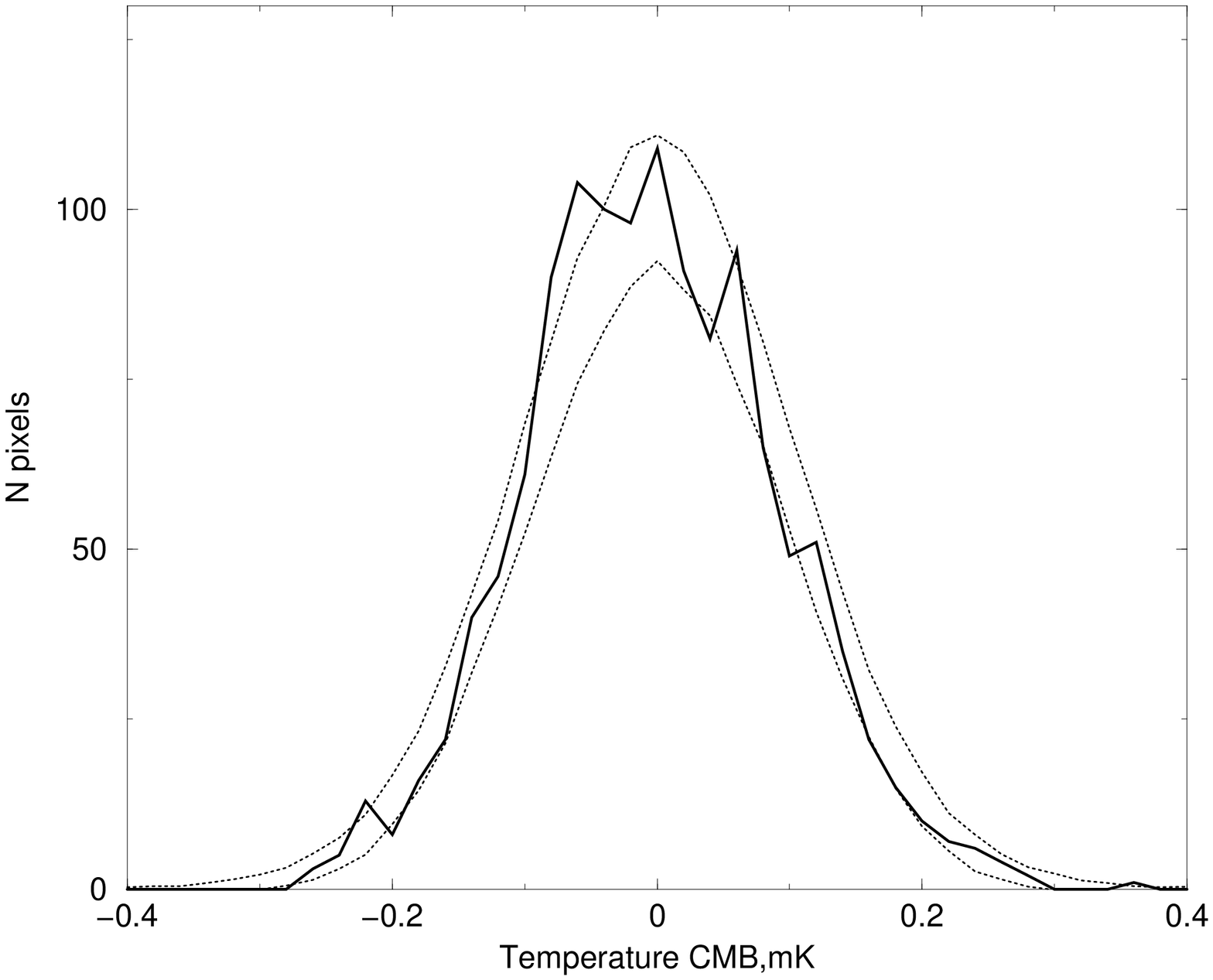,angle=-90,width=7cm}
}
\hbox{
\psfig{figure=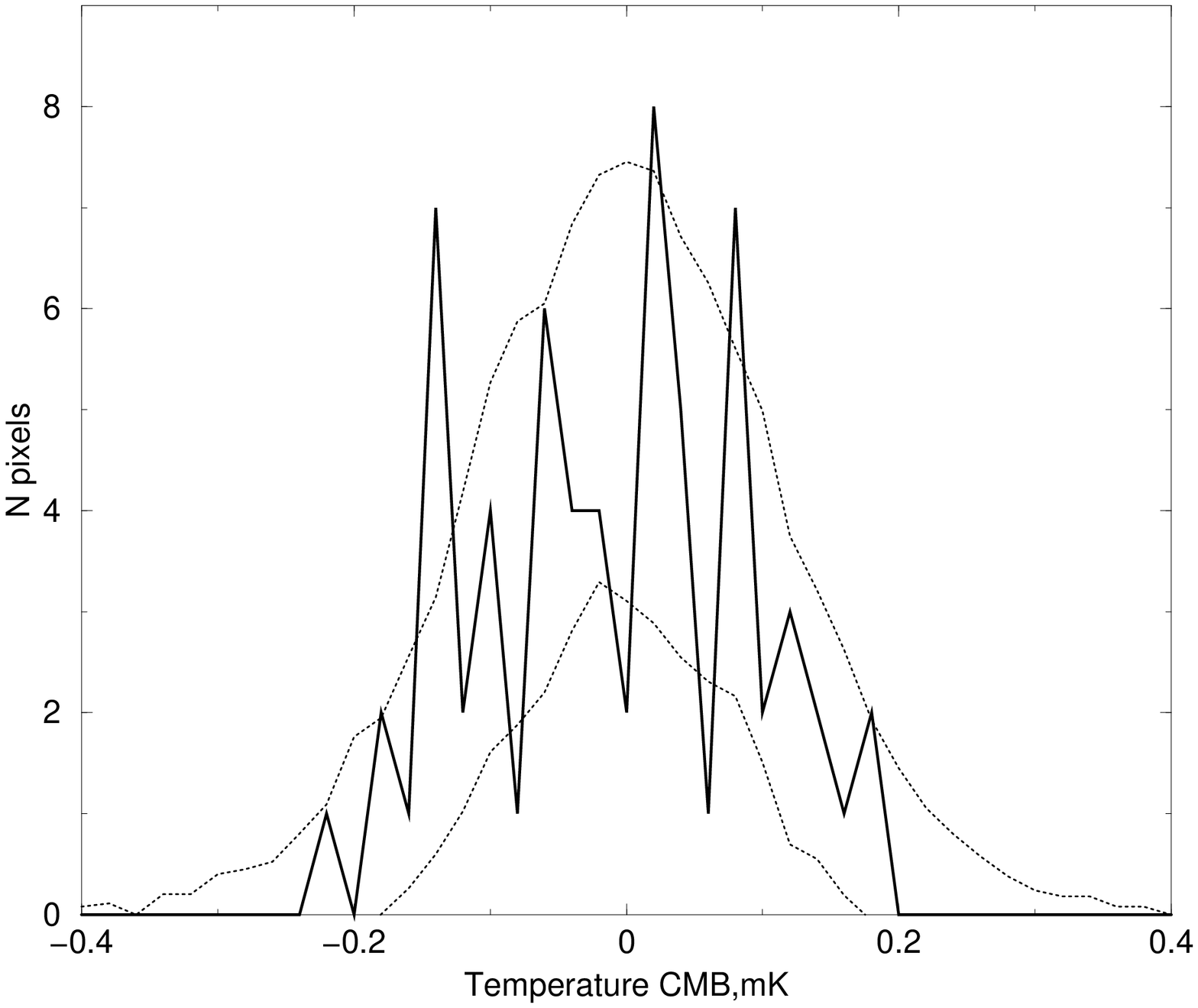,angle=-90,width=7cm}
\psfig{figure=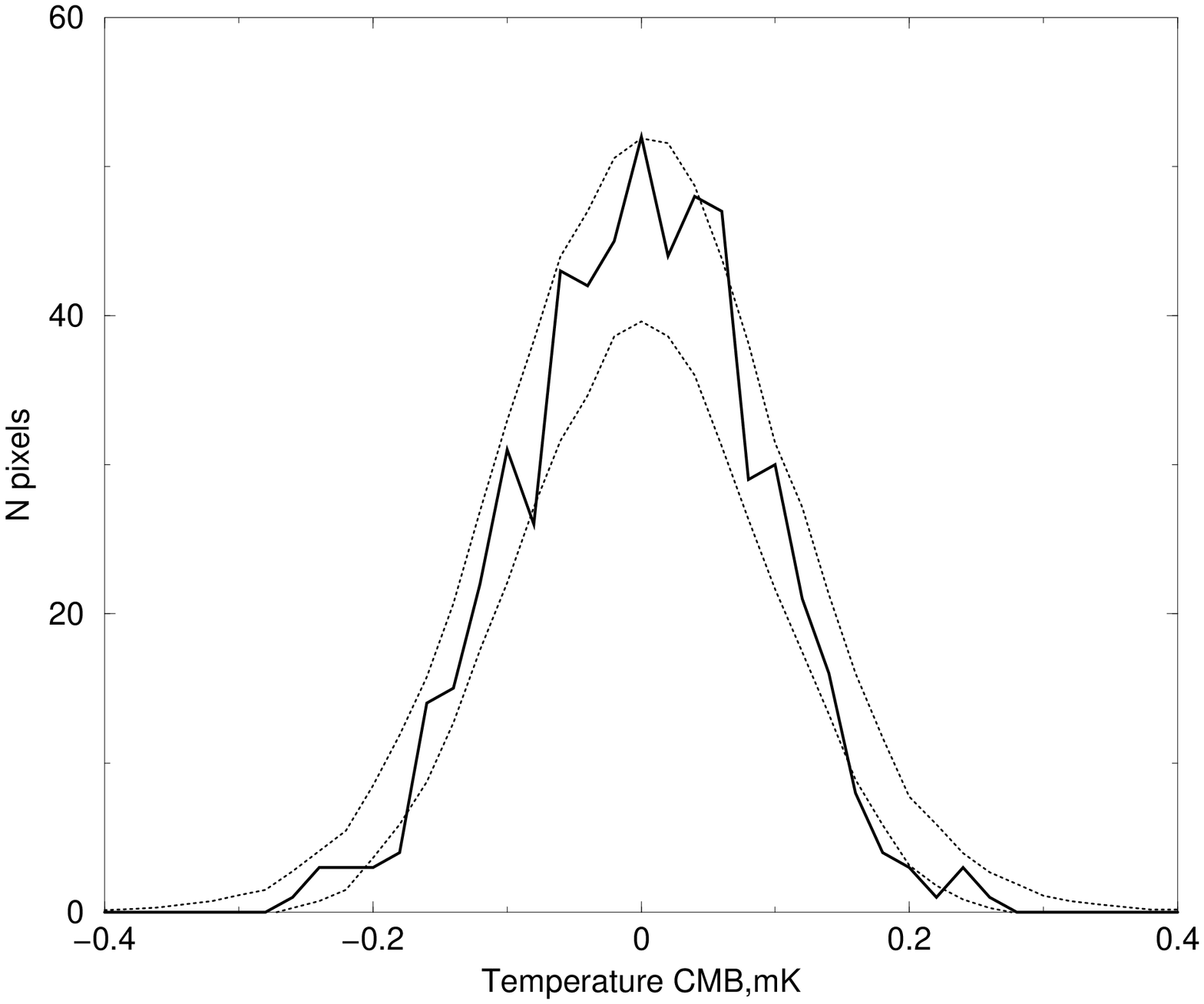,angle=-90,width=7cm}
}
}}
\caption{
Distribution of CMB fluctuation values in the SMICA map pixels
corresponding to GRB locations with the map resolution  $\lmax=300$
for different GRB subsamples. The results are given with the account
for the mask in the SMICA map. The left upper diagrams show the
statistics for short ($t<2$\,sec)
BATSE GRBs. The right upper diagram presents those for long ($t>2$\,sec)
BATSE GRBs. The left bottom diagram presents results for short BeppoSAX
GRBs. The right bottom diagram is for long BeppoSAX events. The dashed
lines show the 1$\sigma$-dispersion of CMB values in the $\Lambda$CDM
cosmological model.
}
\label{hist_grb_cmb_L300}
\end{figure}

\begin{figure} [!th]
\setcaptionmargin{5mm}
\onelinecaptionstrue
\centerline{\vbox{
\hbox{
\psfig{figure=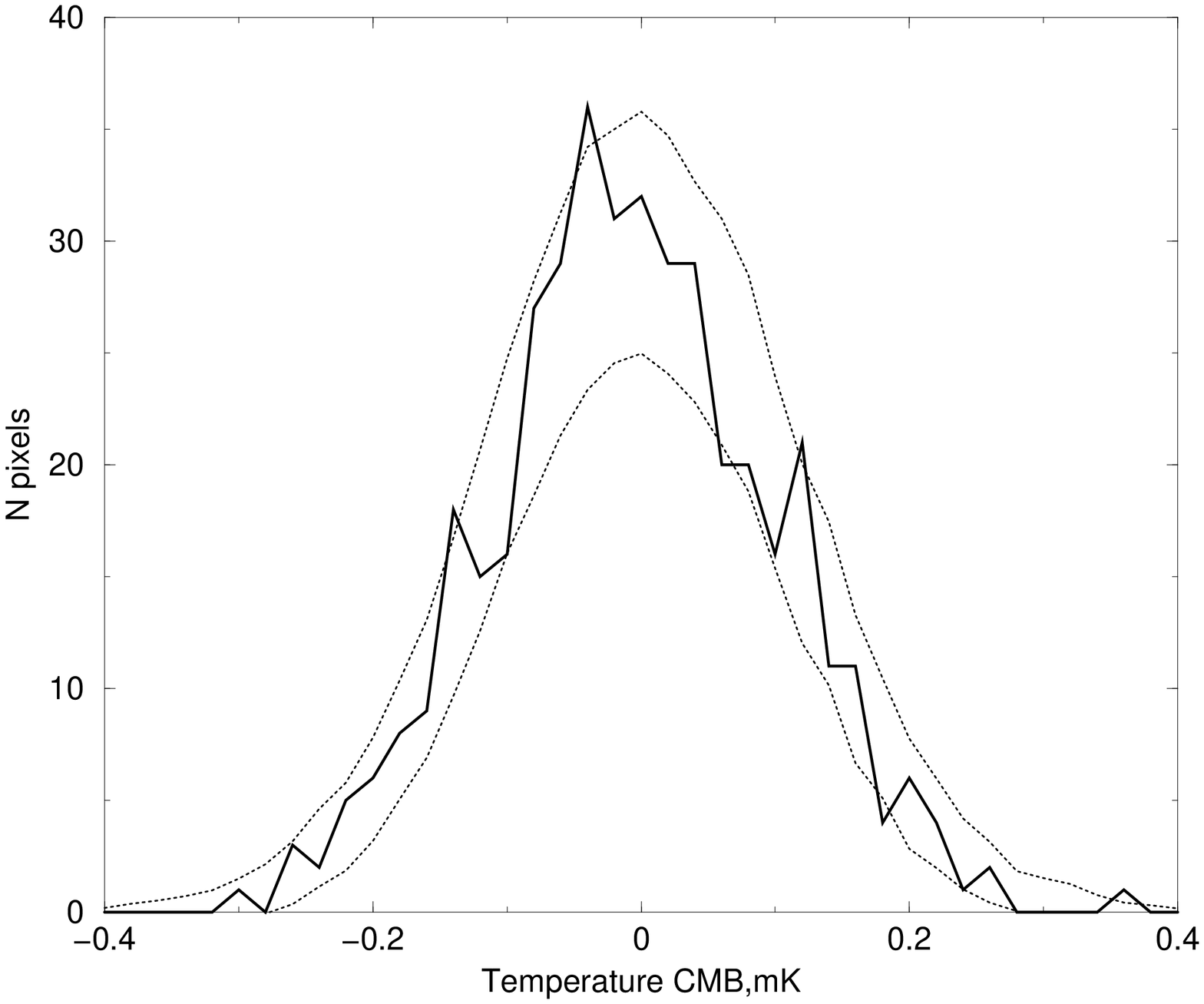,angle=-90,width=7cm}
\psfig{figure=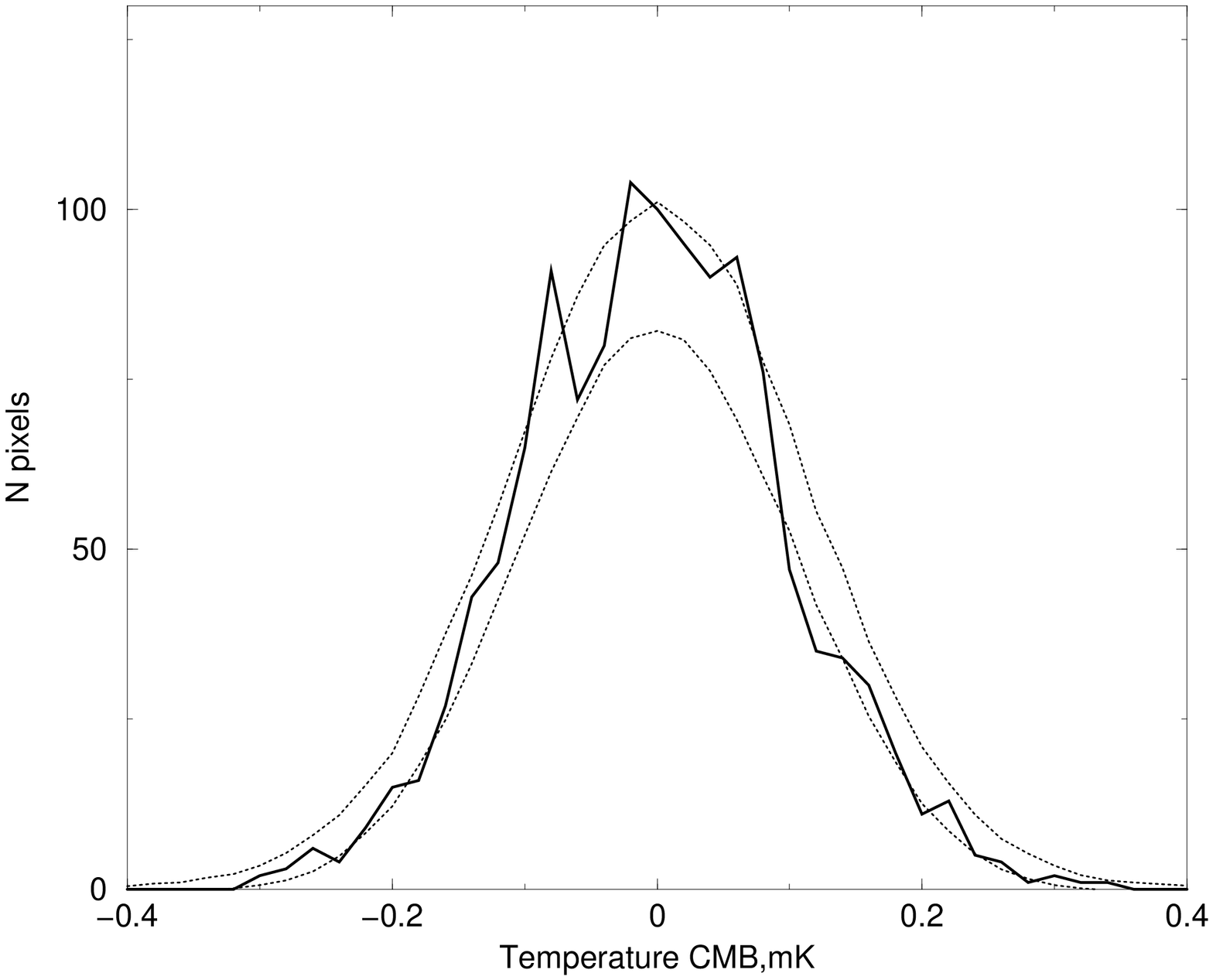,angle=-90,width=7cm}
}
\hbox{
\psfig{figure=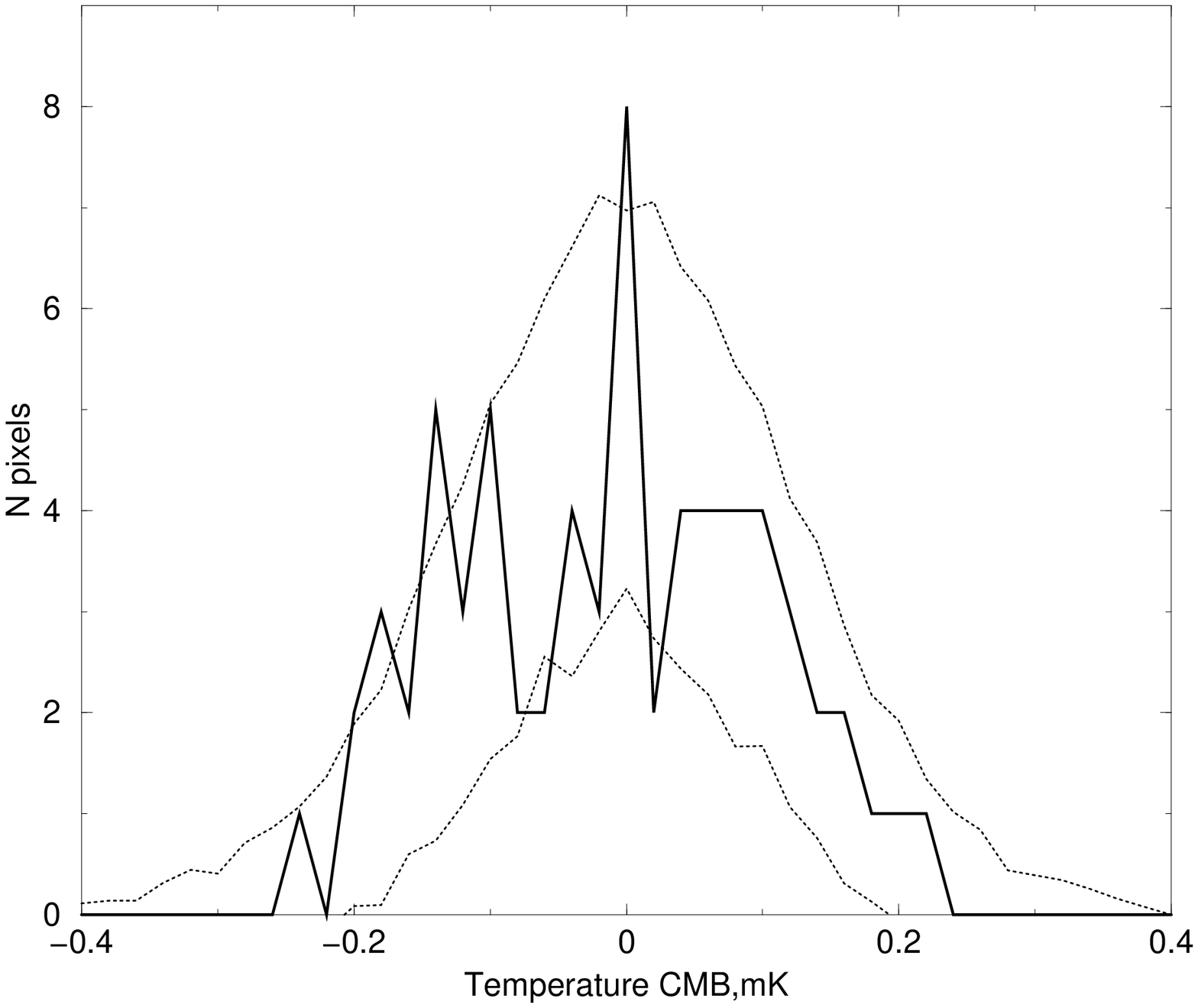,angle=-90,width=7cm}
\psfig{figure=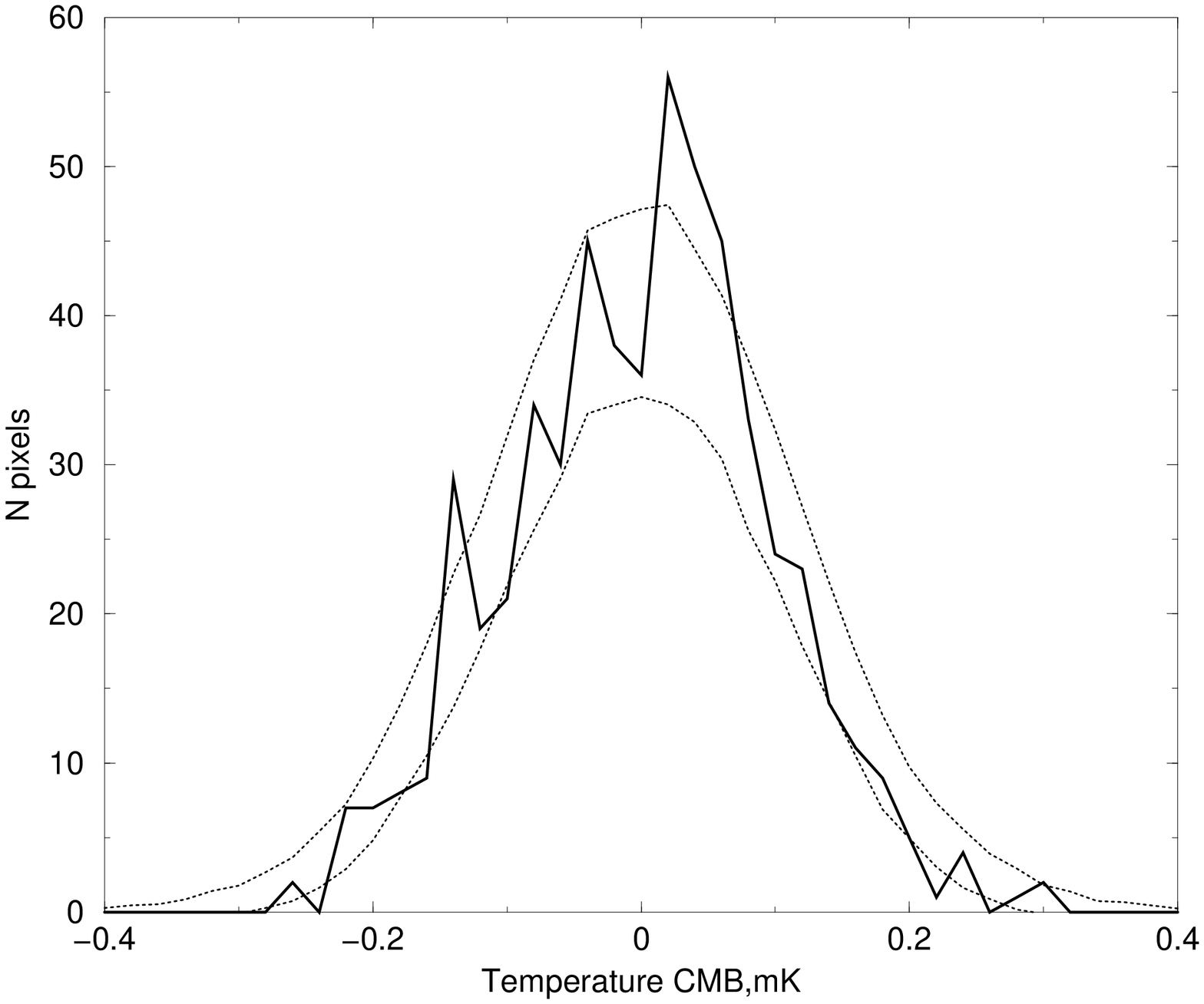,angle=-90,width=7cm}
}
}}
\caption{
Distribution of CMB fluctuation values in the SMICA map pixels
corresponding to GRB locations with the map resolution $\lmax=600$
for different GRB subsamples. The results are given with the account
for the mask in the SMICA map. The left upper diagram shows the
statistics for short ($t<2$\,sec)
BATSE GRBs. The right upper diagram presents those for long ($t>2$\,sec)
BATSE rGRBs. The left bottom diagram presents results for short
BeppoSAX GRBs. The right bottom diagram is for long BeppoSAX events.
The dashed lines show the 1$\sigma$-dispersion of CMB values in the
$\Lambda $CDM cosmological model.
}
\label{hist_grb_cmb_L600}
\end{figure}

To analyze the statistics of pixel
values we used the `{\tt mapcut}'
procedure of the GLESP package\footnote{\tt http://www.glesp.nbi.dk}
\cite{glesp2}.
Calculations were made for the maps
smoothed to the resolution 260\arcmin\, ($\lmax=20$),
35\arcmin\, ($\lmax=150$), 20\arcmin\, ($\lmax=300$) É
10\arcmin\, ($\lmax=600$).
Figures \ref{grb_cmbL20} and \ref{grb_cmbL150} show  the locations of GRBs
from subsamples of BeppoSAX
and BATSE catalogs on CMB maps with the resolution 260\arcmin\ ($\lmax=20$)
and 35\arcmin\, ($\lmax=150$).
The map resolution was selected in accordance with the expected scale
of the Sachs--Wolfe effect manifestations and possible appearance of
some features on the SMICA map.
By way of example, Figure \ref{grb_cmbL150_20} shows GRB locations on the CMB
map within the multipole range $20<\ell\le150$.

\begin{figure} [!th]
\setcaptionmargin{5mm}
\onelinecaptionstrue
\centerline{\vbox{
\hbox{
\psfig{figure=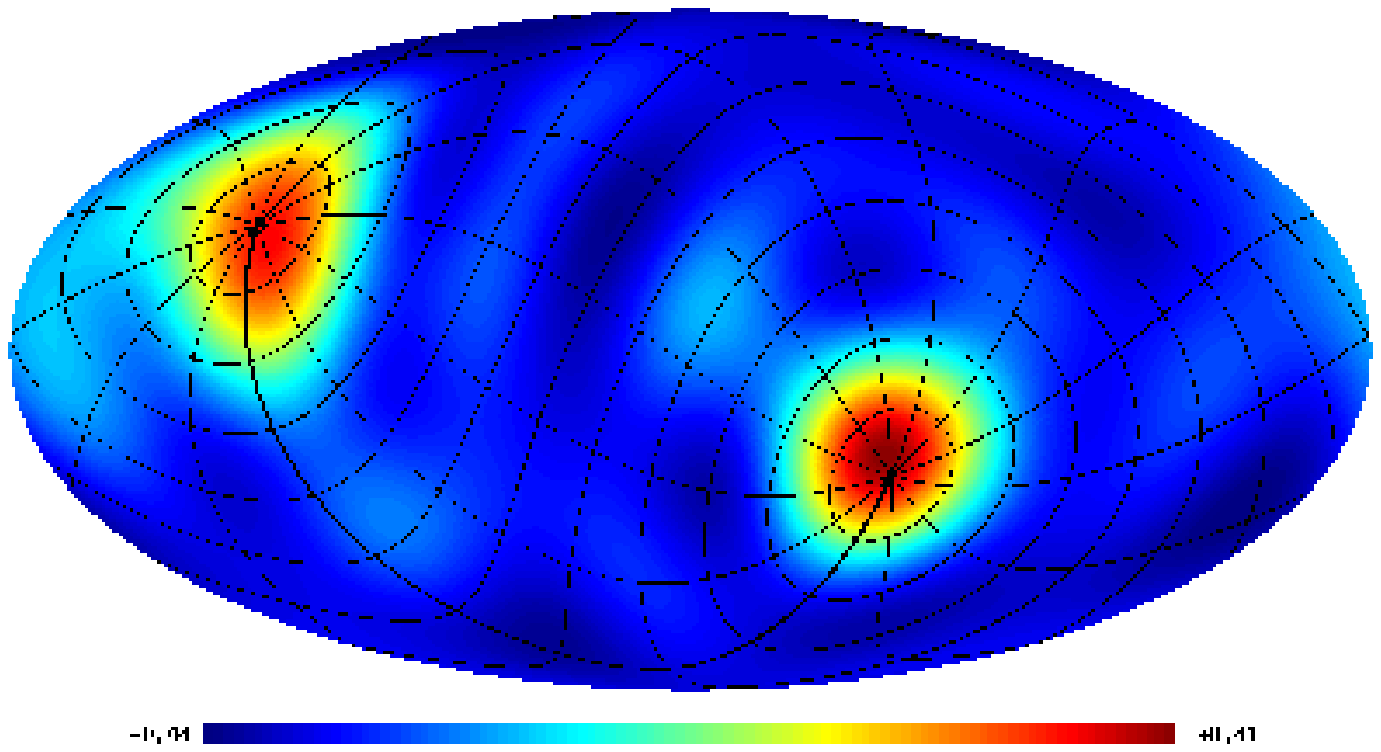,width=8cm}
\psfig{figure=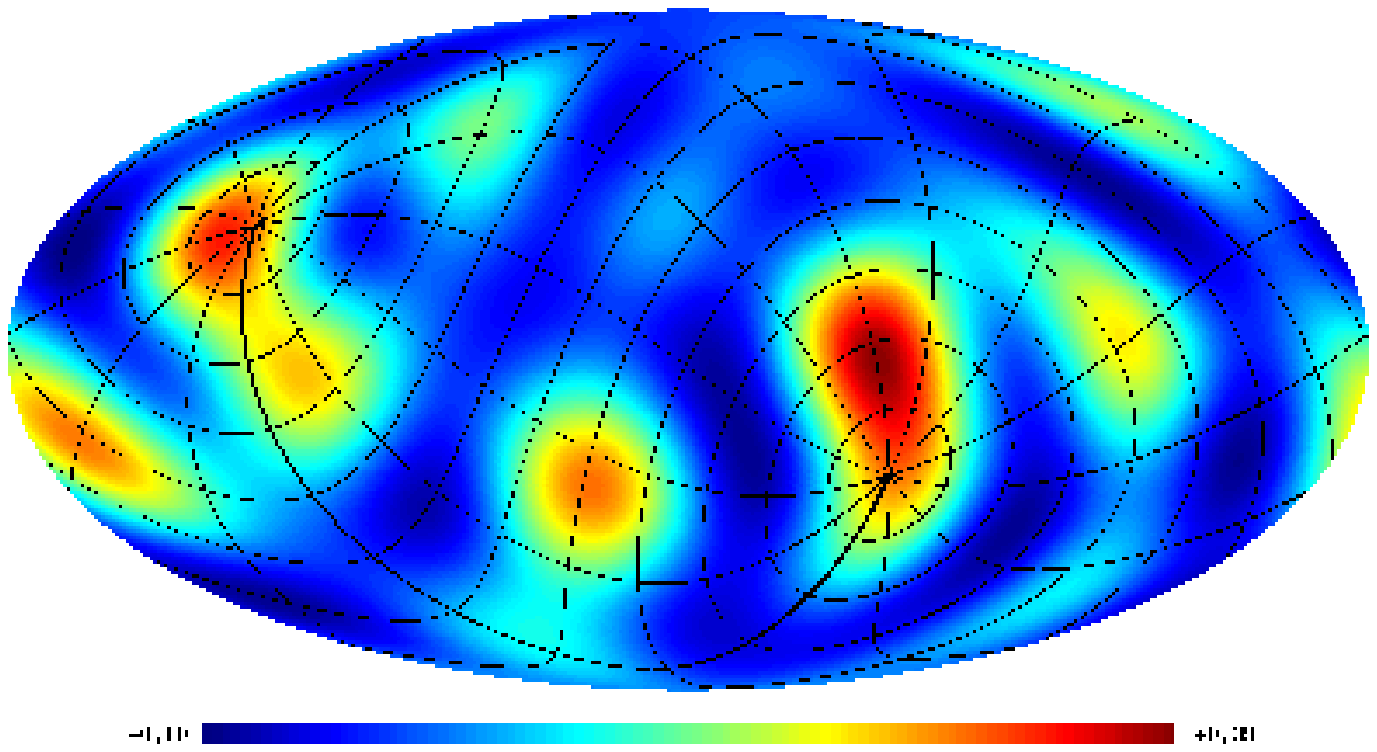,width=8cm}
}
\hbox{
\psfig{figure=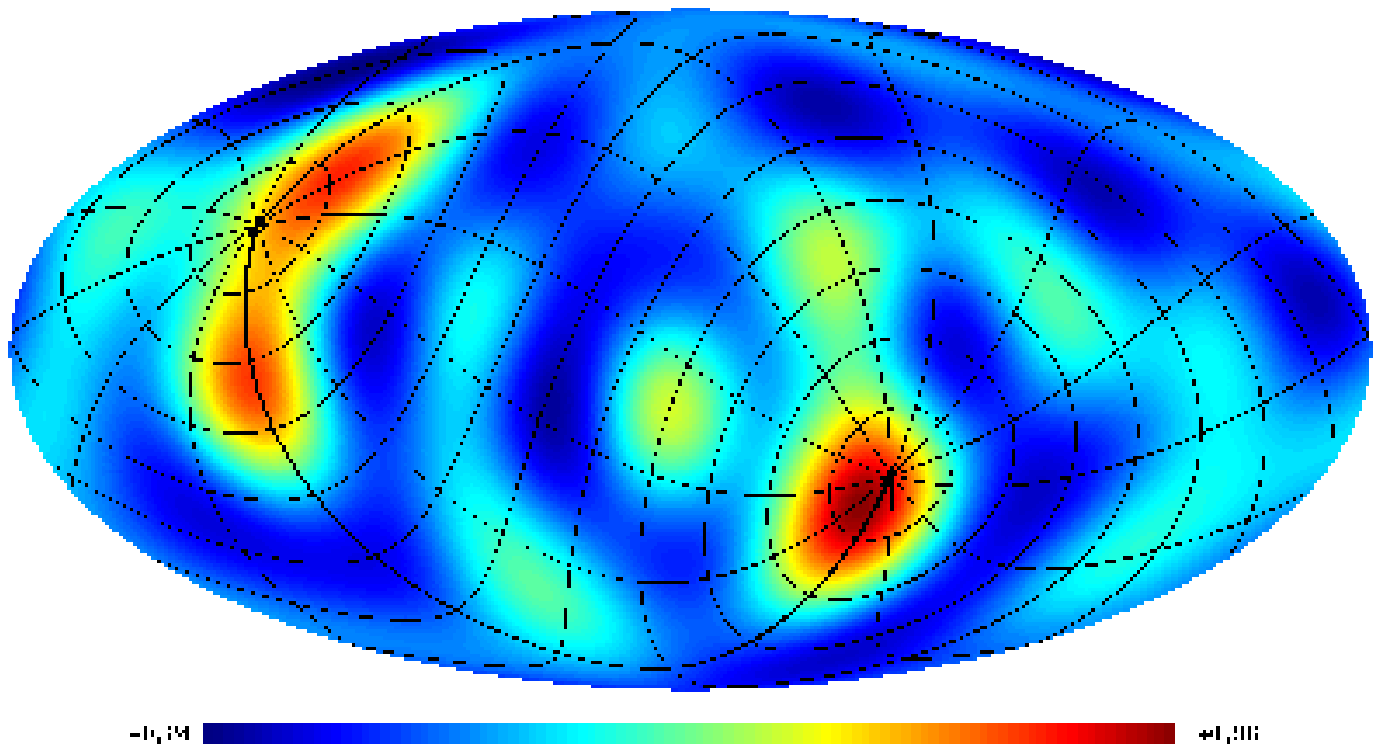,width=8cm}
\psfig{figure=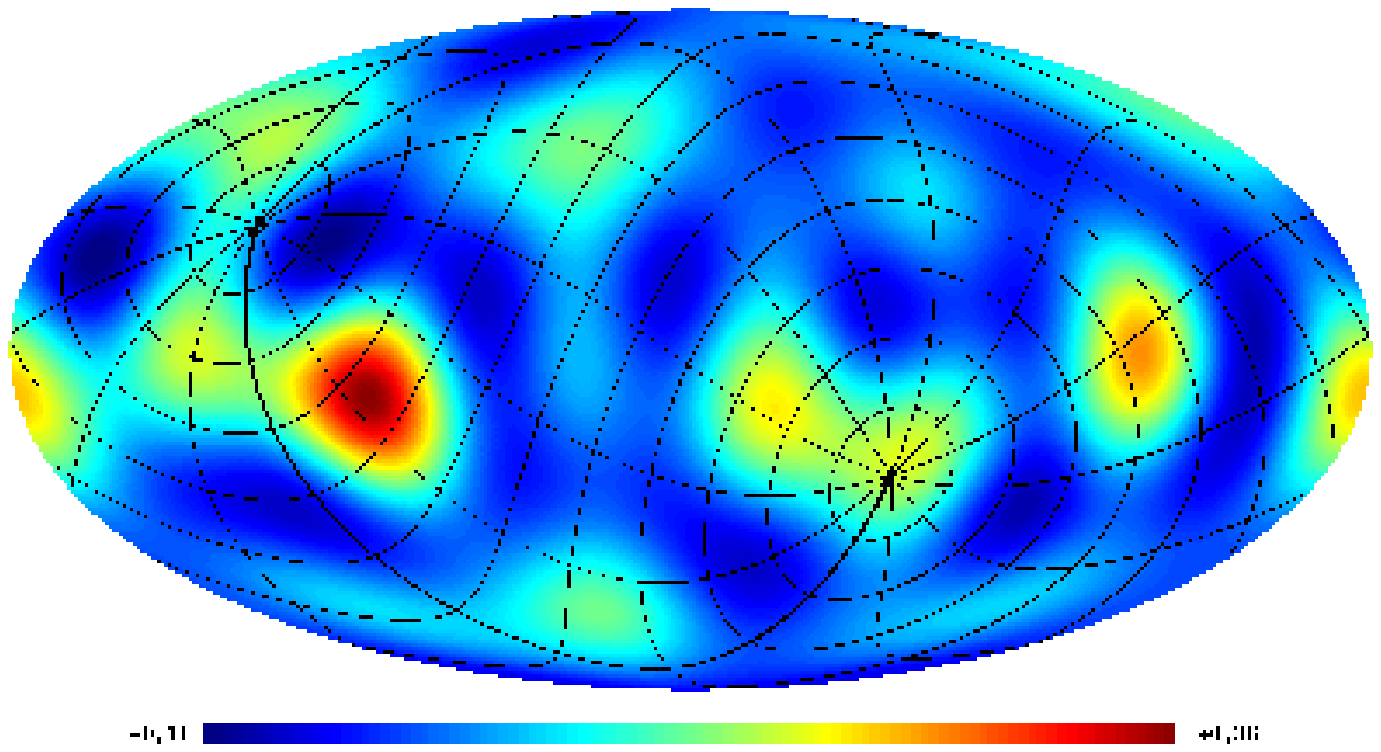,width=8cm}
}
\hbox{
\psfig{figure=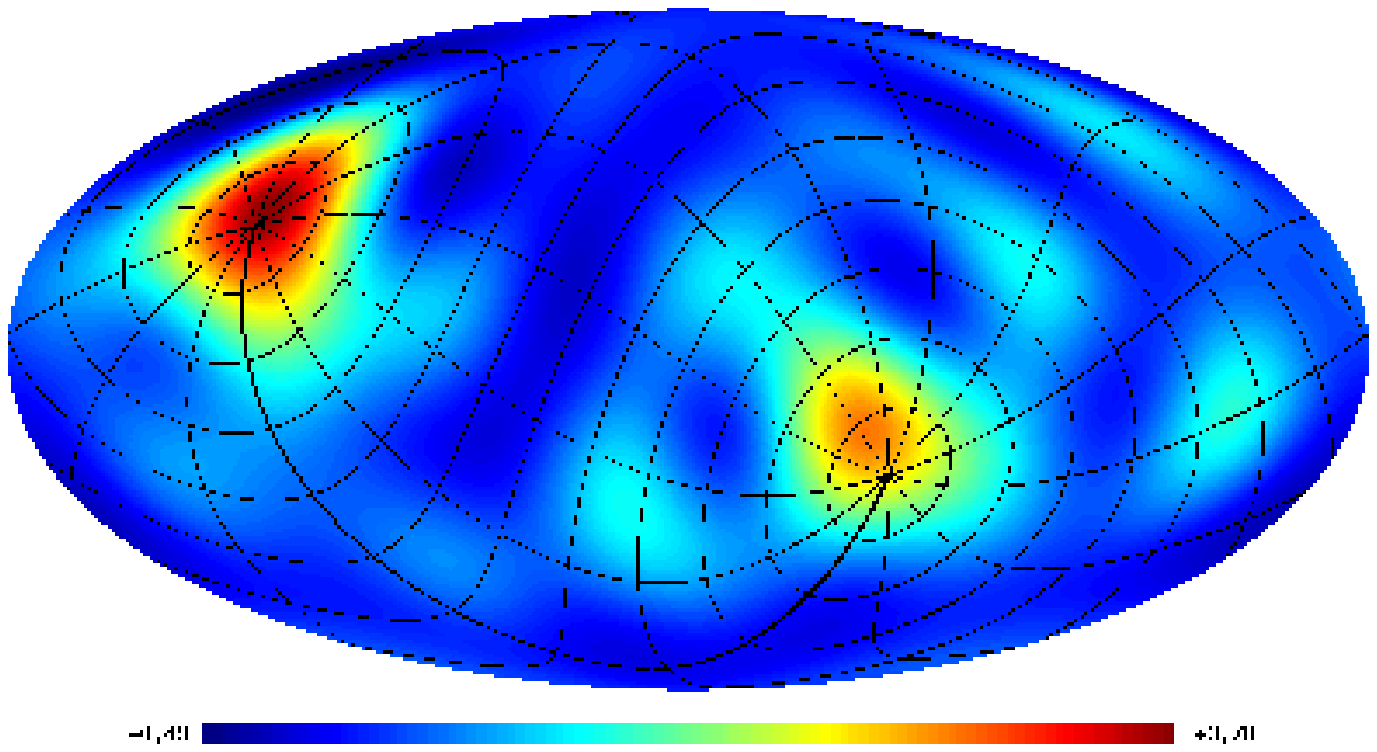,width=8cm}
\psfig{figure=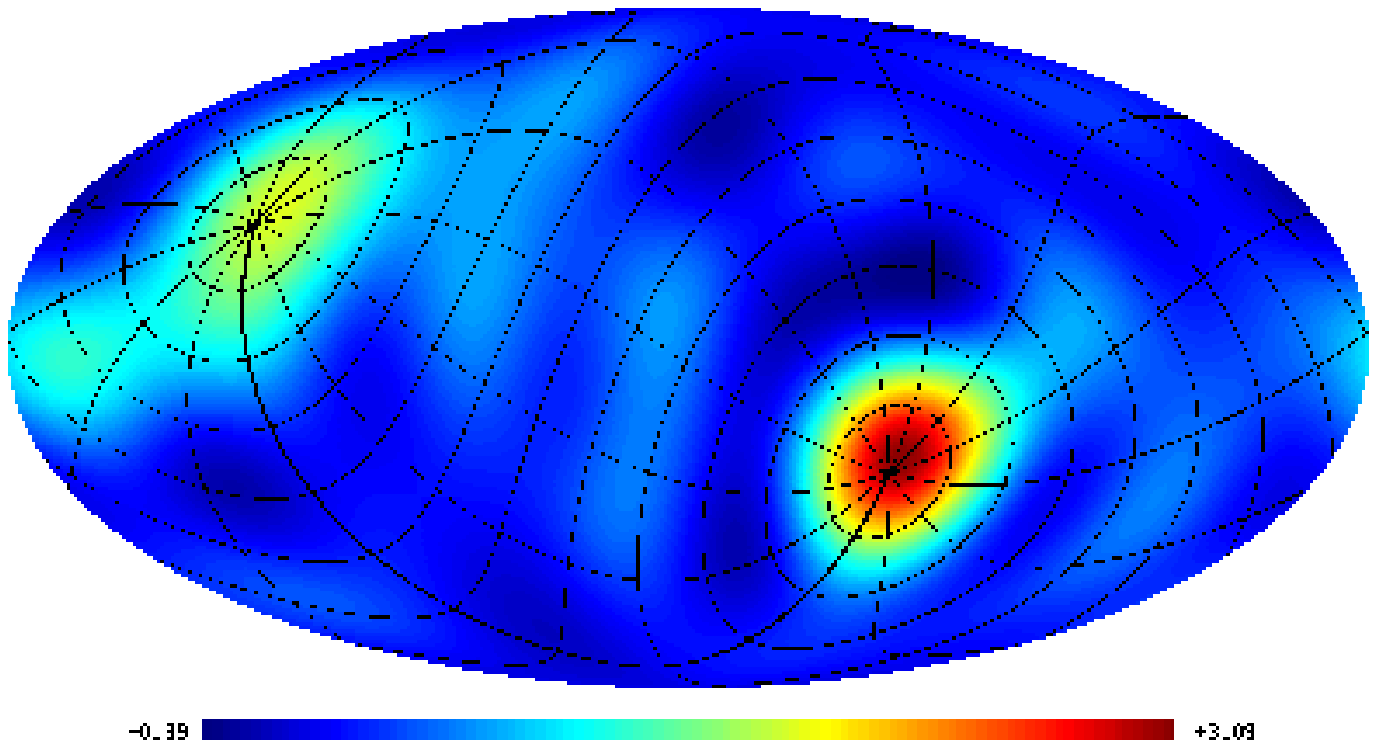,width=8cm}
}
}}
\caption{
The smoothed maps of the sky up to $\lmax=7$
for different GRB subsamples. The maps were built from CMB pixels
corresponding to direction to those GRBs which deflect the histograms
from the expected ones
(see Figs.\,\ref{hist_grb_cmb_L20},
\ref{hist_grb_cmb_L150},\ref{hist_grb_cmb_L300},\ref{hist_grb_cmb_L600}).
All maps were
superimposed with the equatorial coordinate system. The upper images
present data for GRBs compared with CMB in the maps with resolution
$\lmax = 20$ (Fig.\,\ref{hist_grb_cmb_L20});
the left one is for long BATSE GRBs, the right one is for
short BeppoSAX GRBs. In the center image ($\lmax = 150$):
the left one is for data of short BATSE GRBs, the right one is
for short BeppoSAX GRBs. In the bottom pair of images the left one is
for long BATSE GRBs ($\lmax = 300$),
the right one is for long BeppoSAX GRBs ($\lmax = 600$).
}
\label{map_grb_l600}
\end{figure}

To search for potential correlations we
counted how many GRBs get to CMB pixels with negative values of signal
fluctuations, which can be due to above effects on CMB maps with
different resolutions. Table\,1 gives the statistics of CMB pixel values
in GRB locations for subsamples of BATSE and BeppoSAX catalogs for
short and long events. It contains the following parameters: the total
number of sources in subsamples, the amount of sources which get to CMB
pixels with a negative value of fluctuations and also the expected
number of pixels with a negative value of CMB amplitude by model data
of 200 realizations of a Gaussian random CMB signal in the $\Lambda$CDM
cosmological model and the 1$\sigma$-dispersion of these values. Data
were obtained both with consideration for the Planck data and without
it.

\begin{table*}[!th]
\setcaptionmargin{5mm}
\onelinecaptionstrue
\caption{
The statistics of CMB pixel values in GRB locations for BATSE and
BeppoSAX subsamples. The columns are:
the duration $t$ (sec); consideration of mask in the Planck data;
the mission; CMB
map resolution (the multipole number); the total amount of GRB sources
in a subsample ($N_t$),
the number of sources ($N_e$),
which are in the CMB pixels with a negative value of fluctuation;
difference $\Delta N$
between $N_e$
and expected average in random realizations; the expected amount of
pixels with a negative amplitude of CMB values from data of 200
realizations of a random Gaussian signal in the $\Lambda$CDM cosmology
and the 1$\sigma$ dispersion of these values.
}
\begin{tabular}{|cclrrrrc|}
\hline
  $t$ & Mask  & Mission&$\lmax$ & $N_t$ & $N_e$  & $\Delta N$ &Simulation\\
\hline
$>$2 &  n    &  BATSE  &   20        & 1540  & 781  & 12    &  769$\pm$32 \\
$>$2 &  y    &  BATSE  &   20        & 1243  & 632  & 8     &  624$\pm$33 \\
$<$2 &  n    &  BATSE  &   20        & 497   & 247  & $-$1  &  248$\pm$13 \\
$<$2 &  y    &  BATSE  &   20        & 394   & 184  & $-$16 &  200$\pm$13 \\
$>$2 &  n    &  BATSE  &   150       & 1540  & 772  & 4     &  768$\pm$19 \\
$>$2 &  y    &  BATSE  &   150       & 1242  & 638  & 16    &  622$\pm$21 \\
$<$2 &  n    &  BATSE  &   150       & 497   & 248  & $-$1  &  249$\pm$11 \\
$<$2 &  y    &  BATSE  &   150       & 403   & 200  & $-$8  &  208$\pm$11 \\
$>$2 &  y    &  BATSE  &   300       & 1248  & 655  & 33    &  622$\pm$21 \\
$<$2 &  y    &  BATSE  &   300       & 409   & 207  & 2     &  205$\pm$10 \\
$>$2 &  y    &  BATSE  &   600       & 1244  & 625  & 4     &  621$\pm$20 \\
$<$2 &  y    &  BATSE  &   600       & 413   & 220  & 12    &  208$\pm$10 \\
\hline
$>$2 &  n    & Bepposax&   20        & 694   & 343  & $-$5  &  348$\pm$17 \\
$>$2 &  y    & Bepposax&   20        & 555   & 272  & $-$15 &  287$\pm$19 \\
$<$2 &  n    & Bepposax&   20        & 87    & 44   & 0     &  44 $\pm$5  \\
$<$2 &  y    & Bepposax&   20        & 67    & 34   & 1     &  33 $\pm$4  \\
$>$2 &  n    & Bepposax&   150       & 694   & 327  & $-$30 &  347$\pm$15 \\
$>$2 &  y    & Bepposax&   150       & 562   & 271  & $-$9  &  280$\pm$15 \\
$<$2 &  n    & Bepposax&   150       & 87    & 45   & 1     &  44 $\pm$5  \\
$<$2 &  y    & Bepposax&   150       & 66    & 34   & 1     &  33 $\pm$4  \\
$>$2 &  y    & Bepposax&   300       & 559   & 279  & $-$3  &  282$\pm$13 \\
$<$2 &  y    & Bepposax&   300       & 65    & 32   & $-$1  &  33 $\pm$4  \\
$>$2 &  y    & Bepposax&   600       & 563   & 266  & $-$15 &  281$\pm$13 \\
$<$2 &  y    & Bepposax&   600       & 68    & 34   & 0     &  34 $\pm$4  \\
\hline
\end{tabular}
\end{table*}

\begin{figure}[!h]
\setcaptionmargin{5mm}
\onelinecaptionstrue
\centerline{\vbox{
\hbox{
\psfig{figure=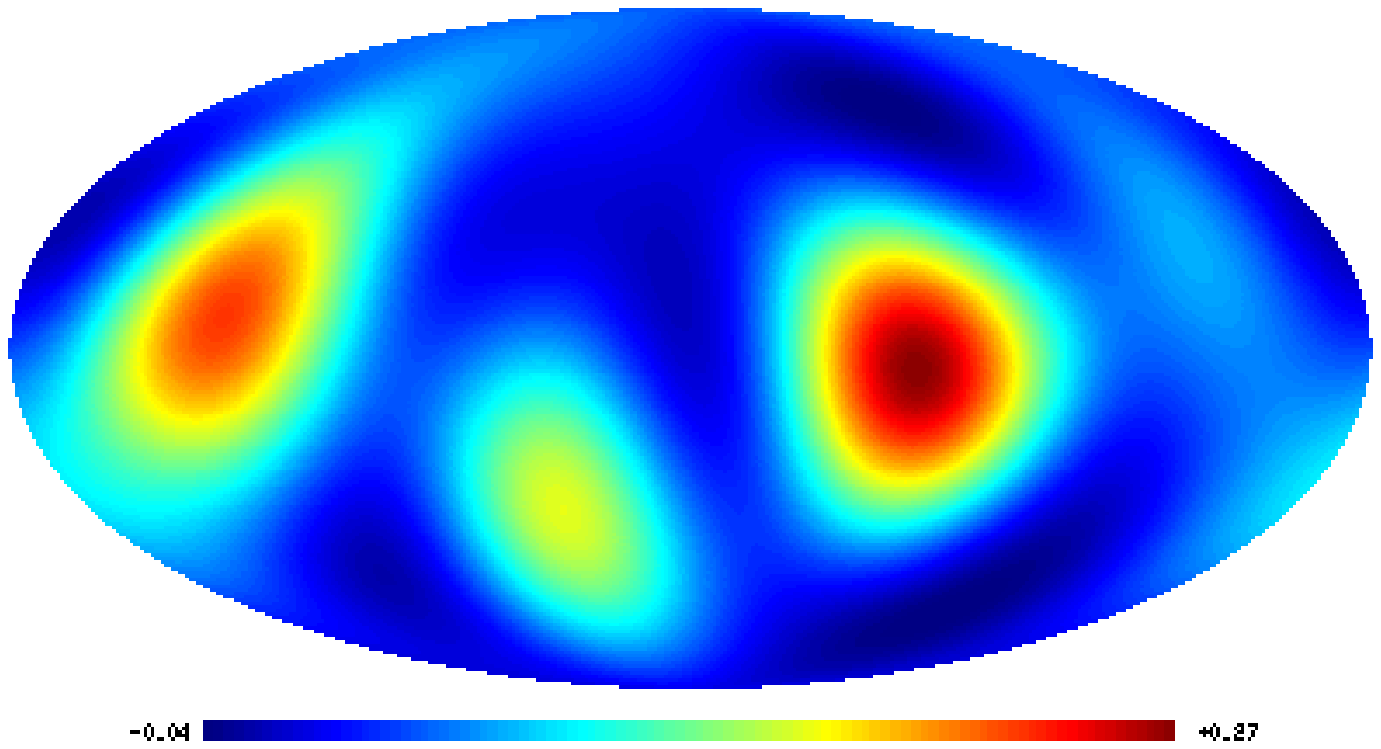,width=8cm}
\psfig{figure=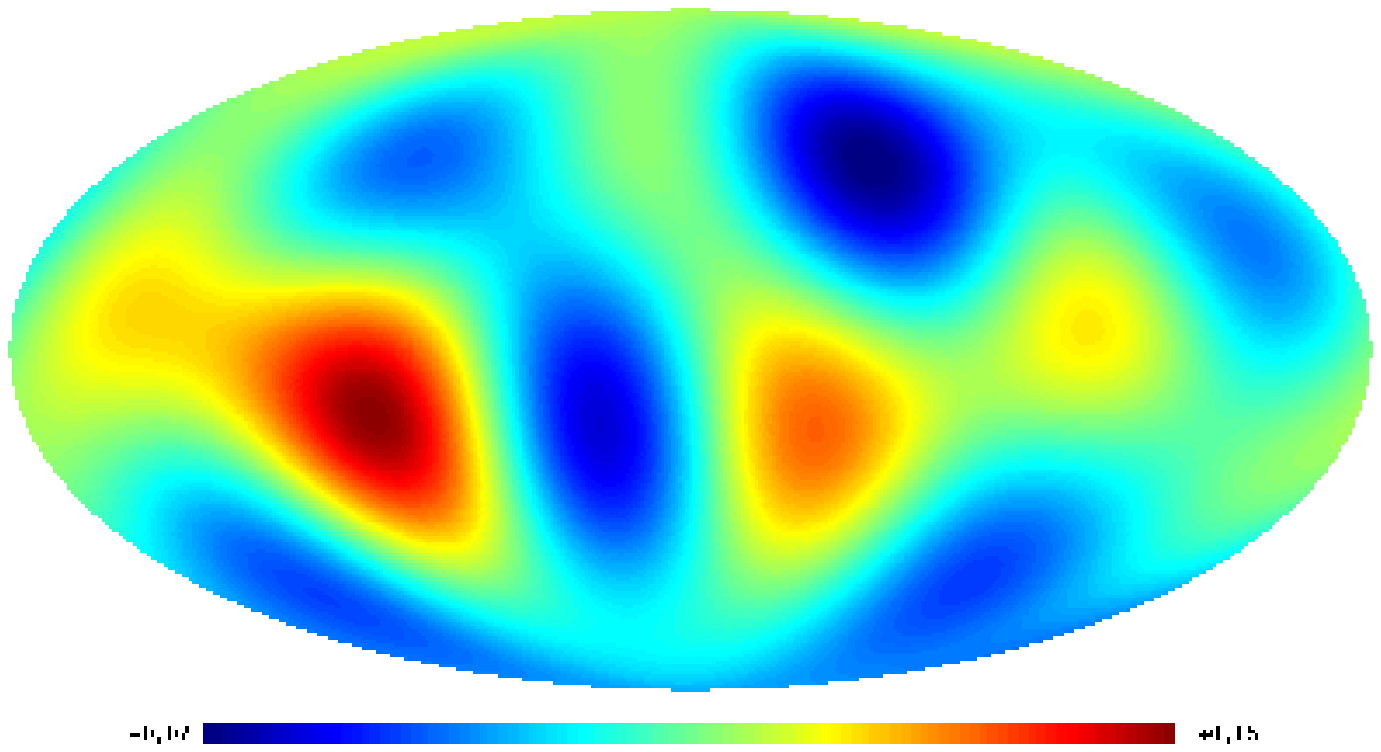,width=8cm}
}
}}
\caption{
The octupole of the smoothed maps of the sky
(Fig.\,\ref{map_grb_l600}) of short BeppoSAX
GRBs locations for \lmax=20 and \lmax=150.
}
\label{map_grb_l600_L2}
\end{figure}

Figures
\ref{hist_grb_cmb_L20}, \ref{hist_grb_cmb_L150},
\ref{hist_grb_cmb_L300}, \ref{hist_grb_cmb_L600}
present diagrams of
distribution of CMB fluctuation values for four subsamples of GRBs and
CMB maps with different resolutions. The dashed lines show the expected
1$\sigma$-dispersion of CMB values in the $\Lambda$CDM cosmological
model. In our previous work \cite{v_grb}  we discovered a deviation from what
was expected with the Gaussian random CMB signal in distribution of
fluctuation values with the resolution $\lmax=150$
in regions of short BATSE GRBs. In the Planck data the deviations are
also observed for short GRBs (see two upper diagrams in
Fig.\,\ref{hist_grb_cmb_L150}).
Besides, there are small deviations from the models for long BATSE GRBs
and short BeppoSAX GRBs at $\lmax=20$
(the second and thirds pairs of diagrams in Fig.\ref{hist_grb_cmb_L20}), short
BeppoSAX GRBs at $\lmax=150$
(the third diagram in Fig.\,\ref{hist_grb_cmb_L150}), long BATSE GRBs at
$\lmax=300$
(the right upper diagram in Fig.\,\ref{hist_grb_cmb_L300}),
and long BeppoSAX GRBs at
$\lmax=600$
(the right bottom diagram in Fig.\,\ref{hist_grb_cmb_L600}).
To analyze the sphere
distribution of those GRBs which are in directions where the detected
signal deviates from the expected one, we made pixelization with the
`{\tt mappat}' procedure from the
software package GLESP \cite{glesp2}. The pixel size
700\arcmin$\times$700\arcmin\
was chosen in such a way that the
maximum pixel value (the number of events in a corresponding area)
would be not less than 3 and the significant dynamic range for harmonic
analysis would be provided. Fig.\,\ref{map_grb_l600}
shows so pixelized and smoothed maps
of GRBs whose contribution into histogram exceeds the expected
1$\sigma$-dispersion. The equatorial coordinate system is plotted on all maps.
All images demonstrate a non-uniform sphere distribution of events
concentrated near the equatorial poles. In many cases, the hot spots are
located directly in the equatorial poles. In maps of $\lmax=20$
and $\lmax=150$
with short BeppoSAX GRBs we observe clusterization of hot spots in the
Galaxy plane, which is more noticeable on octupoles of these maps
(Fig.\,\ref{map_grb_l600_L2}).
In some cases, the presence or absence of events in the pole
regions practically repeat the pole regions of the equatorial
coordinate system.

Note that the peculiarities of CMB data
of the Planck mission are identical to those of WMAP data as compared
with GRB locations. Namely, the deviations in the pixel statistics are
related with signal in a GRB direction oriented in the equatorial
coordinate system. In paper \cite{v_grb}, we estimated probability of the
getting of quadrupole minimums in regions of the 5\degr\, radius  around
the equatorial poles. To do that, we generated 10000 random
realizations of the Gaussian signal to pixelize GLESP with 102 pixels in
the equator. Probability of the random getting to the pole zone is
0.0035.

For a more detailed study of correlation
effects, the mosaic correlation method presented in paper
\citep{cormap} was applied.

\section{Correlation maps of distribution of
gamma-ray bursts and Planck SMICA data}
\label{cmb_corr_grb}

\begin{figure} [!h]
\setcaptionmargin{5mm}
\onelinecaptionstrue
\centerline{\vbox{
\hbox{
\psfig{figure=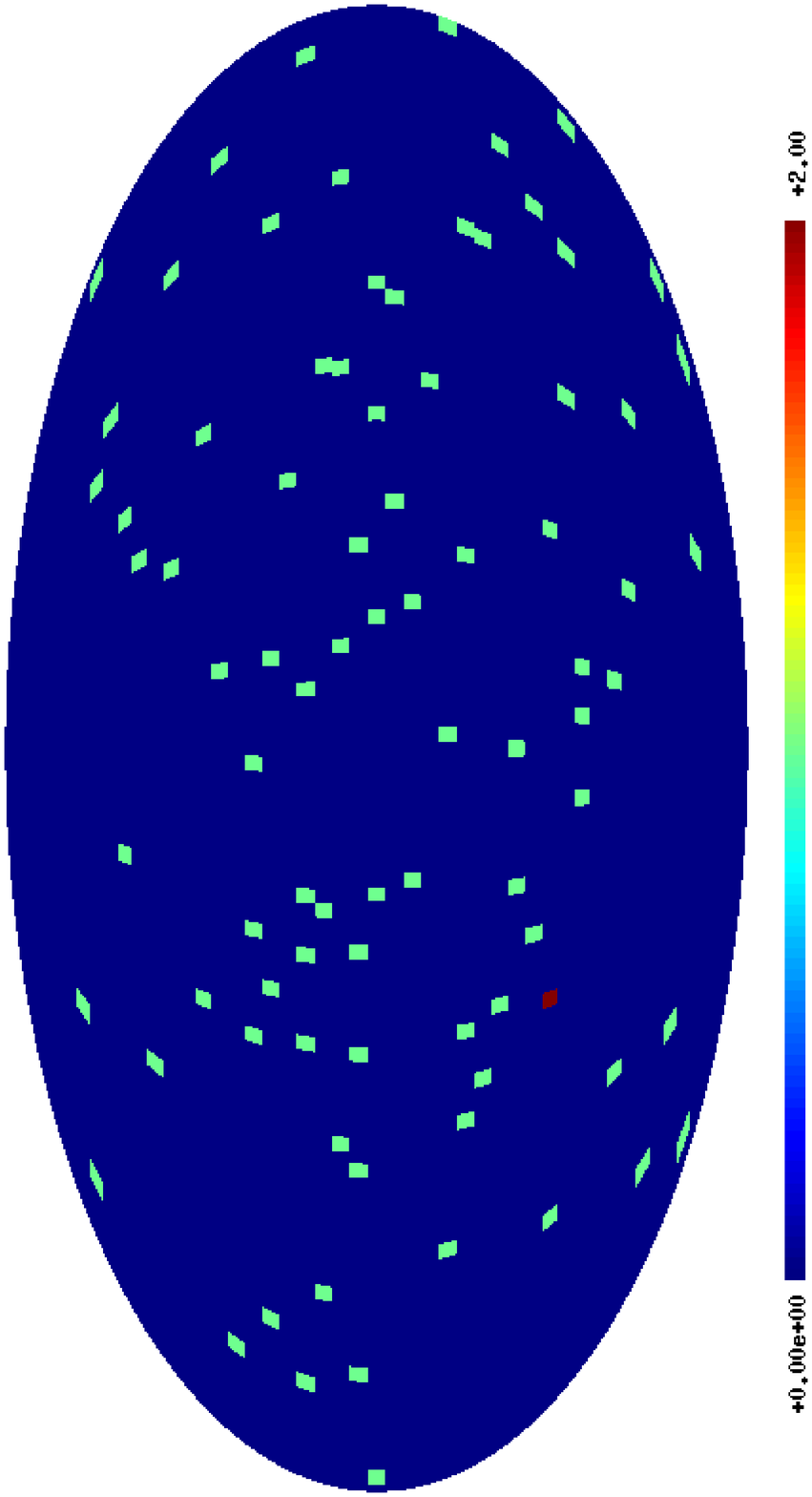,angle=-90,width=7cm,%
bbllx=0pt,bblly=0pt,bburx=500pt,bbury=820pt,clip=}
\psfig{figure=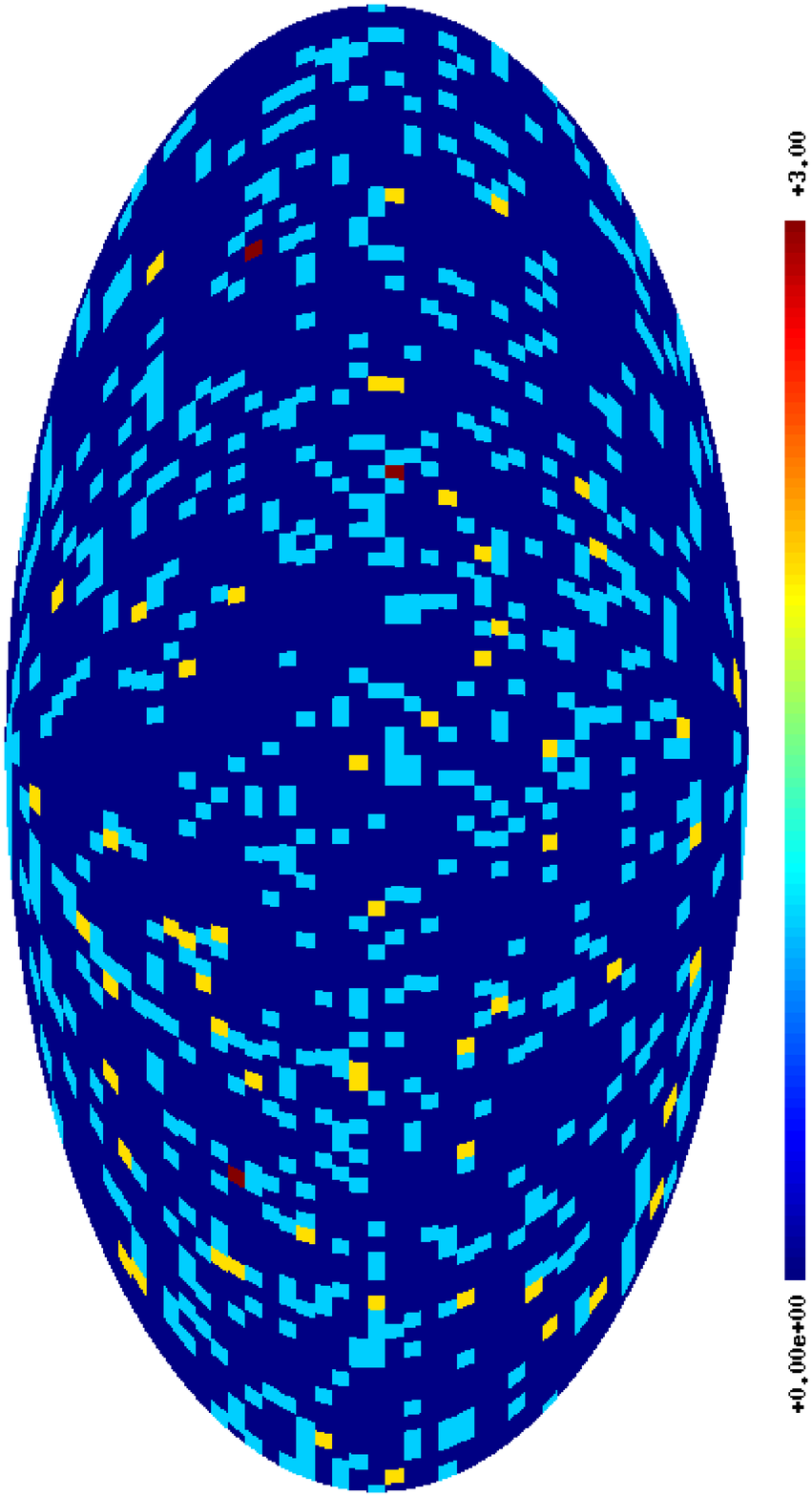,angle=-90,width=7cm,%
bbllx=0pt,bblly=0pt,bburx=500pt,bbury=820pt,clip=}
}
\hbox{
\psfig{figure=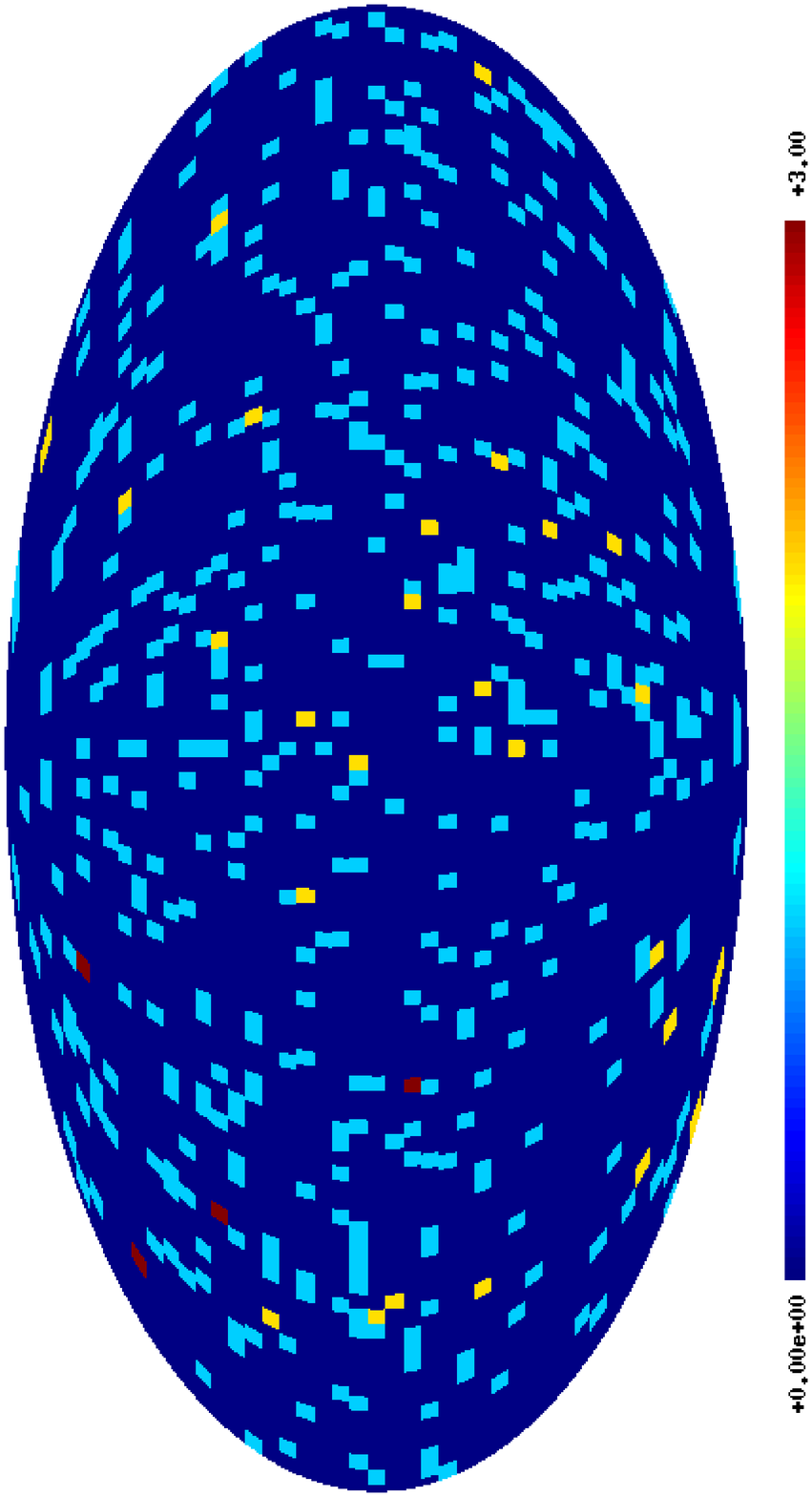,angle=-90,width=7cm,%
bbllx=0pt,bblly=0pt,bburx=500pt,bbury=820pt,clip=}
\psfig{figure=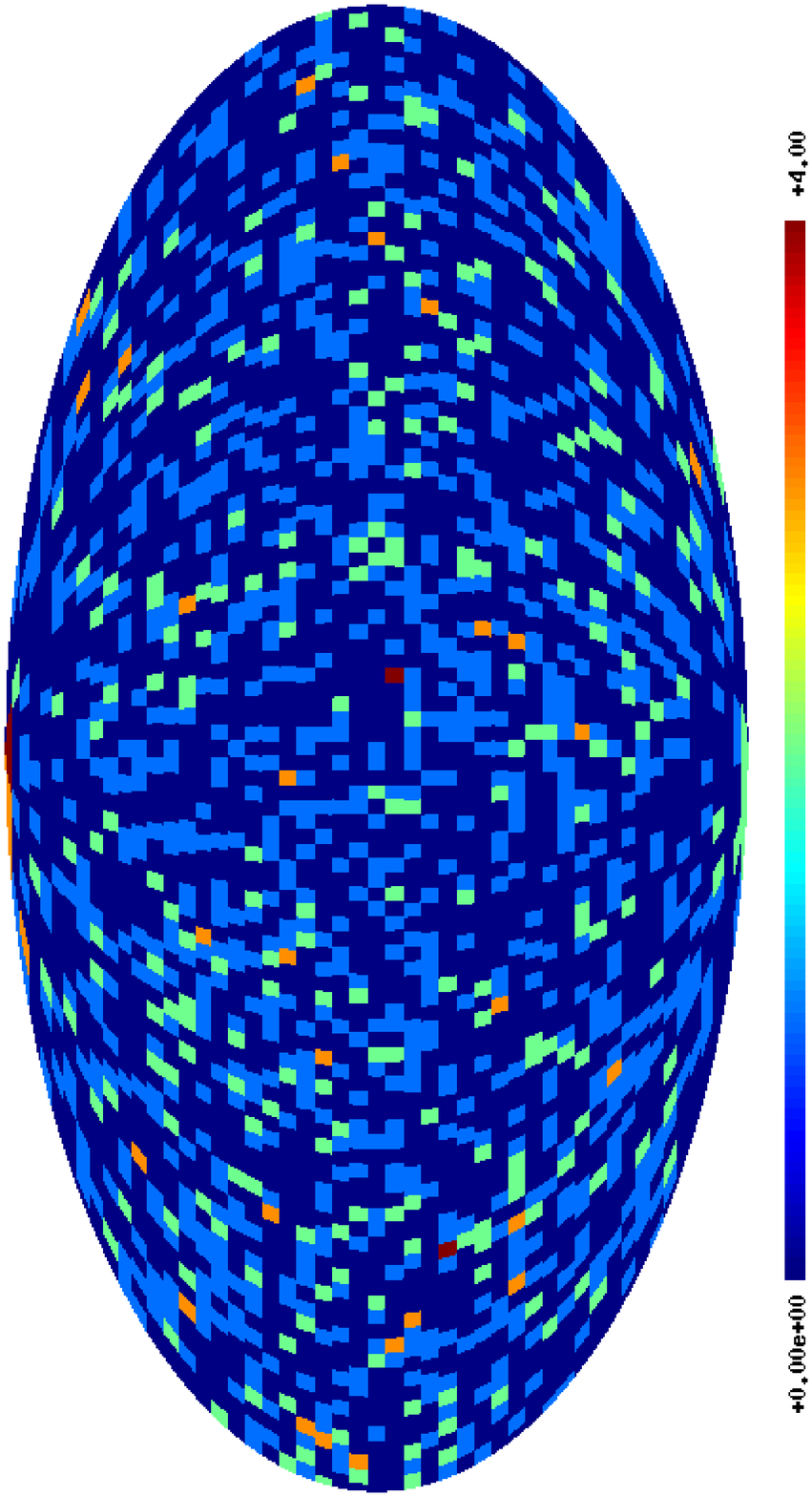,angle=-90,width=7cm,%
bbllx=0pt,bblly=0pt,bburx=500pt,bbury=820pt,clip=}
}}}
\caption{
Pixelized maps of subsample GRB locations. The pixel size is
200'$\times$200'. The top left image shows the BeppoSAX
data, $t<2$\,sec.
The top right image presents the BeppoSAX data, $t>2$\,sec.
The bottom left image shows the BATSE data, $t<2$\,sec.
The bottom right image is for the BATSE data,  $t>2$\,sec.
}
\label{grb_pix_map}
\end{figure}

\begin{figure} [!h]
\setcaptionmargin{5mm}
\onelinecaptionstrue
\centerline{\vbox{
\hbox{
\psfig{figure=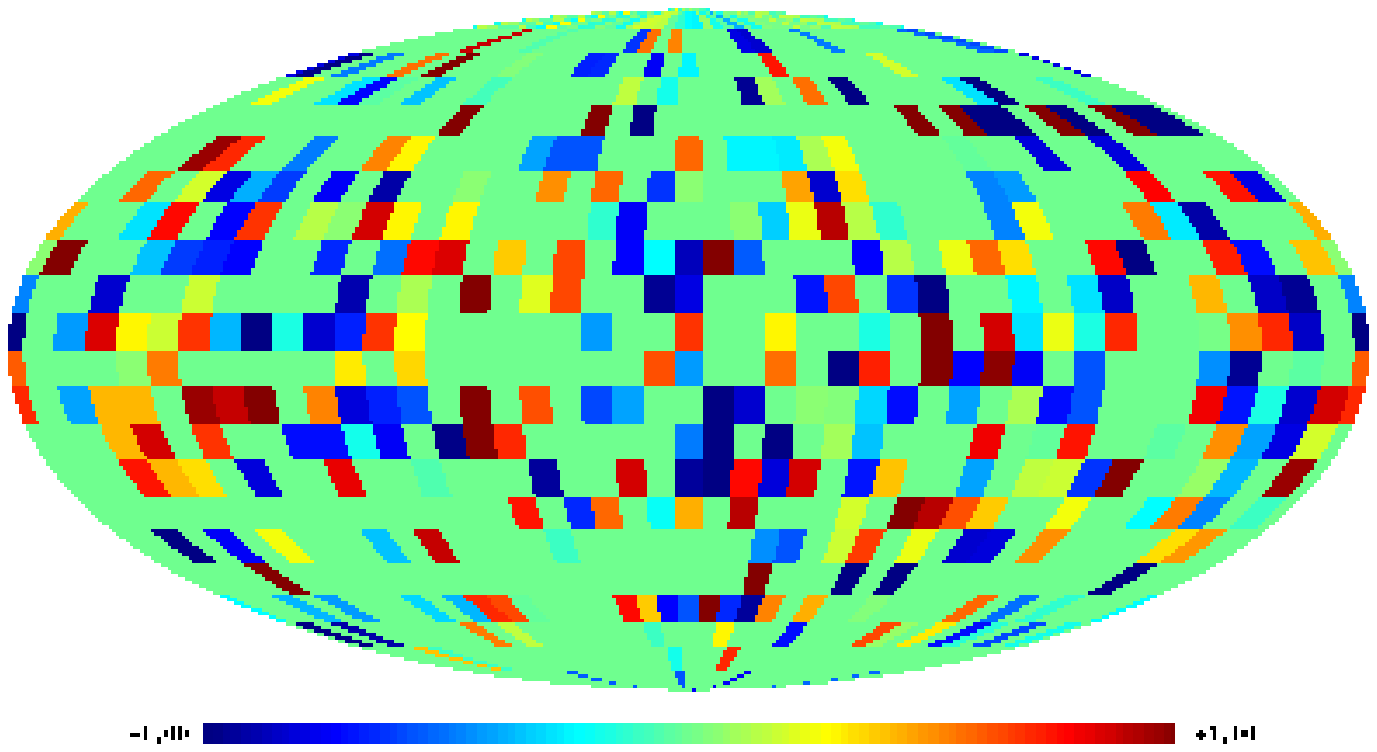,width=7cm}
\psfig{figure=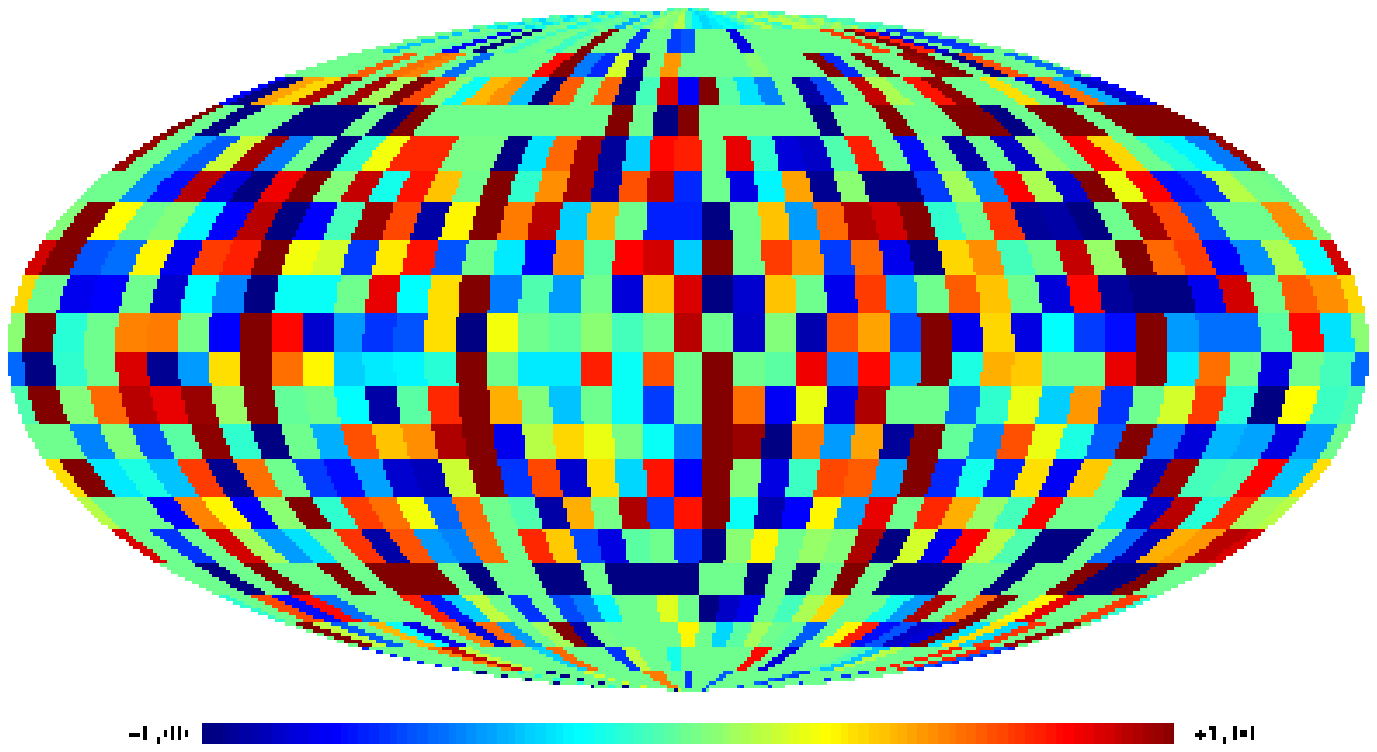,width=7cm}
}
\hbox{
\psfig{figure=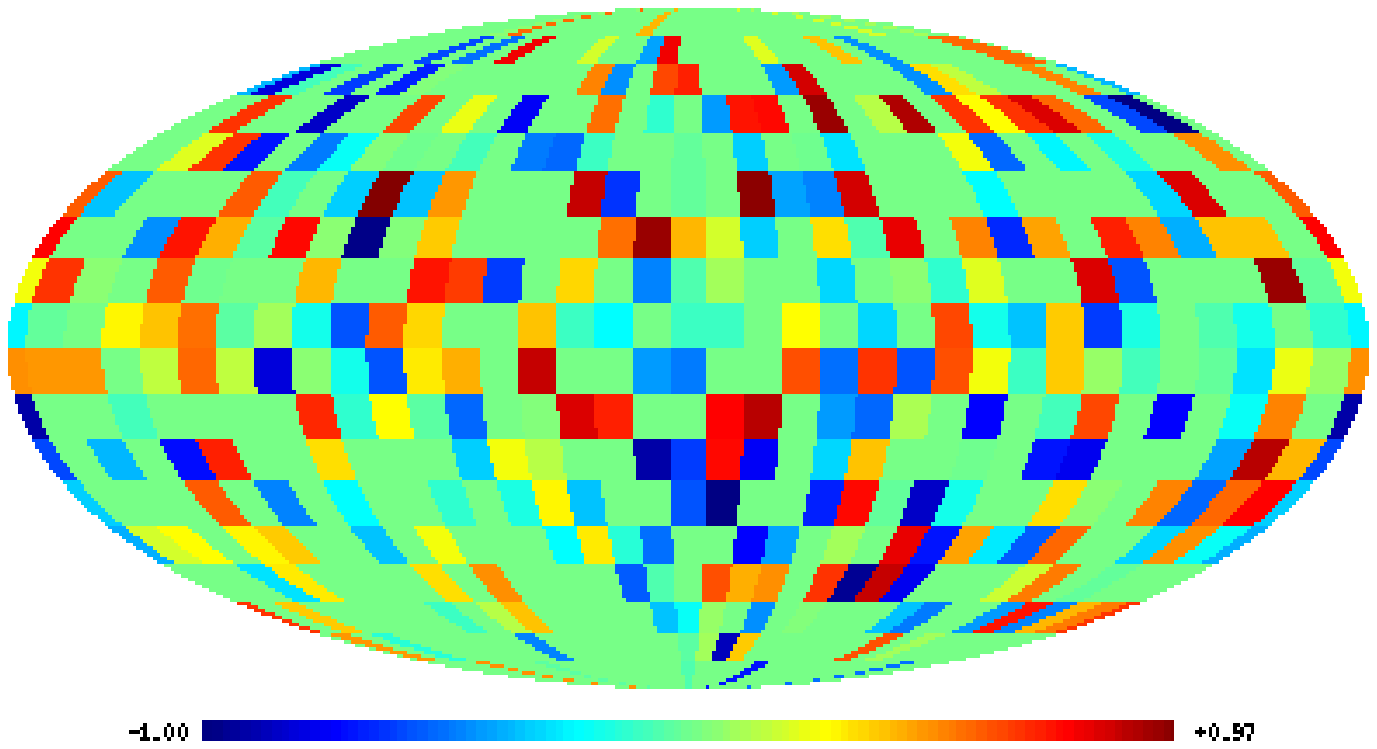,width=7cm}
\psfig{figure=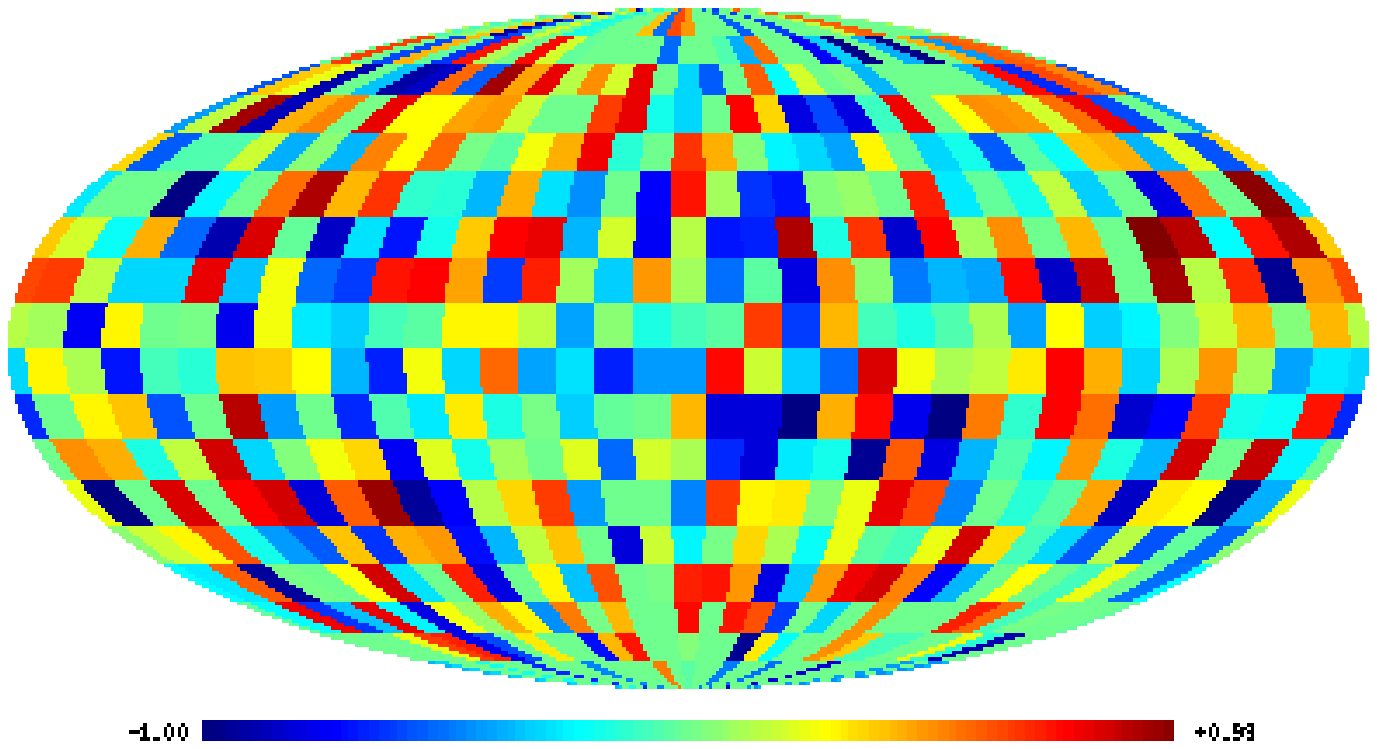,width=7cm}
}
\hbox{
\psfig{figure=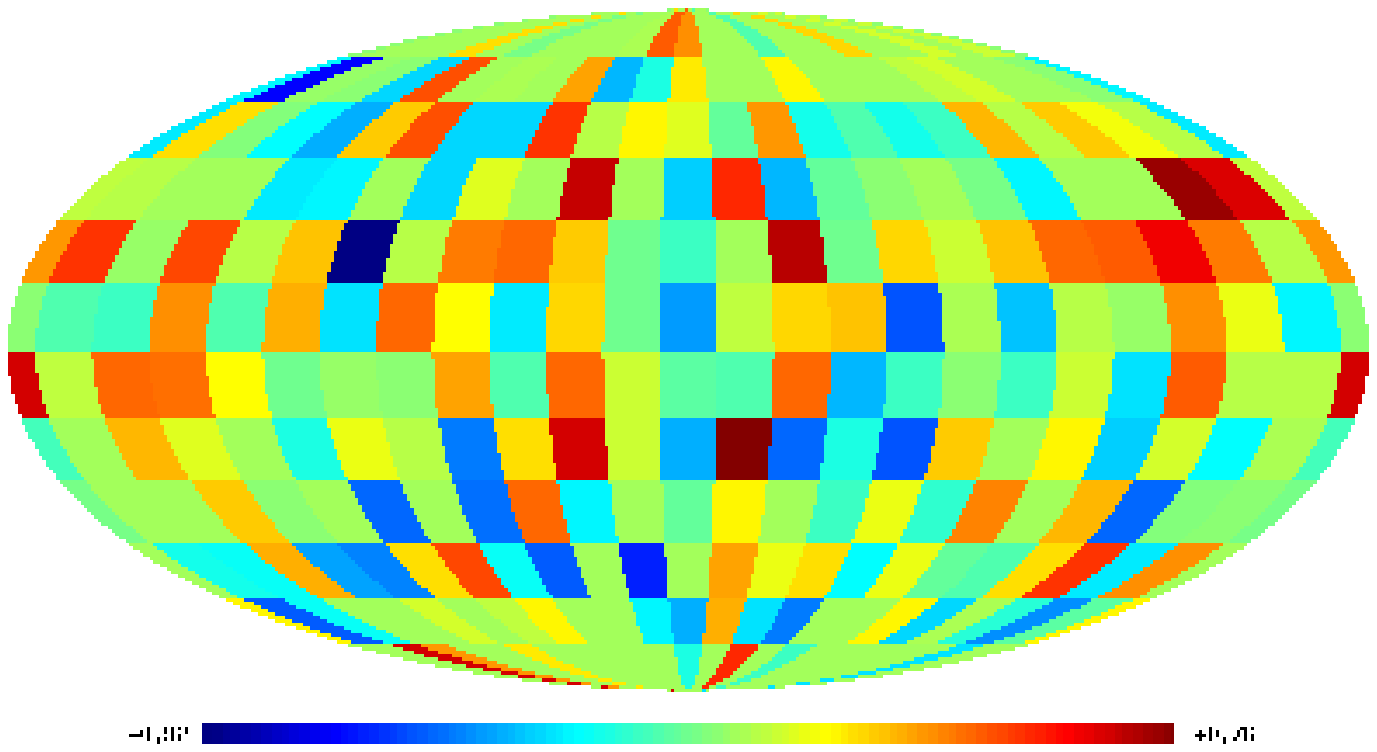,width=7cm}
\psfig{figure=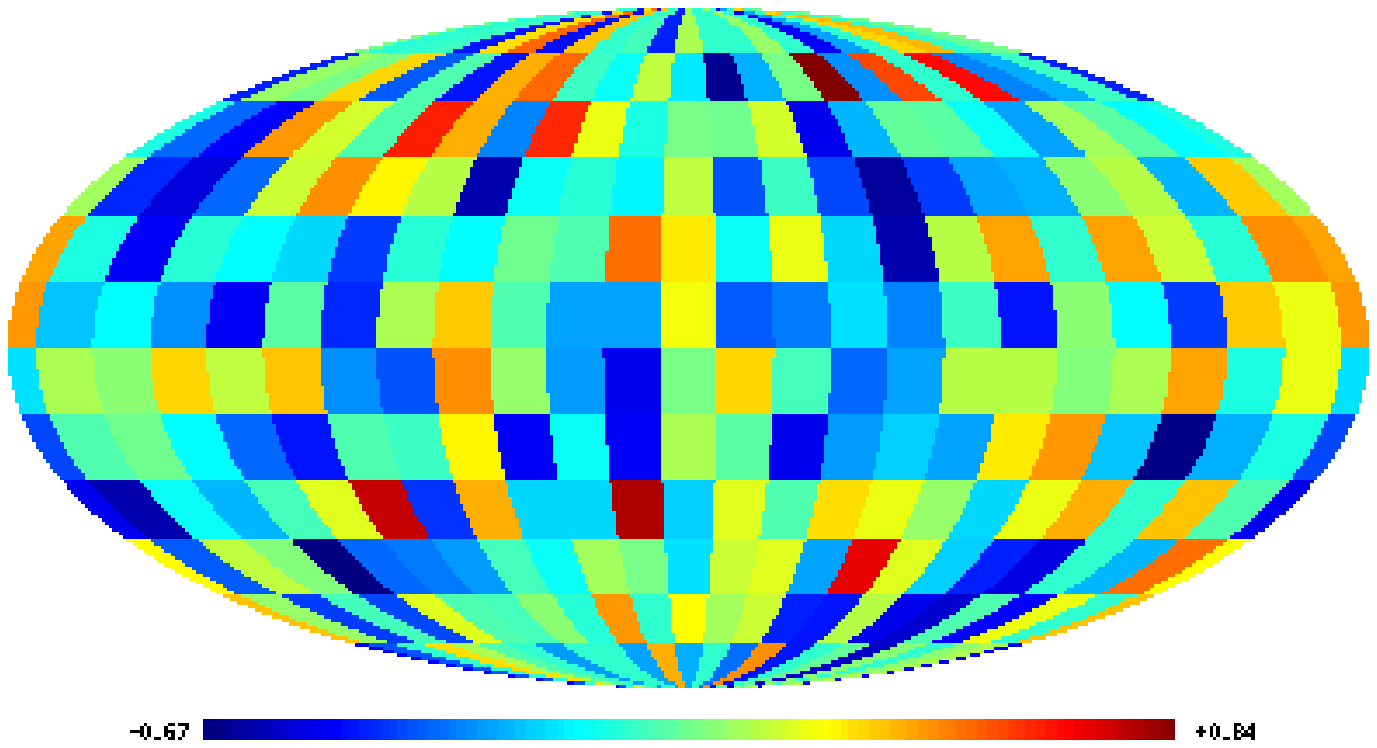,width=7cm}
}}}
\caption{
Correlation maps of CMB and BATSE GRB locations in the galactic
coordinate system. The left column presents results of the CMB and
BATSE ($t<2$\,sec)
data correlations, and the right one is for CMB and BATSE ($t>2$\,sec).
The upper pair of images demonstrates maps of $\lmax=26$
and the mosaic correlation pixel size of 500'$\times$500'.
The central pair of images shows the maps of $\lmax=8$
and the mosaic correlation pixel size of 600'$\times$600'.
The lower pair of images is for $\lmax=5$
and the mosaic correlation pixel size of 900'$\times$900'.
}
\label{grb_cor_map}
\end{figure}

To study properties of maps of GRB
locations and CMB fluctuations we fulfilled the mosaic correlation of
``BATSE --- CMB'' maps with pixels of
different sizes
500\arcmin$\times$500\arcmin, 600\arcmin$\times$600\arcmin and
900\arcmin$\times$900\arcmin
covering areas within which the correlation factor was calculated. To
do that, first we pixelized the maps of GRB locations in all four
subsamples (Fig.\ref{grb_pix_map}). As in the previous stage, the pixel size
of 200'$\times$200'
was chosen in such a way that the
maximum values (the number of events in a corresponding area) would be
not less than 3. The correlation results are shown in Fig.\,\ref{grb_cor_map}.

To analyze the obtained result we
calculated the angular power spectrum of map (\ref{eq2}) using the spherical
harmonics (multipoles) expansion of signal distributed on a sphere
(\ref{eq1}):
\begin{equation}
\Delta S(\theta,\phi)= \sum_{\ell=1}^{\infty}\sum_{m=-\ell}^{m=\ell}
	   a_{\ell m} Y_{\ell m} (\theta, \phi)\,,
\label{eq1}
\end{equation}

\begin{equation}
C(\ell) =
   \frac{1}{2\ell+1}\left[|a_{\ell 0}|^2 +2\sum_{m=1}^\ell
	 |a_{\ell,m}|^2\right]\,.
\label{eq2}
\end{equation}

\begin{figure} [!h]
\setcaptionmargin{5mm}
\onelinecaptionstrue
\centerline{\vbox{
\hbox{
\psfig{figure=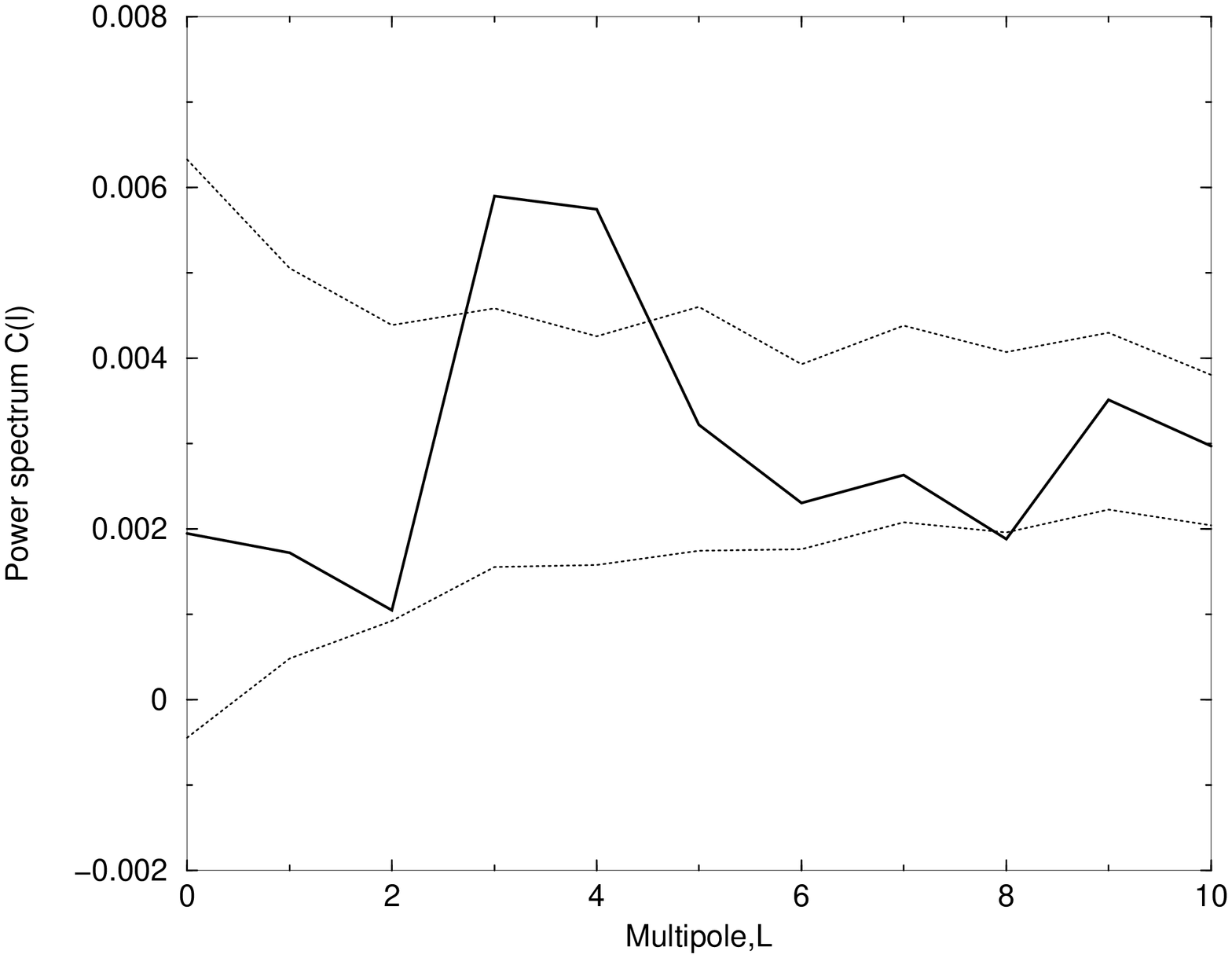,angle=-90,width=7cm}
\psfig{figure=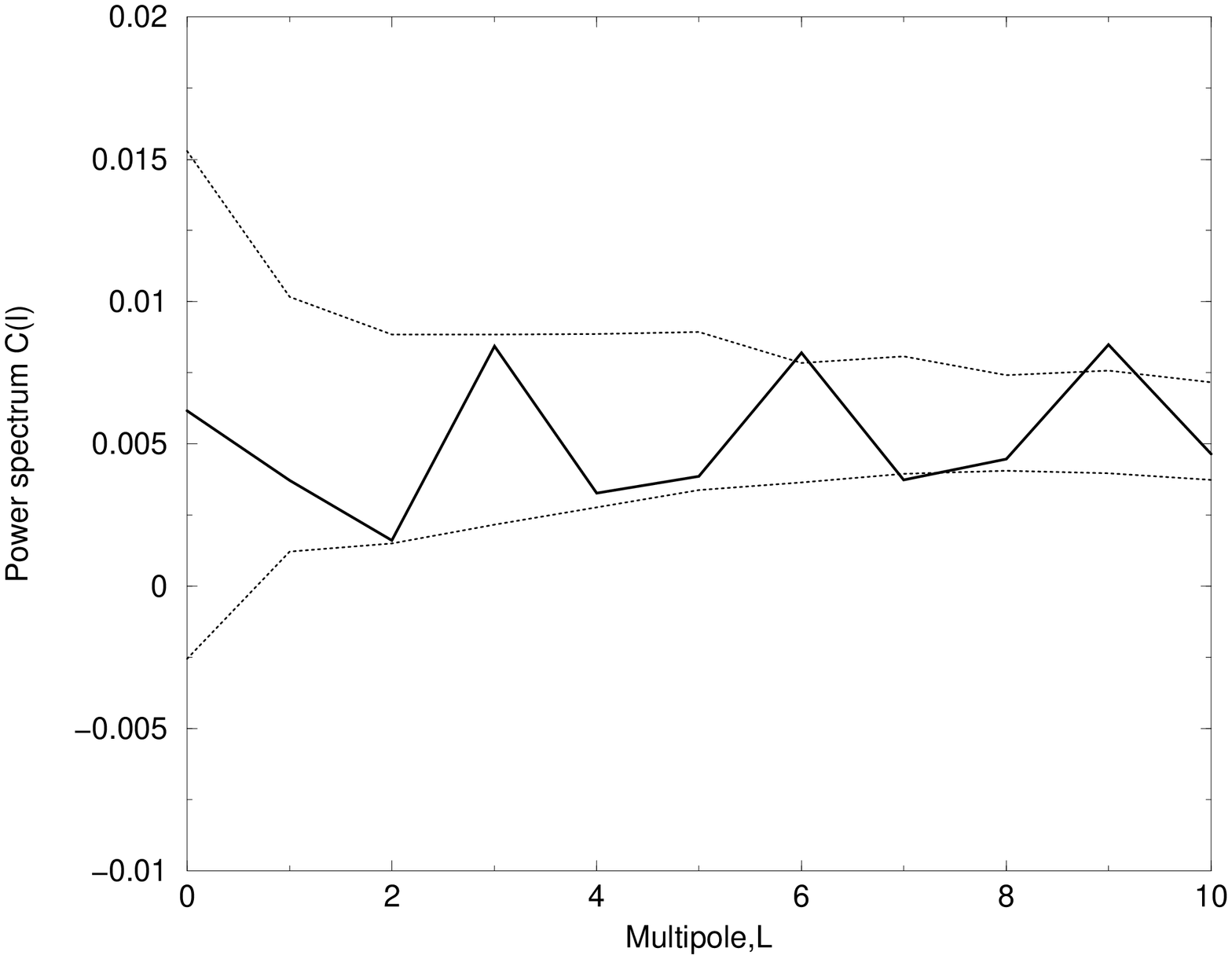,angle=-90,width=7cm}
}
\hbox{
\psfig{figure=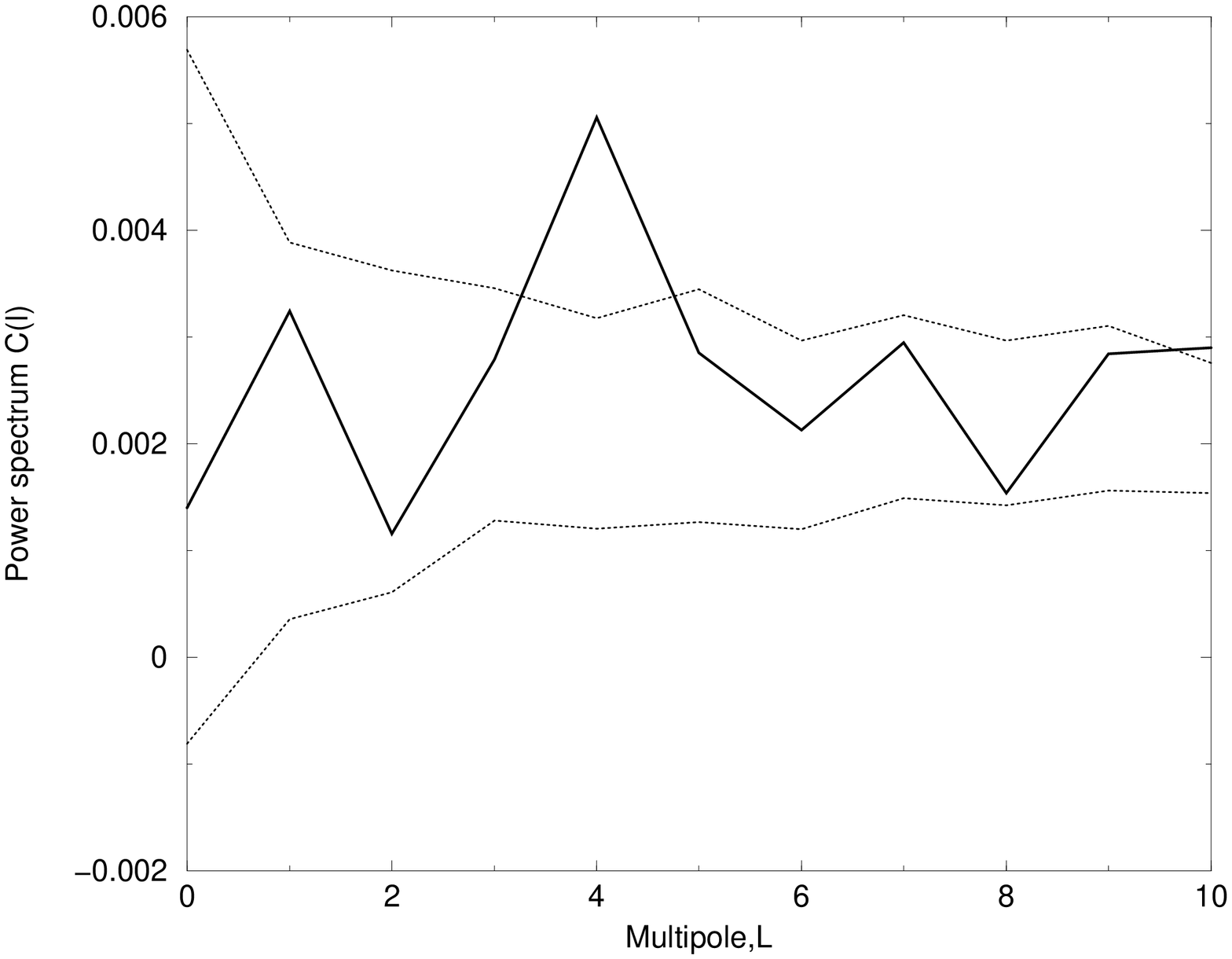,angle=-90,width=7cm}
\psfig{figure=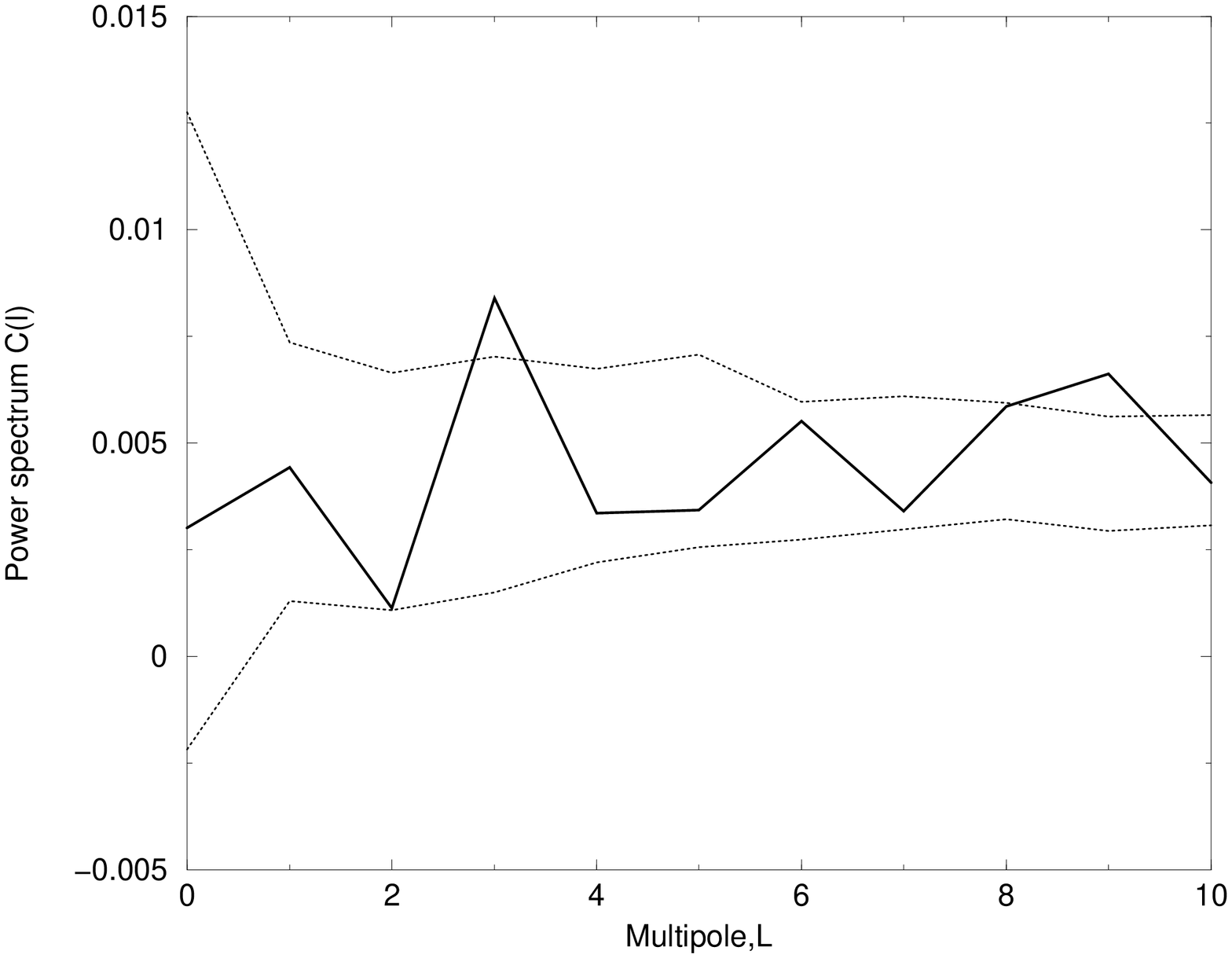,angle=-90,width=7cm}
}
}}
\caption{
Power spectra of of correlation factor maps (with the resolution of
$\lmax=26$)
calculated for the maps of BATSE GRBs locations and CMB distribution
(the solid line). The correlation pixel size is
500\arcmin$\times$500\arcmin. The left top image shows the
correlation spectrum of BATSE data for $t<2$\,sec
and CMB without consideration for the mask. The right top image
contains the correlation spectrum of the BATSE data with $t>2$\,sec
and CMB without consideration for the mask. The bottom left image
demonstrates the correlation spectrum of the BATSE data for $t<2$\,sec
and CMB with consideration for the mask. The right bottom image
shows the correlation spectrum of the BATSE data with $t>2$\,sec
and CMB with consideration for the mask. The images show the
1$\sigma$--dispersion obtained from results of analysis of 200 Gaussian
random realizations of CMB.
}
\label{sp_cor_l26}
\end{figure}

\begin{figure} [!h]
\setcaptionmargin{5mm}
\onelinecaptionstrue
\centerline{\vbox{
\hbox{
\psfig{figure=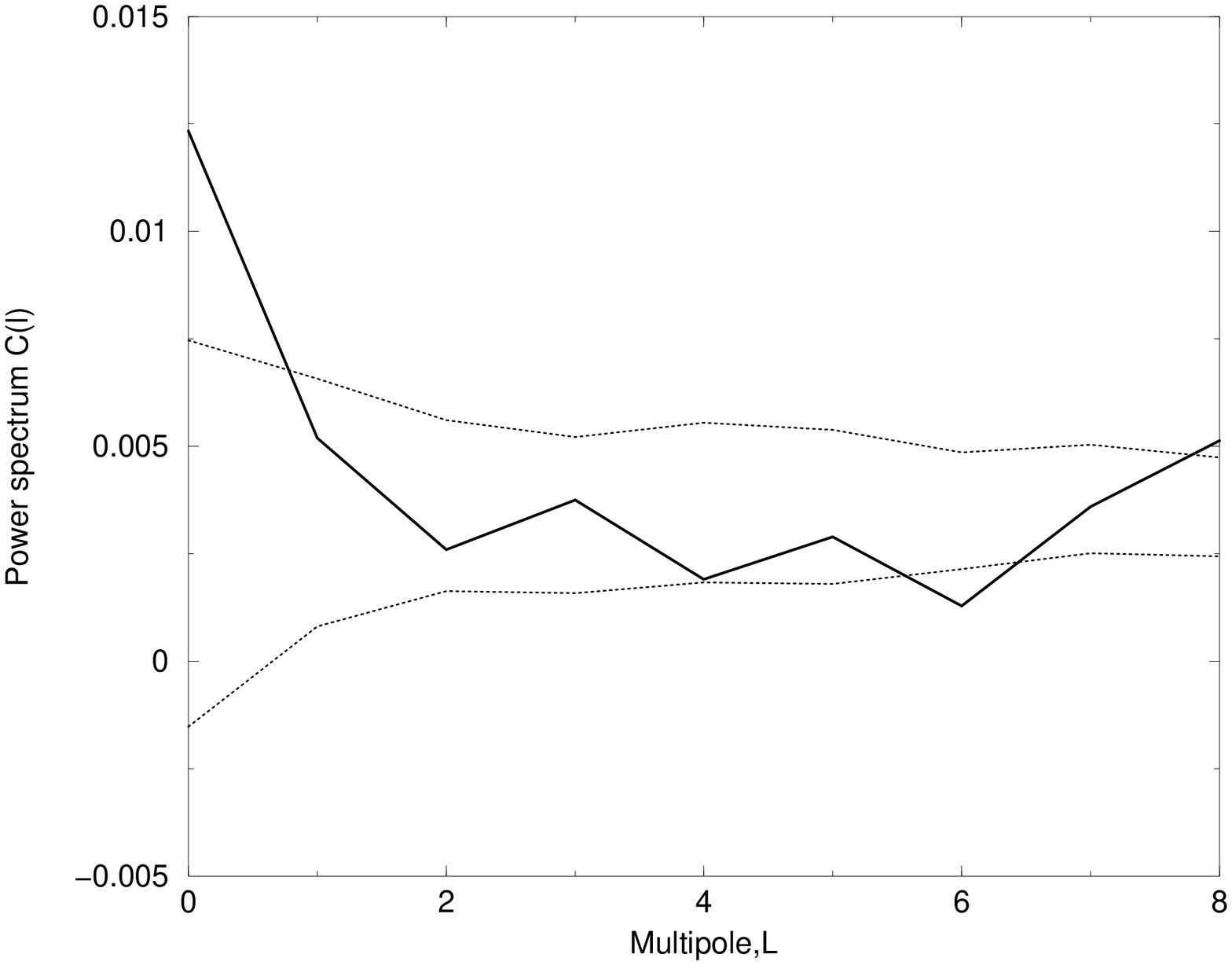,angle=-90,width=7cm}
\psfig{figure=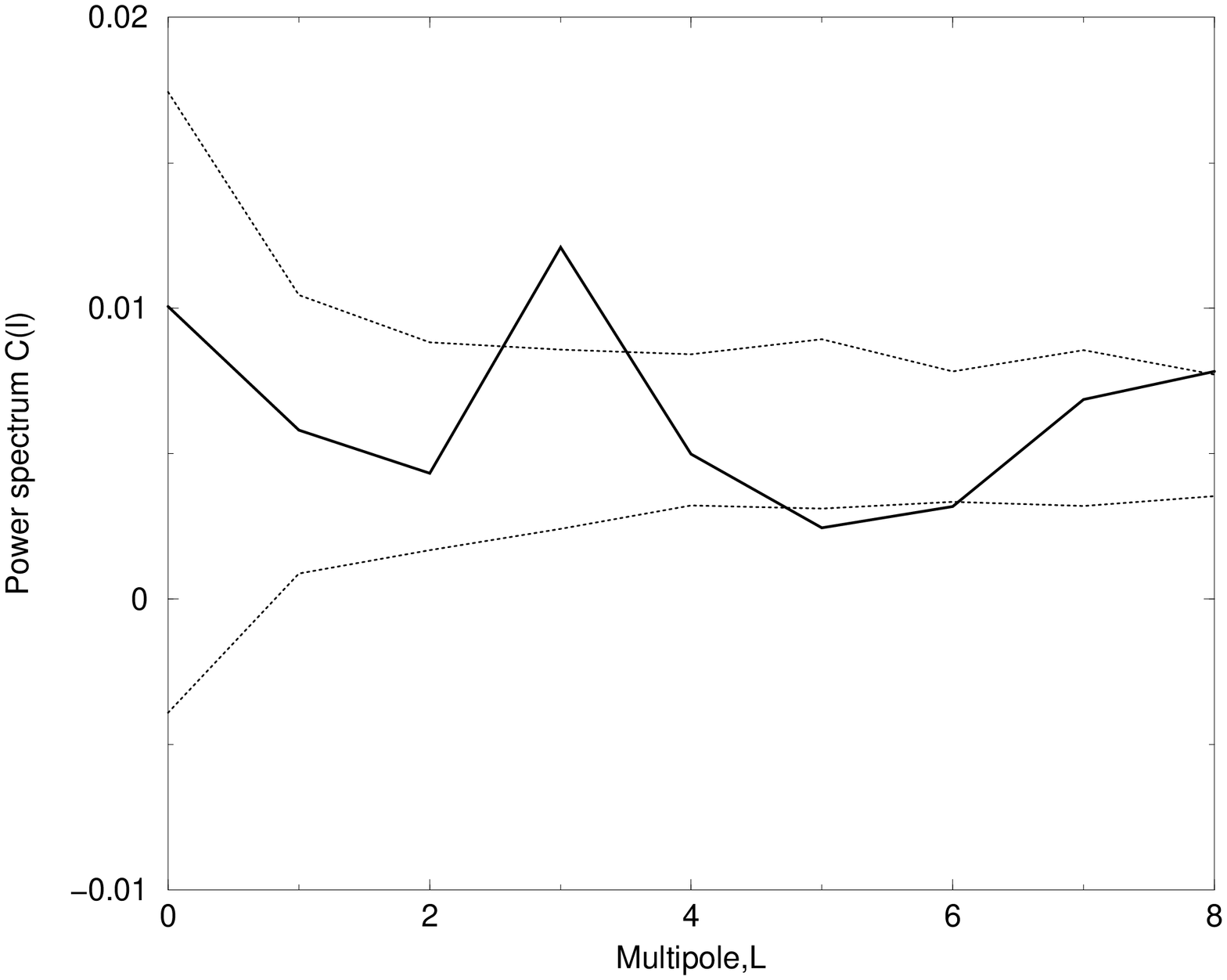,angle=-90,width=7cm}
}
\hbox{
\psfig{figure=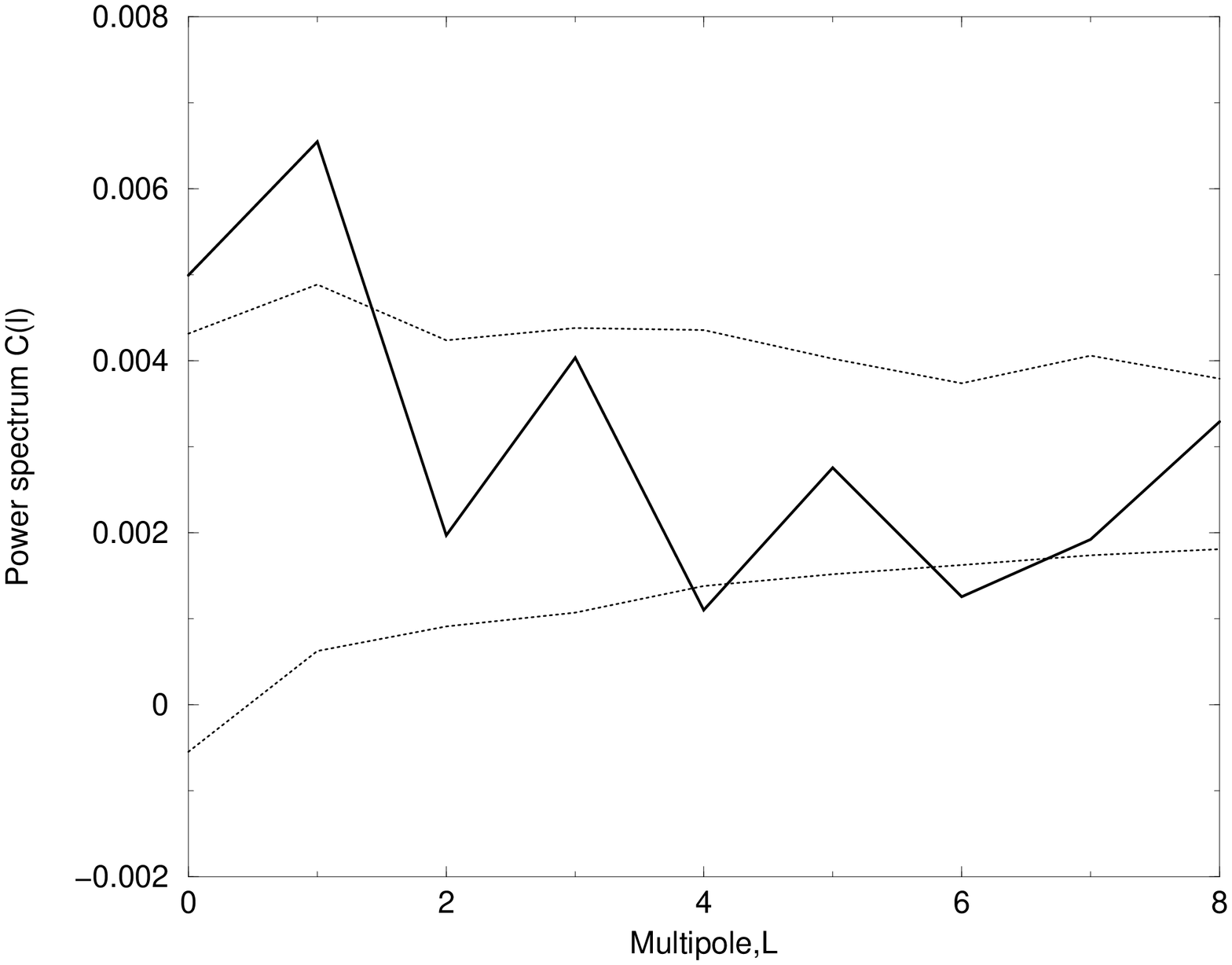,angle=-90,width=7cm}
\psfig{figure=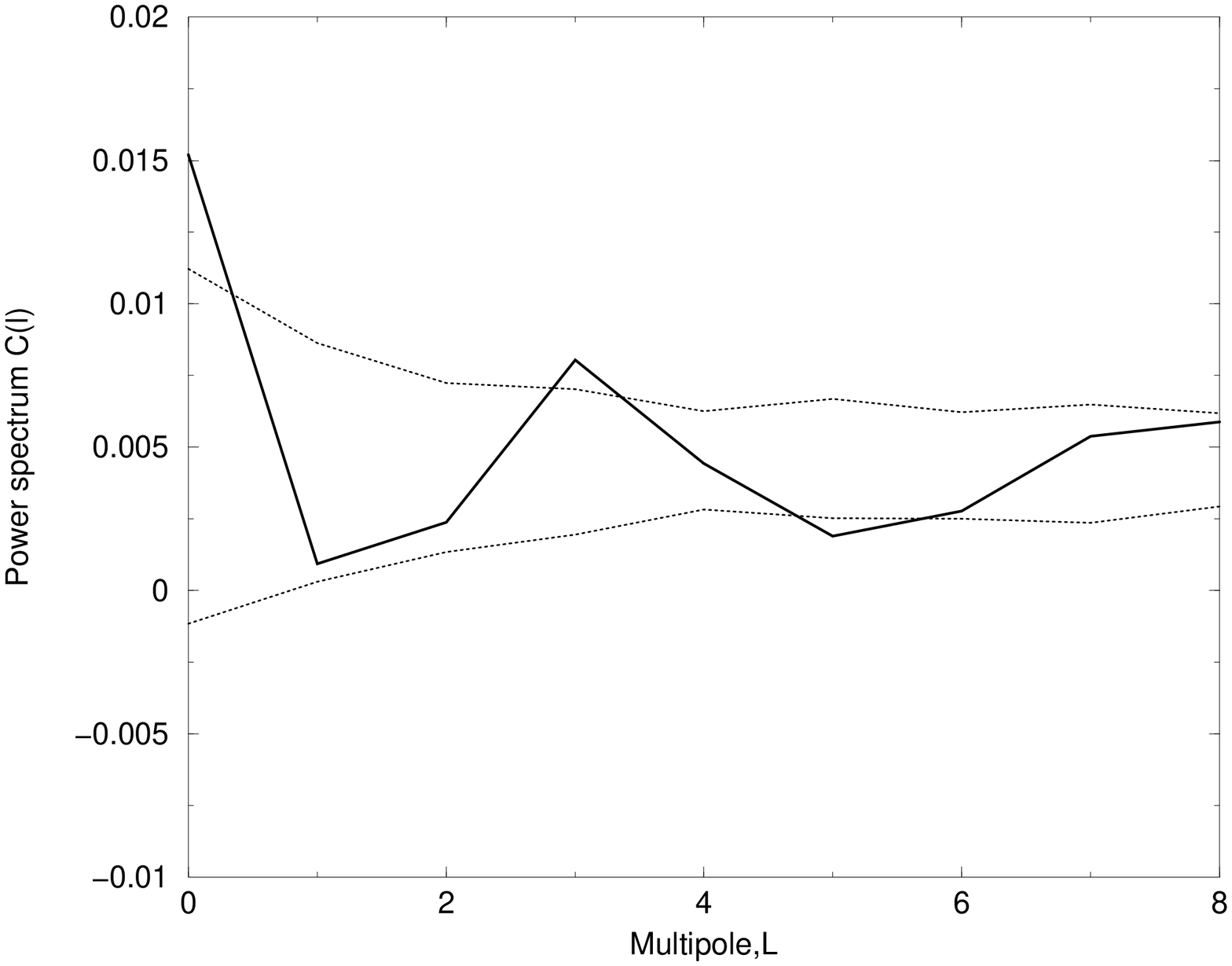,angle=-90,width=7cm}
}
}}
\caption{
Power spectra of correlation maps (with the resolution of $\lmax=8$)
calculated for the maps of BATSE GRBs locations and CMB distribution
(the solid line). The correlation pixel size is
600\arcmin$\times$600\arcmin.
The left top image shows the spectrum
of correlation map for the BATSE data (GRBs with $t<2$\,sec)
and CMB without masking. The right top image
contains the correlation spectrum of the BATSE data for $t>2$\,sec
and CMB without masking. The bottom left image
demonstrates the correlation spectrum of the BATSE data for $t<2$\,sec
and CMB with masking. The right bottom image
shows the correlation spectrum of BATSE data for $t>2$\,sec
and CMB with masking. The images show the
1$\sigma $ dispersion obtained from results of analysis of 200 Gaussian
random realizations of CMB.
}
\label{sp_cor_l8}
\end{figure}

\begin{figure} [!h]
\setcaptionmargin{5mm}
\onelinecaptionstrue
\centerline{\vbox{
\hbox{
\psfig{figure=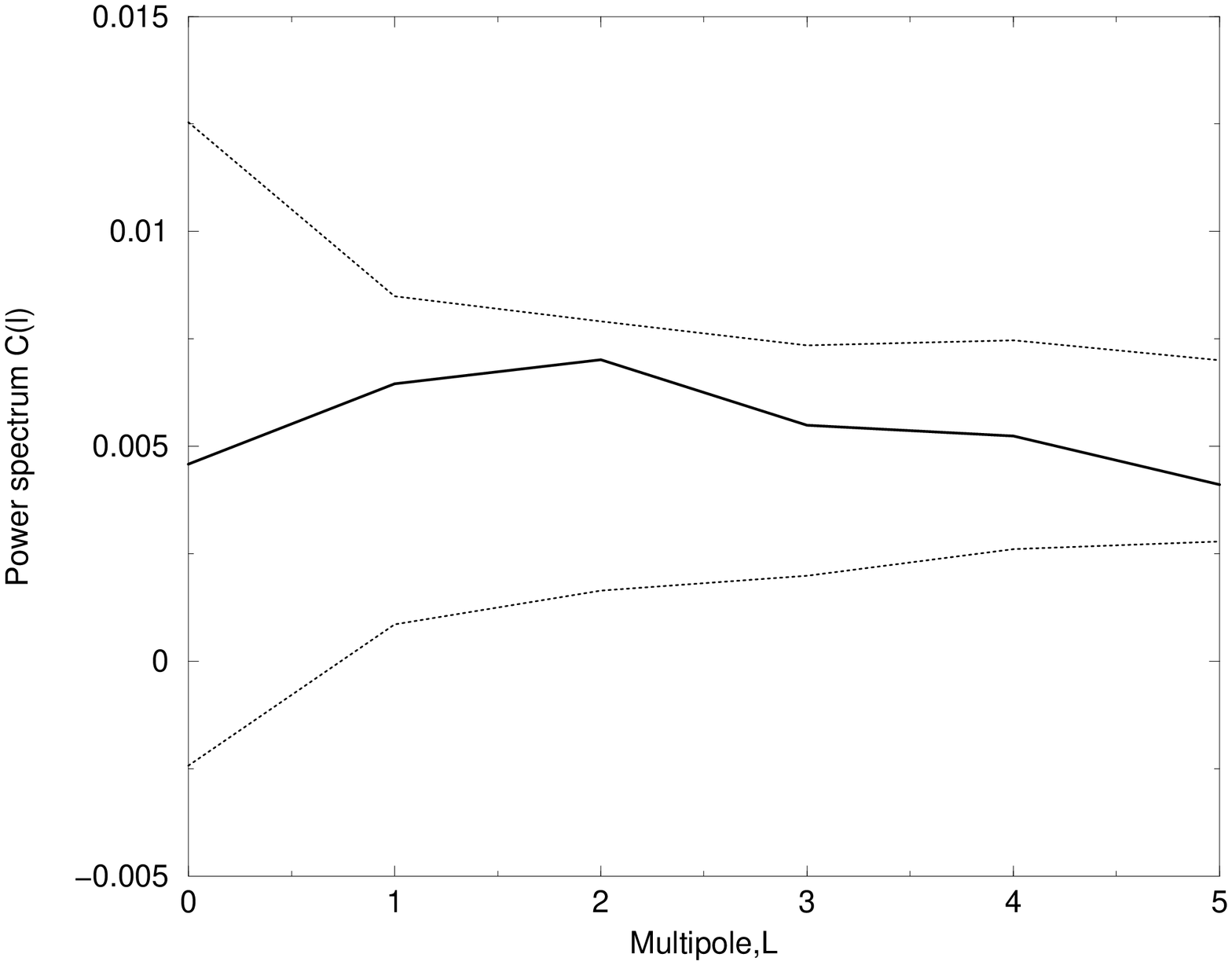,angle=-90,width=7cm}
\psfig{figure=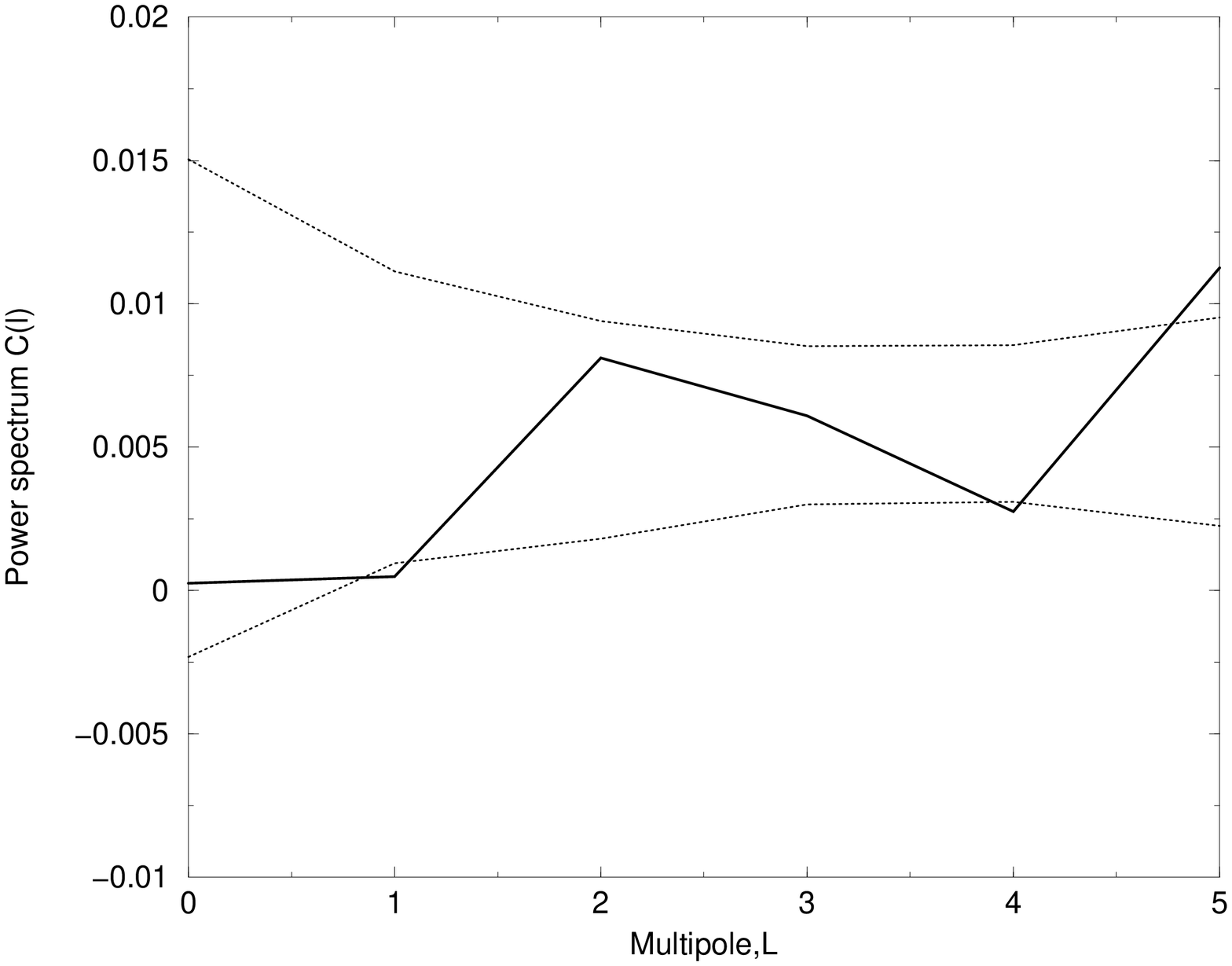,angle=-90,width=7cm}
}
\hbox{
\psfig{figure=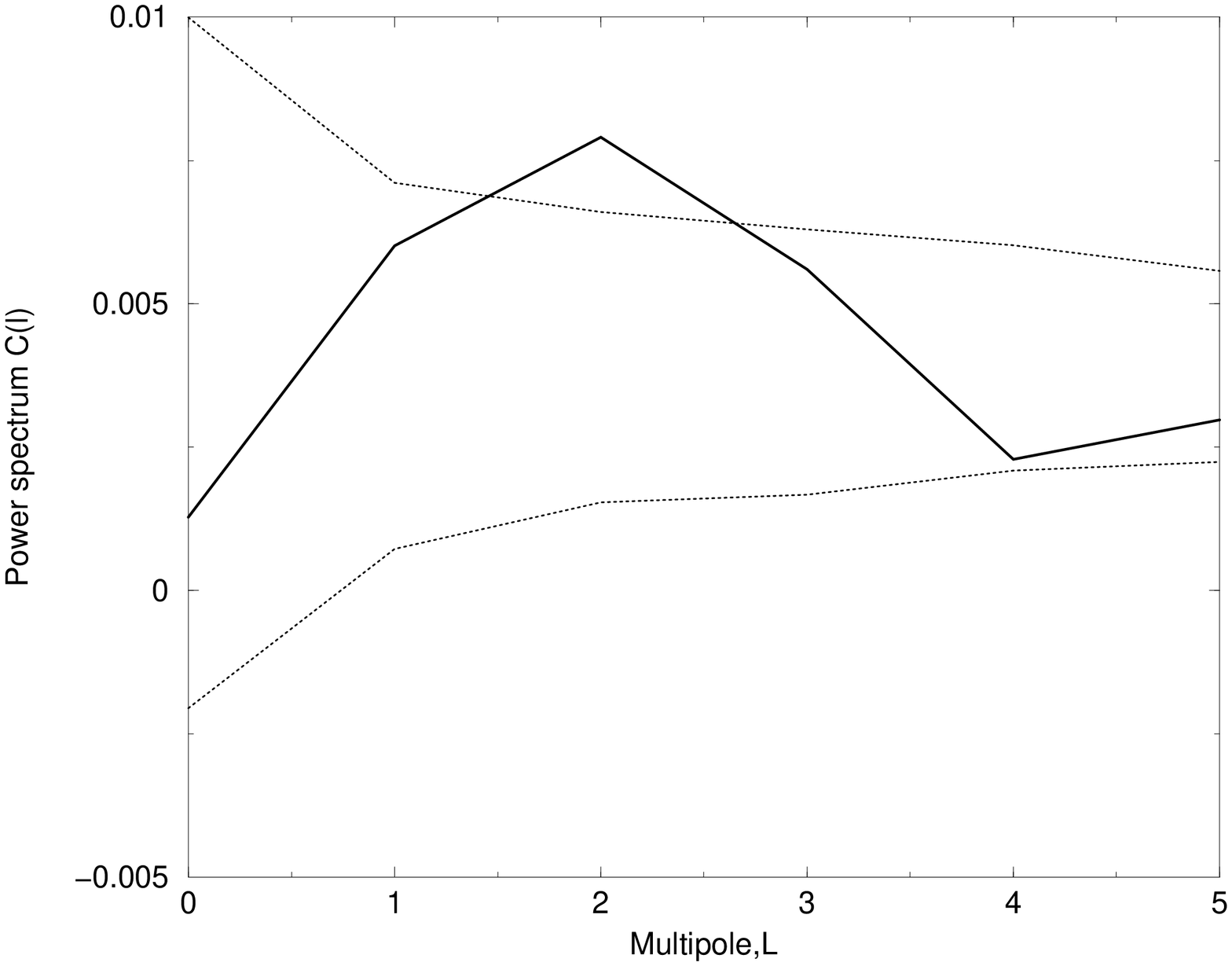,angle=-90,width=7cm}
\psfig{figure=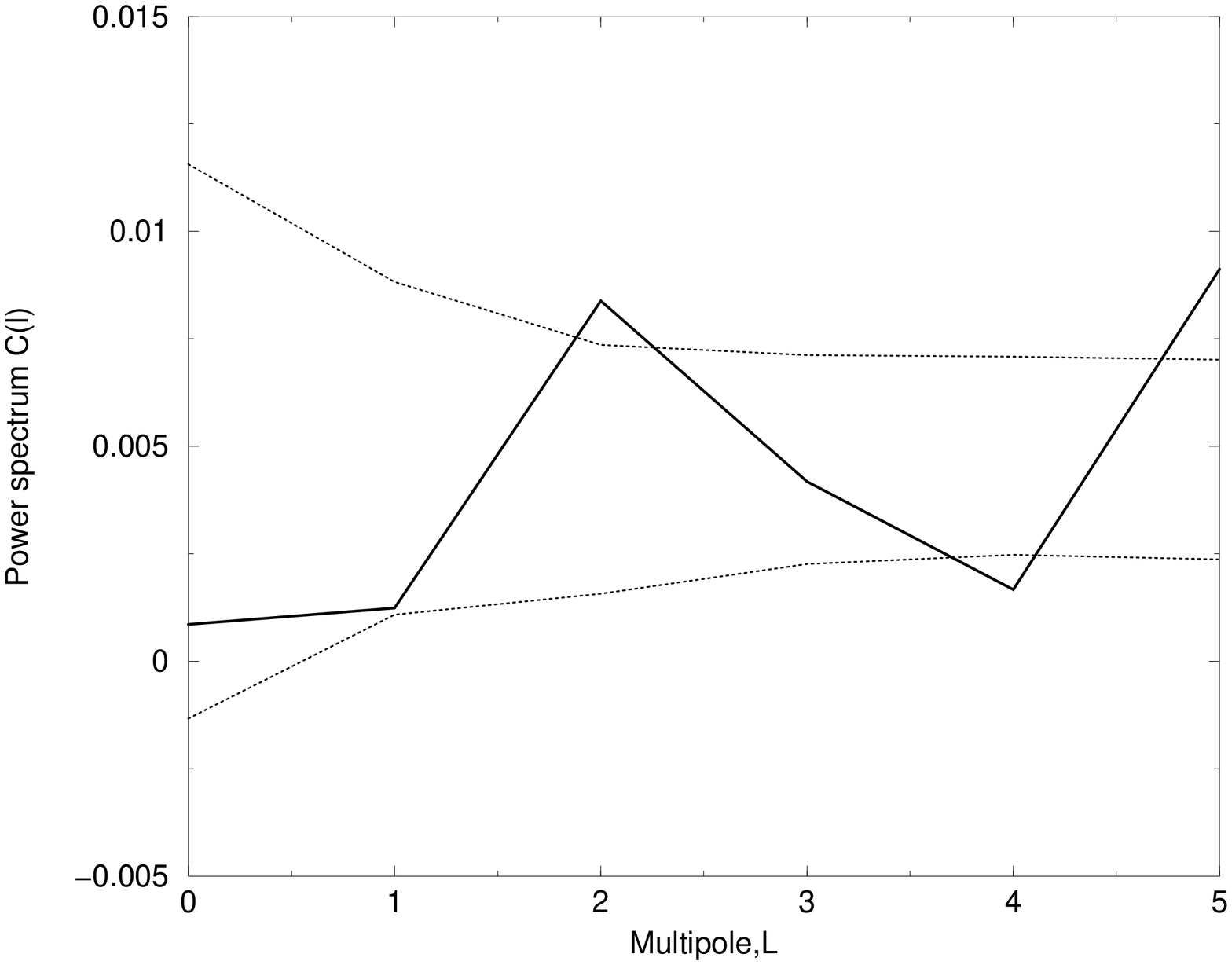,angle=-90,width=7cm}
}
}}
\caption{
Power spectra of mosaic correlation maps with the resolution of $\lmax=5$
for the maps of BATSE GRB locations and CMB distribution (the solid
line). The correlation pixel size is
900\arcmin$\times$900\arcmin.
The left top image shows the correlation spectrum of the BATSE data
($t<2$\,sec)
and CMB without masking. The right top image
contains the correlation spectrum of the BATSE data with $t>2$\,sec
and CMB without masking. The bottom left image
demonstrates the correlation spectrum of the BATSE data with $t<2$\,sec
and CMB with masking. The right bottom image
shows the correlation spectrum of the BATSE data for
and CMB with masking. The images show the
1$\sigma$ dispersion obtained from results of analysis of 200 Gaussian
random realizations of CMB.
}
\label{sp_cor_l5}
\end{figure}

The angular power spectrum permits us
selecting harmonics contributing to the correlation map.
Figs.\,\ref{sp_cor_l26},\ref{sp_cor_l8},\ref{sp_cor_l5}
show power spectra of maps of correlation factors calculated by the
mosaic correlation method for the BATSE and CMB data.

\begin{figure} [!h]
\setcaptionmargin{5mm}
\onelinecaptionstrue
\centerline{\vbox{
\hbox{
\psfig{figure=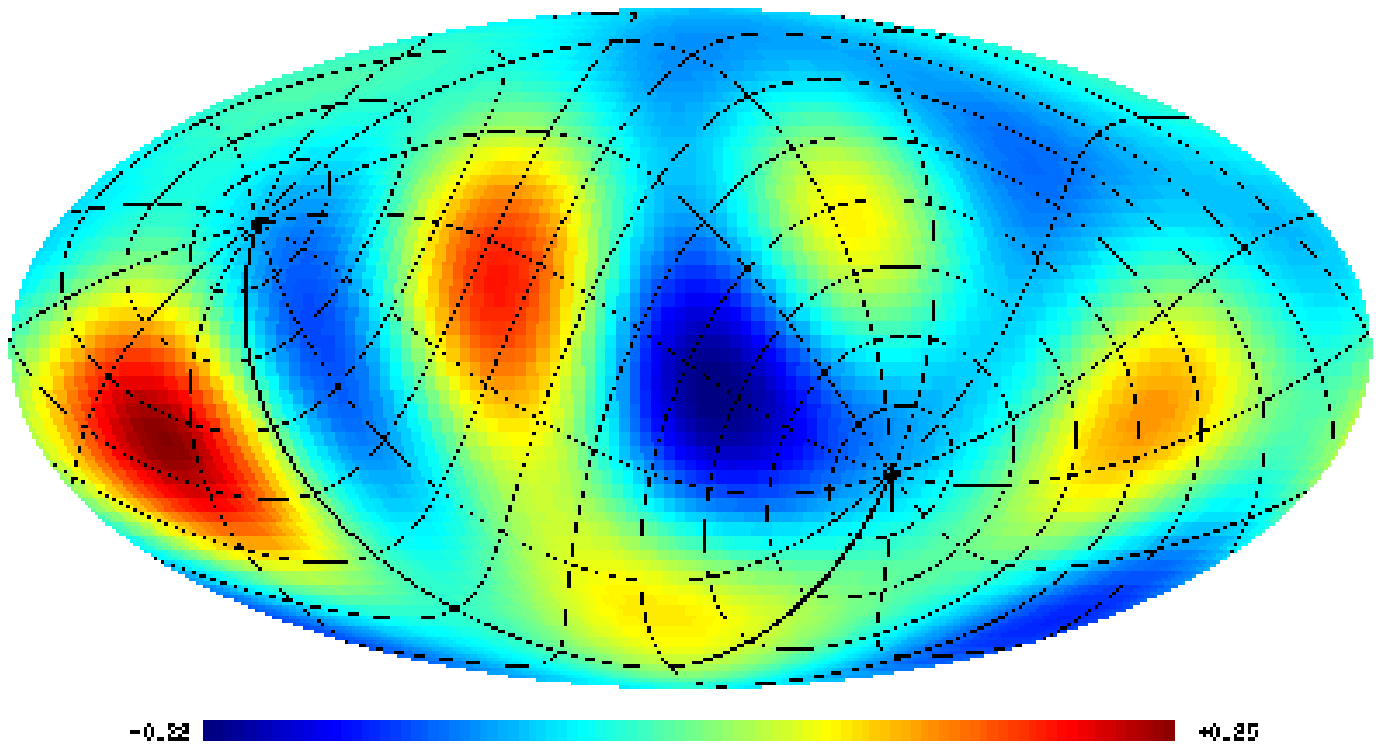,width=8cm}
\psfig{figure=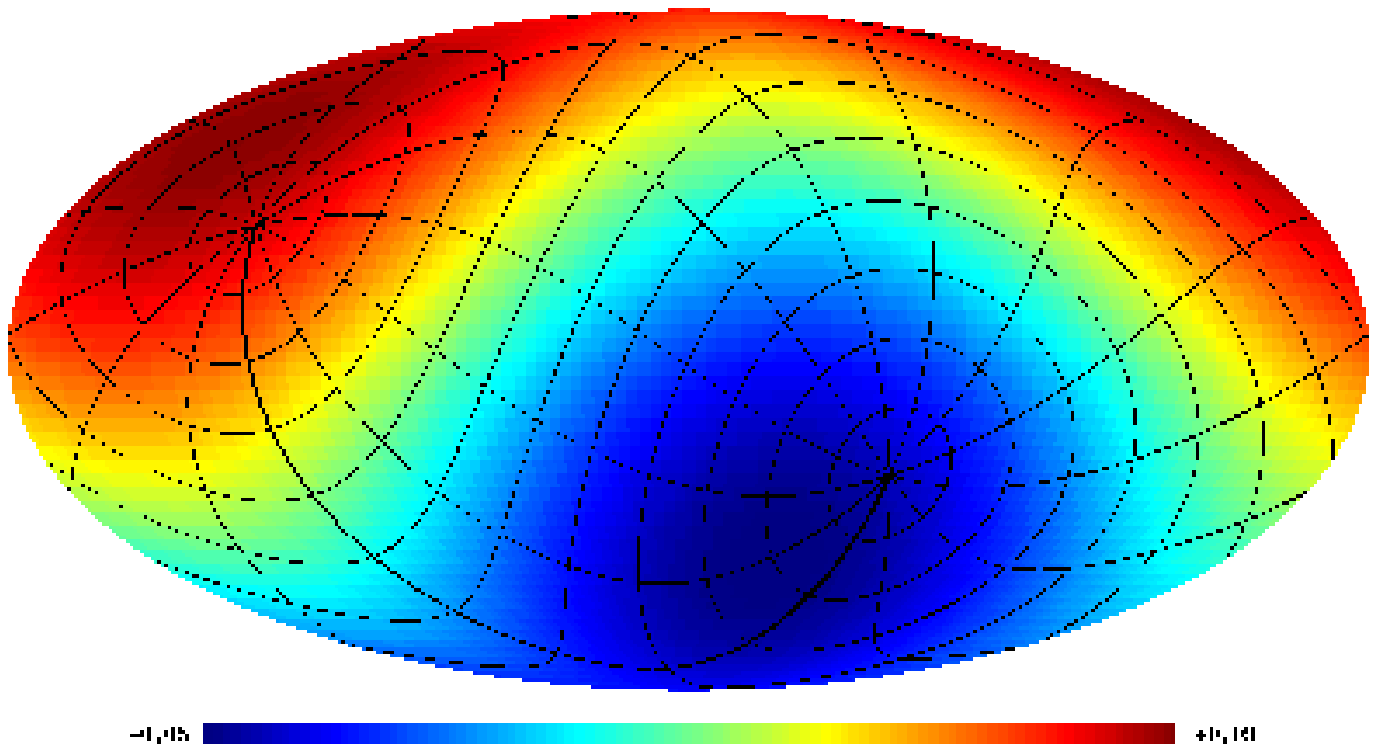,width=8cm}
}
\hbox{
\psfig{figure=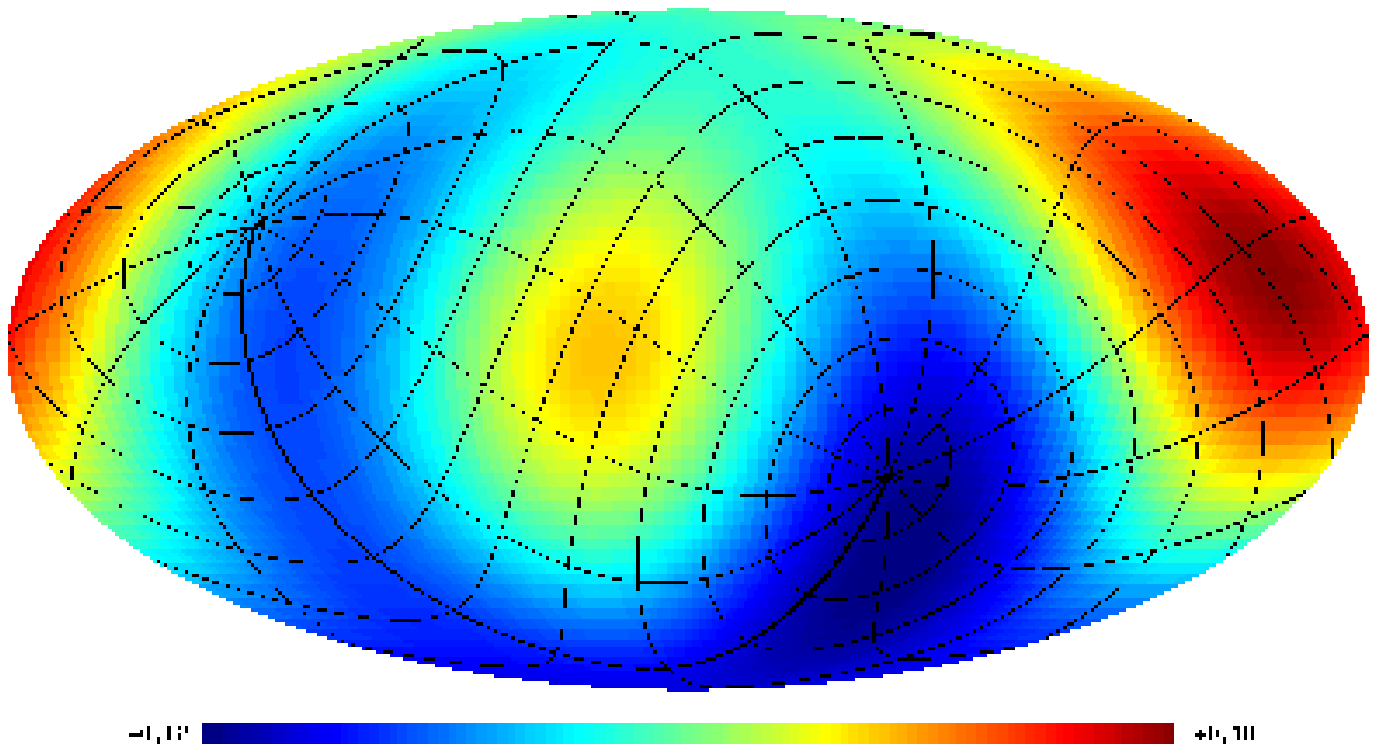,width=8cm}
\psfig{figure=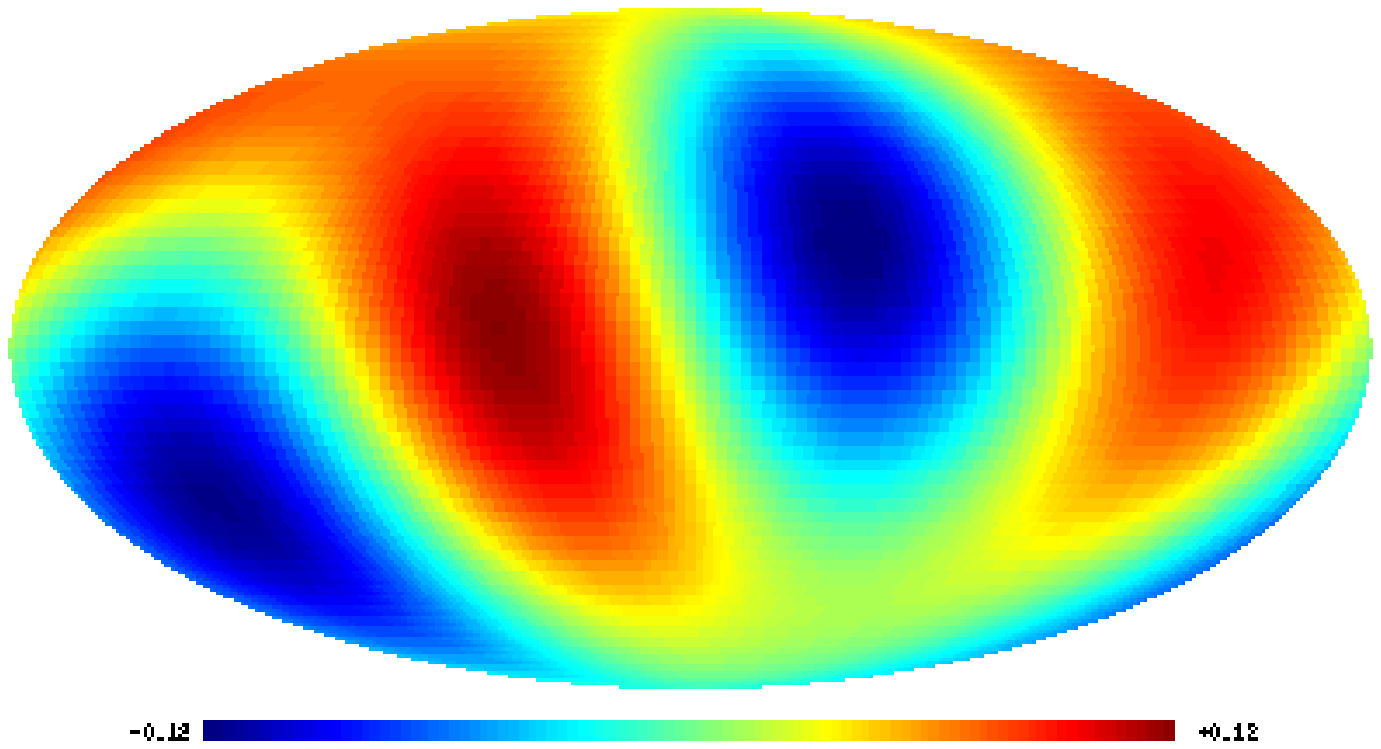,width=8cm}
}
}}
\caption{
Maps of distinguished harmonics in power spectra
(Fig.\,\ref{sp_cor_l26},\ref{sp_cor_l8},\ref{sp_cor_l5}).
The images show multipoles of mosaic
correlations of CMB and GRB locations. Top left: the distinguished
harmonics $\ell=4$
for the correlation window 500\arcmin$\times$500\arcmin for BATSE
data ($t<2$\,sec);
top right: the distinguished harmonic $\ell=1$
for the correlation window 600\arcmin$\times$600\arcmin. Bottom left:
the correlation map for BATSE GRBs ($t<2$\,sec)
and CMB with the correlation window 900\arcmin$\times$900\arcmin ($\ell=2$).
Bottom right: the correlation map for BATSE GRBs ($t>2$\,sec)
and CMB with the correlation window 900\arcmin$\times$900\arcmin ($\ell=2$).
The third map contains the overlaid equatorial coordinate grid.
}
\label{map_sep_mult}
\end{figure}

As is shown in Figs.\ref{sp_cor_l26}, \ref{sp_cor_l8} and \ref{sp_cor_l5},
application of the mask retains location of local maximums in the power
spectra of mosaic maps. In a number of cases, the application of the
mask even amplifies the amplitude of a distinguished harmonic.
Fig.\,\ref{map_sep_mult}
demonstrates examples of such harmonics. The fourth multipole of mosaic
correlation map with the window 500\arcmin$\times$500\arcmin
calculated for the BATSE data ($t<2$\,sec)
contains a feature --- the coldest central spot in the
galactic plane (Fig.\,\ref{map_sep_mult}, left top).
Quadrupole of the correlation map for the BATSE data ($t>2$\,sec)
with the window 900\arcmin$\times$900\arcmin\, (Fig.\,\ref{map_sep_mult}, left
bottom) is sensitive to the equatorial coordinate system. Note that the
change of correlation scale (namely, the size of area in which the
correlation factor is calculated and attributed to a pixel of mosaic
map) changes the power spectrum. Thus, e.g., at transition from the
pixel side of size 500\arcmin\, to that of 600\arcmin\, the harmonic
$\ell=4$
amplitude in the power spectrum passes from position of a local maximum
to a local minimum. This can be caused by the increase of amount of GRB
events in the area of corresponding size.

\section{The averaging of fields}
\label{cmb_stack_grb}

The Plank data allow us using the maps of
higher resolution than the WMAP archives. They can be applied to
estimate a potential signal from ``an average
population GRB''.

\begin{figure} [!th]
\setcaptionmargin{5mm}
\onelinecaptionstrue
\centerline{\vbox{
\hbox{
\psfig{figure=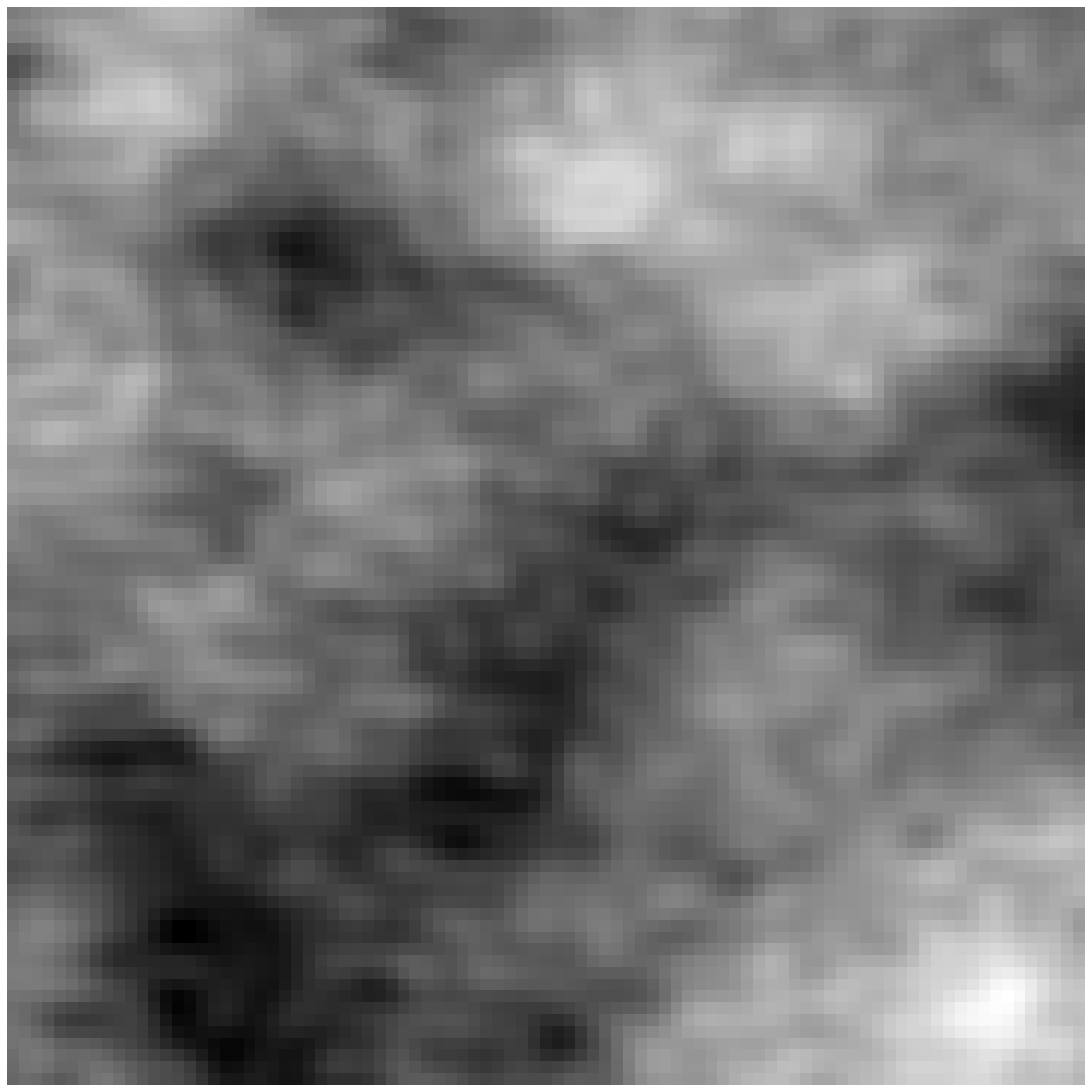,width=7cm}
\psfig{figure=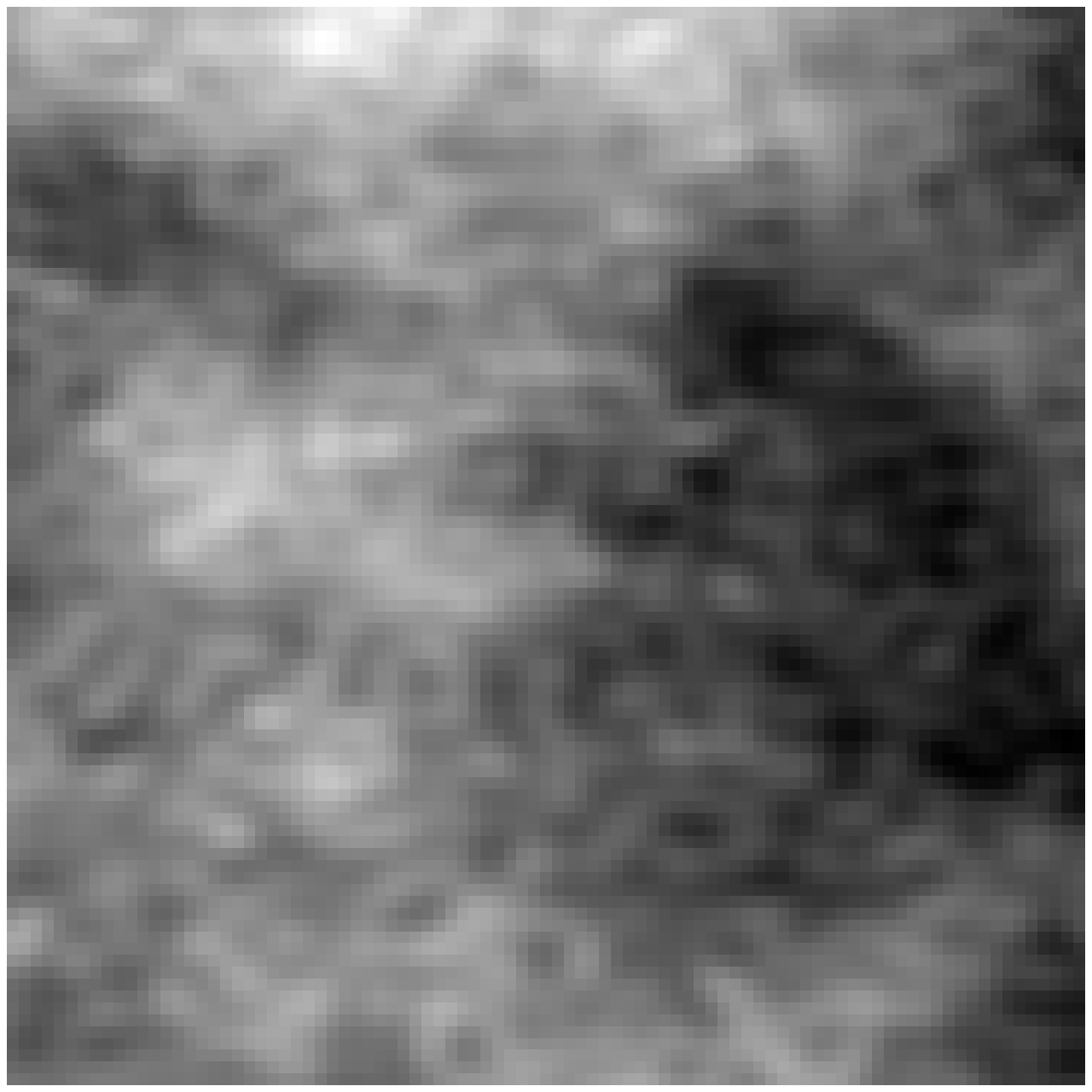,width=7cm}
}
\hbox{
\psfig{figure=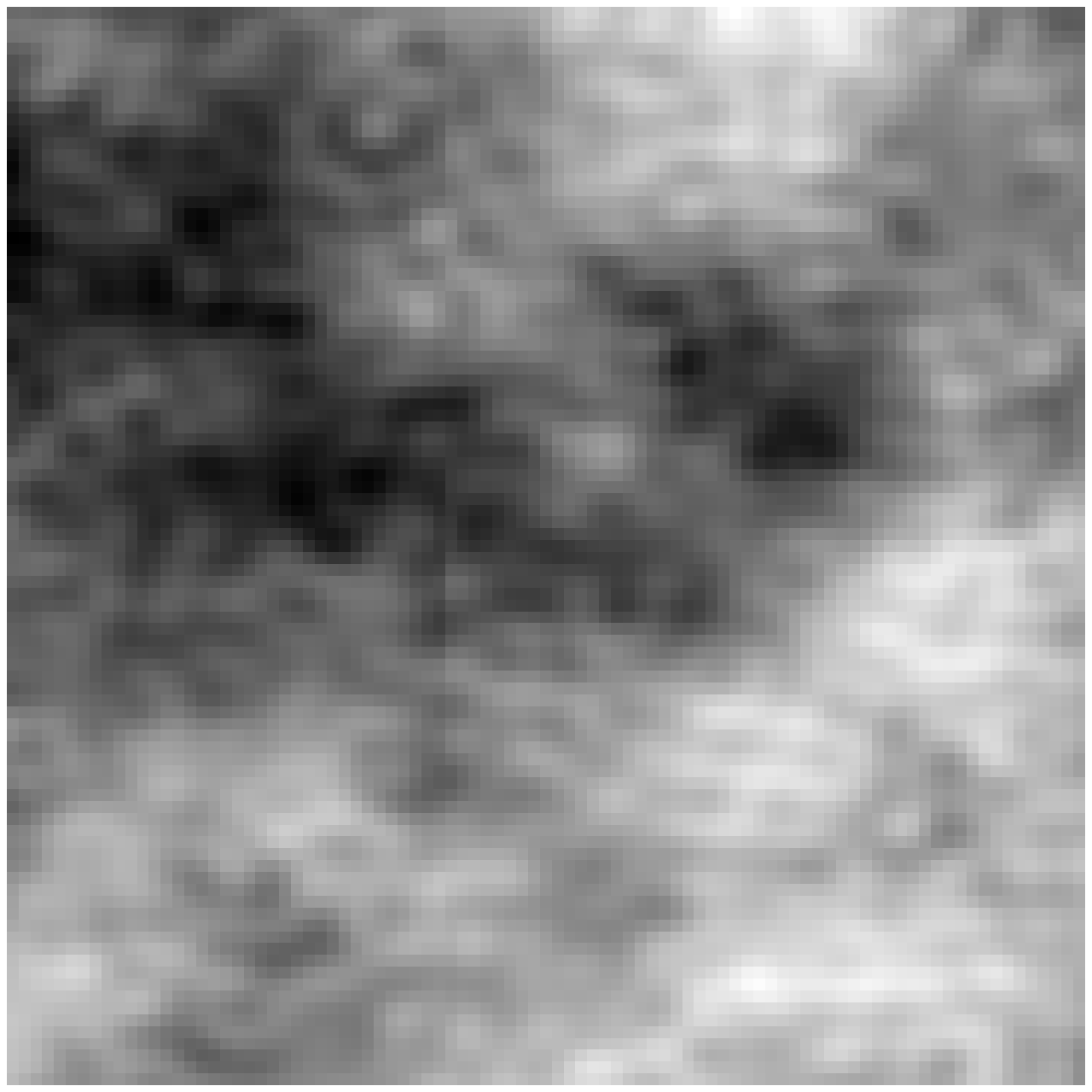,width=7cm}
\psfig{figure=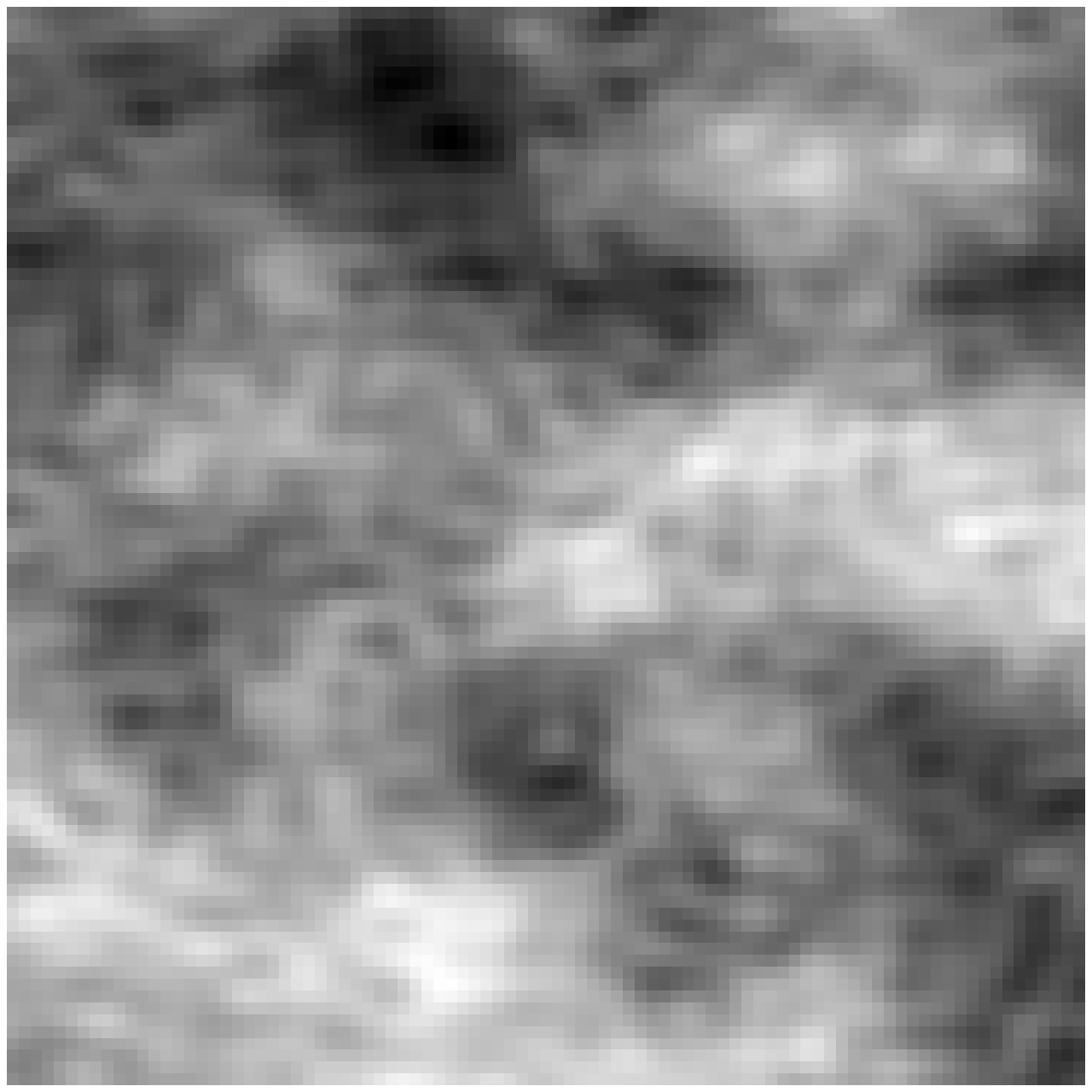,width=7cm}
}}}
\caption{
Results of the averaging of CMB fields of size 2\degr$\times$2\degr\, in GRB
directions. From left to right downward: fields of short BATSE GRBs;
long BATSE GRBs; short and long BeppoSAX GRBs.
}
\label{grb_cmb_stack}
\end{figure}

To do that, areas of an identical linear or
angular size around objects under investigation are selected in
different directions on the celestial sphere. Then they are summed up
to reveal an average signal. Because data on red shifts are not
available, we used areas with identical angular sizes. To avoid
influence of a possible hard-to-consider signal of the Galaxy we
limited ourselves only to regions around GRBs with galactic latitudes
$|b|>20\degr$.
Among BATSE and BeppoSAX samples, this range includes 338 short (68\%
of the initial short BATSE GRBs) and 990 (64\%) long events of the
BATSE catalog, and 51 short (59\% of the BeppoSAX list) and 454 (65\%)
long sources of the BeppoSAX catalog. For every GRB from our subsample
we have chosen a field in the Planck SMICA map of size 2\degr$\times$2\degr\,
in the tangential projection (the pixel size in the area is
$\sim$80\arcsec$\times$80\arcsec). The selected areas were averaged.
Results are shown in Fig.\,\ref{grb_cmb_stack}. Note that the addition
of data from the region $|b|<20\degr$
leads to degradation (blurring)
of images.

The center of each averaged field in Fig.\,\ref{grb_cmb_stack}
gets to a region of a local extremum. Short BATSE and BeppoSAX GRBs are
in regions of a background minimum, and the long ones -- in regions of
a maximum. Ratios of a level of averaged fluctuation of an extremum to
which a generalized GRB gets to a level of noise on averaged maps are
$dS/N$ = -1.65, 1.40, -1.43, 2.01
1.40, -1.43, 2.01
for short and long BATSE GRBs and short and long
BeppoSAX GRBs respectively.

\section{Discussion}

In this paper we investigated the CMB signal
statistics in direction of GRBs from the BATSE and BeppoSAX cataloges.
The Planck SMICA map was used as the CMB map. We applied three
approaches to the study of properties of the sphere distribution of
GRBs. They includes: 1)~analysis of the histogram of Planck CMB signal
values in direction to GRBs, 2) the study of mosaic maps built for GRB
locations and CMB distribution, 3) the study of an average response on
the CMB map in the region of an ``average population
GRB''.

Application of the first two methods
demonstrates that the correlation between GRB and CMB is caused, at
least partially, by a signal in the equatorial coordinate system. This
agrees with results of our previous work \cite{v_grb}. This relation can be
caused by modulation of the CMB signal observed in the point L2 by
microwave radiation of the Earth through the far antenna beam lobes.
Deviations in distribution of GRBs towards the equatorial system are
caused by non-uniform sky sensitivity (the time of signal accumulation)
of the receiving equipment of gamma-ray satellite observatories
rotating around the Earth and always directed at the opposite side from
it. Then a distinguished character of the equatorial coordinate system
appears naturally. Note that indication to the presence of signs of the
equatorial coordinate system (e.g., location of equatorial spots) in
CMB data both for WMAP and Planck maps was already discussed in a
number of papers \cite{v_grb,planck_wmap,planck_est}.
Besides, radiation of the Earth can be
not a single factor. Another discussed reason can be the modulation of
the solar wind by the Earth magnetic field passing through the point
L2. It should be added that such effects which are not detectable by
the standard analysis could be a source of
the secondary gaussianity observed at low harmonics
\cite{schwarz2007,kim_nas2010b,nong_ufn,nas_kouper,copi_planck_wmap}.

The third method applied by us has shown
that there is an insignificant difference at the level $|S/N|>1.4$
which can randomly occur in $\lesssim20$\% cases for the Gaussian noises in
distribution of average CMB signal in GRB directions. As this takes
place, short ($t<2$\,sec)
GRBs in an averaged field gets to a local background minimum, and long
($t>2$\,sec)
GRBs -- to a local maximum. If we assume that short
GRBs arise in old elliptic galaxies formed from merging less massive
galaxies and located in galaxy clusters, then a local minimum can be
due to the Zeldovich--Syunyaev effect \cite{zs}.
The getting of long GRBs to a
local maximum of CMB distribution could be caused by another effect. If
long GRBs are related to supernovae explosions, i.e. with starforming
galaxies, then even in spite of location in a galaxy cluster the proper
microwave emission of a galaxy containing dust and gas would prevail
over effects of surrounding and lead to appearance of a local maximum
in CMB maps. This effect could be tested by means of more sensitive
data of the Planck experiment in the next publication release which is
expected in the second half of 2014.

\bigskip
{\small
{\bf Acknowledgments}.
The authors are grateful to EAS for the open access to results of
observations and processing of data in Planck Legacy Archive.
The authors thank T.N.Sokolova for reading, correcting and translating
this text.
For analysis of extended emission on sphere we used the package GLESP
\cite{glesp2,glesp,glesp1}.
M.L.Kh. and O.V.V. thank RFBR for a partial support of the
project study by the RFBR grant 13-02-00027.
}


\begin{thebibliography}{}

\bibitem{sdssIII}
D.~J.~Eisenstein, D.~H.~Weinberg, E.~Agol, et al.,
AJ {\bf 142}, 72 (2011), arXiv:1101.1529.

\bibitem{swe}
R.~K.~Sachs and A.~M.~Wolfe,
ApJ {\bf 147}, 73 (1967).

\bibitem{zs}
R.~A.~Sunyaev and Ya.~B.~Zeldovich,
Astrophys. Sp. Sci. {\bf 7}, 3 (1970).

\bibitem{princ_cosm}
P.~J.~E.~Peebles,
{\it Principles of Physical Cosmology} (Princeton Univ. Press, 1993).

\bibitem{cosm_princ}
~J.~A.~Peacock,
{\it Cosmological Physics} (Cambridge Univ. Press, 2000).

\bibitem{rudnick}
L.~Rudnick, S.~Brown, and  L.~R.Williams,
{\apj} {\bf 671}, 40 (2007).

\bibitem{springel}
V.~Springel, C.~S.~Frenk, and S.~D.~M.~White,
Nature {\bf 440}, 1137 (2006).

\bibitem{yadav}
J.~Yadav, S.~Bharadwaj, B.~Pandey, and T.~R.~Seshadri, MNRAS
{\bf 364}, 601 (2005), astro-ph/0504315

\bibitem{sarkar}
P.~Sarkar, J.~Yadav, B.~Pandey, and S.~Bharadwaj, MNRAS {\bf 399},
L128 (2009).

\bibitem{labini_baryshev}
F.~Sylos Labini and Y.~V.~Baryshev, JCAP {\bf 6}, 021 (2010).

\bibitem{jubilee}
W.~A.~Watson, I.~T.~Iliev, J.~M.~Diego, et al.,
MNRAS {\bf 437}, 3776 (2014), arXiv:1305.1976.

\bibitem{cmb_sdss_cell}
Ya.~V.~Naiden and O.~V.~Verkhodanov,
Astrophys. Bull. {\bf 68}, 471 (2013).

\bibitem{hierar}
J.~F.~Navarro, C.~S.~Frenk, and S.~D.~M.~White,
ApJ {\bf 490}, 493 (1997), astro-ph/9611107.

\bibitem{bh_rita}
M.~L.~Khabibullina and O.~V.~Verkhodanov,
Astronomy Reports {\bf 55}, 302 (2011), arXiv:1108.4506.

\bibitem{amati1}
L. Amati, et al.,
\aaa {\bf 390}, 81 (2002).

\bibitem{amati2}
L. Amati, et al.,
\mnras {\bf 391}, 557 (2008).

\bibitem{bepposax}
D.~Riccia, F.~Fioreb, and P.~Giommia,
Nuclear Physics B - Proc. Suppl.
{\bf 69}, 618 (1999).

\bibitem{batse}
W.~S.~Paciesas, C.~A.~Meegan, G.~N.~Pendleton, et al.
McCollough,
    Astrophys. J. Suppl. {\bf 122}, 465 (1999),
astro-ph/9903205.

\bibitem{vavrek}
R.~Vavrek, et al.,
in ``A Workshop Celebrating the First Year of the HETE Mission'',
AIP Conf. Proc. {\bf 662}, 163 (2003).

\bibitem{grb_apm}
L.~L.~R.~Williams  and N.~Frey,
{\apj} {\bf 583}, 594 (2003).

\bibitem{grg_anis}
A.~M\'esz\'aros and J.~Stocek,
{\aa} {\bf 403}, 443 (2003), astro-ph/0303207.

\bibitem{Bernui}
A.~Bernui, I.~S.~Ferreira, and C.~A.~Wuensche,
{\apj} {\bf 673}, 968 (2008), arXiv:0710.1695.

\bibitem{grb_vor}
A.~M\'esz\'aros, L.~G.~Bal\'azs, Z.~Bagoly and P.~Veres,
GAMMA-RAY BURST: Sixth Huntsville Symposium, AIP Conf.
Proc. {\bf 1133}, 483 (2009), arXiv:0906.4034.

\bibitem{v_grb}
O.~V.~Verkhodanov, V.~V.~Sokolov, M.~L.~Khabibullina and S.~V.~Karpov,
Astrophys. Bull. {\bf 65}, 238 (2010), arXiv:1009.3720.

\bibitem{raikov_orlov}
V.~N.~Yershov, V.~V.~Orlov, and A.~A.~Raikov,
MNRAS {\bf 423}, 2147 (2012).

\bibitem{cormap}
O.~V.~Verkhodanov, M.~L.~Khabibullina and E.~K.~Majorova,
Astrophys. Bull. {\bf 64}, 263 (2009).
\bibitem{corr_ecl}
O.~V.~Verkhodanov and M.~L.~Khabibullina,
 Astrophys. Bull. {\bf 65}, 390 (2010).

\bibitem{wmap7ytem}
N.~Jarosik, C.~L.~Bennett, J.~Dunkley, et al.,
{\apjs}, submitted (2010), arXiv:1001.4744.

\bibitem{planck_rev}
Planck Collaboration: P. A. R. Ade, et al.,
\aaa, submitted (2013), arXiv:1303.5062.

\bibitem{planck_sep}
Planck Collaboration: P. A. R. Ade, et al.,
\aaa, submitted (2013), arXiv:1303.5072.

\bibitem{glesp2}
A.~G.~Doroshkevich, O.~B.~Verkhodanov, P.~D.~Naselsky, et al.,
{Intern. J. Mod. Phys.} {\bf 20}, 1053 (2011), arXiv:0904.2517.

\bibitem{planck_wmap}
O.~V.~Verkhodanov,
Astrophys. Bull. {\bf 69}, accepted (2014).


\bibitem{planck_est}
Ya.~V.~Naiden, O.~V.~Verkhodanov,
Astrophys. Bull. {\bf 69}, accepted (2014).

\bibitem{schwarz2007}
 A.~Rakic and D.~J.~Schwarz,
  {Phys. Rev. D} {\bf 75}, 103002 (2007).

\bibitem{kim_nas2010b}
  Ja.~Kim and P.~Naselsky,
Phys. Rev D {\bf 82}, 063002 (2010).

\bibitem{nong_ufn}
O.V.~Verkhodanov,
Phys. Usp. {\bf 55}, (2012).

\bibitem{nas_kouper}
M.~Hansen, J.~Kim, A.~M.~Frejsel, et al.,
JCAP {\bf 10}, 059 (2012).

\bibitem{copi_planck_wmap}
C.~J.~Copi, D.~Huterer, D.~J.~Schwarz, and G.~D.~Starkman,
arXiv:1311.4562 (2013).

\bibitem{glesp}
A.~G.~Doroshkevich, P.~D.~Naselsky, O.~V.~Verkhodanov et al.,
{Int. J. Mod. Phys.} {\bf 14}, {275} ({2003}), {astro-ph/0305537}.

\bibitem{glesp1}
O.~V.~Verkhodanov , A.~G.~Doroshkevich, P.~D.~Naselsky, et al.,
Bull. SAO {\bf  58}, 40 (2005).

\end{thebibliography}
\end{document}